\documentclass[journal,letterpaper]{IEEEtran}

\usepackage{cite}
\usepackage{graphicx}
\usepackage{subfigure}
\usepackage{amsmath} 
\usepackage{multirow}
\usepackage{amssymb}

\begin{document}

\title{Ionization cross sections for low energy electron transport}

\author{Hee Seo, Maria Grazia Pia, Paolo Saracco and Chan Hyeong Kim% <-this % stops a space
\thanks{Manuscript received May 18, 2011. This research was partly supported by National
Nuclear R\&D Program through the National Research Foundation of Korea (NRF) funded by the Ministry of Education, Science and Technology (No. 2010-0023825, 2010-0028913).}% <-this % stops a space
\thanks{H. Seo and C. H. Kim are with 
	the Department of Nuclear Engineering, Hanyang University, 
        Seoul 133-791, Korea 
	(e-mail: shee@hanyang.ac.kr; chkim@hanyang.ac.kr).}
\thanks{M. G. Pia and P. Saracco are with 
	INFN Sezione di Genova, Via Dodecaneso 33, I-16146 Genova, Italy 
	(phone: +39 010 3536328, fax: +39 010 313358,
	e-mail: MariaGrazia.Pia@ge.infn.it, Paolo.Saracco@ge.infn.it).}
}

\maketitle

\begin{abstract}
Two models for the calculation of ionization cross sections by electron impact
on atoms, the Binary-Encouter-Bethe and the Deutsch-M\"ark models, have been
implemented; they are intended to extend and improve Geant4
simulation capabilities in the energy range below 1 keV.
The physics features of the implementation of the models are described,
and their differences with respect to the original formulations are discussed.
Results of the verification with respect to the original theoretical sources and of
extensive validation with respect to experimental data are reported.
The validation process also concerns the ionization cross sections included in
the Evaluated Electron Data Library used by Geant4 for low energy
electron transport.
Among the three cross section options, the Deutsch-M\"ark model is identified as
the most accurate at reproducing experimental data over the energy range subject
to test.

\end{abstract}

% -----------------------------------------------------------------------

\section{Introduction}
\label{sec_intro}
\PARstart{V}{arious} 
experimental research topics require the capability of simulating electron
interactions with matter over a wide range - from the nano-scale to the
macroscopic one: some examples are ongoing investigations on
nanotechnology-based particle detectors, scintillators and gaseous detectors,
radiation effects on semiconductor devices, background effects on X-ray
telescopes and biological effects of radiation. 

Physics tools for the simulation of electron interactions are available 
%and well established 
in all Monte Carlo codes based on condensed and mixed 
transport schemes \cite{berger}, like EGS \cite{egs5,egsnrc}, FLUKA
\cite{fluka1,fluka2}, Geant4 \cite{g4tns,g4nim}, MCNPX \cite{mcnpx}, Penelope
\cite{penelope} and PHITS \cite{phits}.
General-purpose Monte Carlo codes based on these transport schemes
typically handle particles with energy above 1 keV;
Geant4 and Penelope extend their coverage below this limit.

In the lower energy end below 1 keV, so-called ``track structure'' codes
handle particle interactions based on discrete transport schemes; 
they provide simulation capabilities limited to a single target, or a small number of 
target materials, and are typically developed for specific application purposes.
Some examples of such codes are OREC \cite{orec}, PARTRAC \cite{partrac},
Grosswendt's  Monte Carlo for nanodosimetry \cite{ptbmc}, TRAMOS \cite{tramos},
and Geant4 models for microdosimetry simulation in water \cite{tns_dna}.

The developments described in this paper address the problem of endowing
a general purpose, large scale Monte Carlo system for the first time with the
capability of simulating electron impact ionisation down to the scale of a few
tens of electronvolts for any target element.
For this purpose, models of electron impact ionization cross sections suitable
to extend Geant4 capabilities in the low energy range have been implemented and
validated with respect to a large set of experimental measurements. 

The validation process, which involves experimental data pertinent to 
more than 50 elements, also addresses the ionization cross sections encompassed 
in the Evaluated Electron Data Library (EEDL) \cite{eedl}, which are used in
Geant4 low energy electromagnetic package \cite{lowe_chep,lowe_nss}. 
To the best of the authors' knowledge, this is the first time that EEDL
is subject to extensive experimental benchmarks below 1 keV.

% -----------------------------------------------------------------------

\section{Overview of electron ionization in Geant4}
\label{sec_g4models}

The Geant4 toolkit provides various implementations of electron ionization based
on a condensed-discrete particle transport scheme. Two of them, respectively
based on EEDL \cite{lowe_e} and on the analytical models originally developed
for the Penelope \cite{penelope} Monte Carlo system, are included in the low
energy electromagnetic package; another implementation is available in the
standard \cite{standard} electromagnetic package.
In addition, a specialized ionization model for interactions with thin layers of
material, the photoabsorption-ionization (PAI) model \cite{pai}, is implemented
in Geant4.

The EEDL data library tabulates electron ionization cross sections in the energy
range between 10 eV and 100 GeV; nevertheless, due to intrinsic limitations of
the accuracy of EEDL and its companion Evaluated Photon Data Libray (EPDL)
\cite{epdl97} highlighted in the documentation of these compilations,
the use of Geant4 low energy models based on them was originally recommended for incident
electron energies above 250 eV \cite{lowe_e}.
This limit of applicability was an ``educated guess'' rather than a rigorous
estimate of validity of the theoretical calculations tabulated in EEDL and EPDL. 
%The Penelope-like models currently implemented in Geant4 reflect the status of
%Penelope as in its 2001 version; more recently, cross sections for electron
%impact ionization of inner shells (K, L and M) based on \cite{bote} have been
%included in the 2008 Penelope version.
The lower energy limit of Penelope's applicability is generically indicated by
its authors as ``a few hundred electronvolts'' \cite{penelope2008}.
%both in the most recent
%version of the code  and in the 2001 version \cite{penelope2001}.
The lower limit of applicability of Geant4 standard electromagnetic package is 1 keV.
 
The validation of Geant4 models for electron transport based on the EEDL data
library and on Penelope-like models is documented in \cite{tns_sandia}
for what concerns the energy deposition in extended media.

Ionization models suitable for microdosimetry simulation, which operate in a
discrete particle transport scheme, are available in Geant4 for electron
interactions in water \cite{tns_dna}; they are applicable for energies
down to the electronvolt scale.
The cross section models implemented in that context are specific to one
material (liquid water); due to lack of pertinent experimental data, their
validation is still pending.

%The implemented models in principle have the capability of calculating cross
%sections for the ionization of individual shells; nevertheless, the validation
%process described in this paper is limited to comparison with available total cross
%section measurements, while the validation of cross section calculations for
%the ionization of specific shells will be the object of a dedicated paper.
%can deal with
%multiple ionized atoms and molecules as well; further extensions of the software
%to account for these capabilities will be the object of future development
%cycles, once the validity of the approach has been assessed.

% ------------------------------------------------------------------------
\section{Software development process}
\label{sec_sw}

The developments described in this paper adopt an iterative-incremental process
consistent with the Unified Process \cite{up}.
The features and results documented in the following sections correspond to the
first cycle of a wider project concerning the development and assessment of
models for multi-scale electron transport \cite{nano5_mc2009,nano5_nss2009},
which is motivated by multi-disciplinary experimental applications.
A characterizing feature of the Unified Process, which differentiates it from
other widespread software life-cycle models adopting a waterfall
\cite{waterfall} approach, is the production of concrete deliverables even at
intermediate stages of the project: this development cycle has
enabled the validation and comparative evaluation of different physics models,
and has produced a data library usable in multiple environment.

%will not affect the main features described in the following sections.

%at this stage, the emphasis of the development was focused on
%the assessment of the physics capabilites of the implemented models.

The software described and validated in the following sections is intended
for release in the Geant4 toolkit following the publication of this paper.
%Extensions and refinements will be the object of further
%development cycles based on the feedback from the first operational version.

% -----------------------------------------------------------------------

\subsection{Physics models}

The developed software tools concern the calculation and validation
of cross sections for the ionization of an atom by electron impact at energies
below 10 keV: they are focused on the total cross section for single ionization,
namely the emission of one electron from a neutral atom, irrespective of the
shell from which the electron is emitted.
Collective phenomena and solid state effects are outside their scope, as well as
the treatment of electron interactions with matter other than ionization.

%which adopts an iterative-incremental software process \cite{up}.

%They intend to extend the simulation capabilities currently available in 
%Geant4 by implementing ionization cross section models 

Two ionization cross section models, which specifically address the low energy
range, have been implemented: the Binary-Encounter-Bethe (BEB) model
\cite{bebKim1994} and the Deutsch-M\"ark (DM) \cite{dmDeutsch1987} model.
Their accuracy at reproducing experimental data is extensively investigated in
the following sections, along with the validation of the ionization cross
sections included in EEDL, currently used by Geant4.

The theoretical models adopted in the software implementation have a wider scope
of applicability: they can calculate cross sections for the ionization of
individual shells as well as for multiple ionized atoms and for molecules.
The assessment of these additional capabilities is intended to be the object of 
further cycles in the software process, to be documented in dedicated papers.

% \cite{epistemic}.

% ------------------------------------------------------------------------
\subsection{Software design}
\label{sec_design}

The software adopts a policy-based class design \cite{alexandrescu}; this 
technique was first introduced in a general-purpose Monte Carlo system in
\cite{tns_dna}.
This programming paradigm allows the exploitation of the developed models in
different contexts with great versatility, without imposing the burden of
inheritance from a pre-defined interface, since policies are syntax-oriented,
rather than signature-oriented.

Preliminary evaluations \cite{em_chep2009,em_nss2009} indicate that policy-based
design contributes to achieve better computational performance than conventional
inheritance in the calculation of cross sections, thanks to compile-time
binding.
This feature is relevant to the computationally intensive domain of Monte Carlo
particle transport, especially at low energies, where discrete transport
methods, involving a large number of steps and accounting for individual
collisions with the interacting medium, may be required for precise calculation.
% like the calculation of cross sections, which is
%performed at each step in matter to determine the mean free path associated with
%a physics process.

The classes responsible for the calculation of ionization cross sections conform
to the cross section policy defined in \cite{em_chep2009,em_nss2009}.
% through duck typing.
The policy consists of a \textit{CrossSection} function, whose arguments are
associated with characteristics of the incident particle and the target atom; it
returns the value of the cross section calculated in the conditions specified
by the arguments.

% ------------------------------------------------------------------------
\subsection{Implementation}
\label{sec_implem}

The Binary-Encounter-Bethe and Deutsch-M\"ark models are implemented
according to the analytical formulations devised by their original authors.
The implementation is based on the latest revisions of the models
available in the literature.

%equations (\ref{eq_beb}) and (\ref{eq_dms}) respectively.

Both cross section calculations involve a few atomic parameters; the
Deutsch-M\"ark model also involves some empirical parameters derived from fits
to experimental data.
The software implementation is based on the parameters documented in the
literature by the original authors of the theoretical models; alternative
sources were used in the software implementation, when the original sources are
not publicly accessible, or not specified.
The differences of the implementation with respect to the original models and
their implications are discussed in detail in the following sections.

% -----------------------------------------------------------------------
\subsection{Software Verification and Validation}
\label{sec_verification}

The verification process ascertained whether the cross sections calculated by
the software implementation of the BEB and DM models are consistent with the
original values published by the respective authors.
The validation process involves comparisons with experimental data
to ascertain whether the two new models and the EEDL data library 
describe electron ionization cross sections accurately.
%EEDL cross sections, which consist of tabulated data, are input directly 
%to the validation process.

%The correctness of the implementation has been verified by comparing
%cross section values calculated by the software with original references
%of the physics models documented in the literature.
%The verification process was necessarily limited to the sample of target
%elements and incident electron energies for which cross section values
%calculated by the original authors are reported in the literature.

%The validation of the physics models and the comparative evaluation of their
%accuracy are a major part of the study; the method and results are documented in
%detail in section \ref{sec_validation} and \ref{sec_results}.

%and the identification of epistemic uncertainties.

The software verification and validation follow the guidelines of the
pertinent IEEE Standard \cite{ieee_vv}.
Nevertheless, these two processes are intertwined: the verification of compatibility
with original calculations cannot be completly disjoint from the assessment
whether any detected discrepancies would affect the model accuracy 
significantly with respect to experimental data.
%similarly, the analysis of possible sources of systematic effects takes
%experimental data as a reference to estimate the relevance of any detected
%deviation.

Some of the reference cross section values for the verification and validation
process, which are not available in numerical format, were digitized from
published plots by means of the Engauge \cite{engauge} software.
The uncertainty associated with the digitization process was evaluated by
digitizing plots whose entries were known a priori; it is smaller than 1\%.
%In the majority of the verification cases the implementations of the
%Binary-Encounter-Bethe and Deutsch-M\"ark models reproduced the published
%reference values within the uncertainties associated with the digitization
%process.

%The evaluation of the simulation models includes an analysis of possible 
%systematic effects originating from the options adopted in the implementation 
%and of possible sources of bias in the validation process.

%is reported in section \ref{sec_systematic}.

%The compliance of the cross sections calculated by the software with published
%values of the original theoretical models is analyzed in section
%\ref{sec_verification}; 

% ------------------------------------------------------------------------
\section{The Binary-Encounter-Bethe model}
\label{sec_beb}

\subsection{Theoretical background}

The Binary-Encounter-Bethe model is a simplified version of the
Binary-Encounter-Dipole (BED) model \cite{bebKim1994} proposed by Kim and Rudd
to calculate electron impact ionisation cross sections.

%The BED model is an impact theory for an incident electron interacting with a bound target electron;
%it combines the Mott cross section \cite{mott} modified by the binary-encounter
%theory \cite{vriens} for hard or close collision 
%%for low incident energies 
%with the Bethe theory \cite{bethe} for soft or distant collision.
%%for high energies.

The BED model combines a modified form of the Mott cross section \cite{mott}
with the Bethe theory \cite{bethe}.
Mott theory describes the collision of two free electrons: it is expected to
give good results for small impact parameters, or hard collisions, but it must be
corrected for large impact parameters, or soft collision, where dipole
interaction is prevalent, especially at high incident electron energies.
Several attempts have been made to simultaneously describe hard and soft
collisions \cite{jain,khare_meath,khare_kumar,kaushik}, but they generally
failed in finding the proper mixing between these two different physical
situations.

The BED model was proposed to describe in a parameter-free fashion the impact of
a free electron on a bound one: it is able to determine the proper mixing by
requiring the asymptotic behaviour of the ionization cross section to coincide
with the one obtained in Bethe's theory, but some issues remain open on how to
describe within the model the fact that the outgoing primary and secondary
electrons are undistinguishable.
This crucial feature is included only in the Mott cross section. 
These shortcomings can explain the observed difference between the predictions of the BED
and Deutsch-M\"ark model, which, as discussed in the following section, is to a
large extent a phenomenological description of ionization processes.

The BED model prescribes procedures to evaluate the energy distribution of the
ejected electron for each subshell using the binding energy, average kinetic
energy and dipole oscillator strength for each subshell.
The agreement of BED with known experimental data is of the order of 10\% in the
region from the threshold to some keV.

The oscillator strengths required in the BED formula can be obtained by
theoretical calculations or experimental photoionization cross sections;
nevertheless, they are not easily available for every atom and for each
subshell.
Although the BED model shows qualitatively good agreement with
experimental data for many atoms (e.g. H, He, Ne, Rb) \cite{bebKim1994,
bebKim1998}, the difficult availability of these components of its formulation
limits its practical use.

The BEB model \cite{bebKim1994} was proposed as a simplification of the BED
model, when the differential dipole oscillator strength is unknown.
It assumes a simple form for the oscillator strengths, which approximates the
shape of the oscillator strength for ionization of the ground state of hydrogen.

The BEB cross section for the ionization of subshell \textit{i} is given by:
\begin{equation}
%\sigma_{BEB, i} = \frac{S}{t+(u+1)/n}\left[\frac{\text{log}(t)}{2}\left(1-\frac{1}{t^2}\right)
%+1-\frac{1}{t}-\frac{\text{log}(t)}{t+1} \right] (Hee: is there an extra /n in the first coeffcient?)
\sigma_{BEB, i} = \frac{S}{t+(u+1)/n}\left[\frac{\text{ln}(t)}{2}\left(1-\frac{1}{t^2}\right)
+1-\frac{1}{t}-\frac{\text{ln}(t)}{t+1} \right]
\label{eq_beb}
\end{equation}
where:
\begin{equation}
 t=\frac{T}{B}, ~~~~~ u=\frac{U}{B}, ~~~~~ S=4\pi a^2_0 N \left(\frac{R}{B}\right)^2
\label{eq_bebsub}
\end{equation}
In the above equations \textit{T} is the incident electron energy, \textit{t}
and \textit{u} are normalized incident and kinetic energies, $n$ is the principal
quantum number (only taken into account when greater than 2), \textit{a$_0$} is
the Bohr radius 
%(5.292×10$^{-9}$ cm), 
and \textit{R} is the Rydberg constant.
%(13.6057 eV).
The BEB model involves three atomic parameters for each subshell of the target
atom, as shown in (\ref{eq_beb}): the electron binding energy \textit{B}, the
average electron kinetic energy \textit{U} and the occupation number \textit{N}.
%The constant $m$ appearing in the denominator of the first term of 
%(\ref{eq_beb}) takes different values depending on the principal quantum number
%$n$: $m = 1$ when $n = 1,2$ and $m = n$ when $n \geq 3$.
%The first term in the square brackets in \ref{eq_beb} is related to the
%Bethe theory of soft collisions for high-\textit{T} electron, and the residual
%terms are related to binary-encounter theory of hard collisions for
%low-\textit{T} electron.
The sum over all subshells \textit{i} gives the total cross section;
in practice, only the valence shell and a few outer subshells 
%below it 
contribute significantly to determine the total cross section value.

In equation (\ref{eq_beb}), the term associated with the first logarithmic function 
%on the right-hand side 
represents distant collisions (i. e.  large impact parameters) 
dominated by the dipole interaction, and the rest of the terms represent close
collisions  described by the Mott cross section;
the second logarithmic function originates from the interference of direct
and exchange scattering.
%also described by Mott cross section.

% ------------------------------------------------------------------------

\subsection{Implementation of the BEB model}
\label{sec_implbeb}

The BEB cross section model is implemented according to (\ref{eq_beb}).

The atomic parameters originally used by the authors of the model have been
documented in the literature only for a small number of target elements;
therefore, to satisfy the requirement of general applicability in a large scale
Monte Carlo system, alternative compilations of parameters, covering the whole
periodic system, are utilized in the software implementation.

%The sources of the  parameters used in the original calculations are
%not entirely accessible or documented.

The electron binding energies (except for the valence electron) and average
electron kinetic energies appearing in the original formulation of the
Binary-Encounter-Bethe model derive from relativistic Dirac-Fock calculations
\cite{desclaux_metecc}; since the original numerical values are not publicly
available, they were replaced in the code implementation by the values reported
in in the Evaluated Atomic Data Library (EADL) \cite{eadl}. 
EADL was also used in the software implementation to retrieve the occupation
numbers of each subshell, since the source of these parameters in the original
BEB calculations is not explicitly documented in the literature.

The choice of EADL in the software implementation as an alternative source of
the atomic parameters was mainly dictated by the limited availability of
compilations of mean electron energies covering the whole periodic system; other
parameters, such as atomic binding energies and occupation numbers, were taken
from the same source for consistency.

Both in the original formulation of the model and in the software implementation
the binding energy for the valence electron is obtained from the compilation of
experimental ionization energies \cite{nist_ionipot} included in the NIST
(National Institute of Standards and Technology) Physics Reference Data.

%have been taken from EADL.
%\textit{Hee Seo: is this correct? Are any complete alternative compilations of U 
%available?}.

%The effects of the differences in the software implementation with respect to
%the original formulations are discussed in section \ref{sec_verification}.

% ------------------------------------------------------------------------
\subsection{Verification of the BEB model implementation}
\label{sec_bebveri}

\begin{figure}
\centerline{\includegraphics[angle=0,width=8cm]{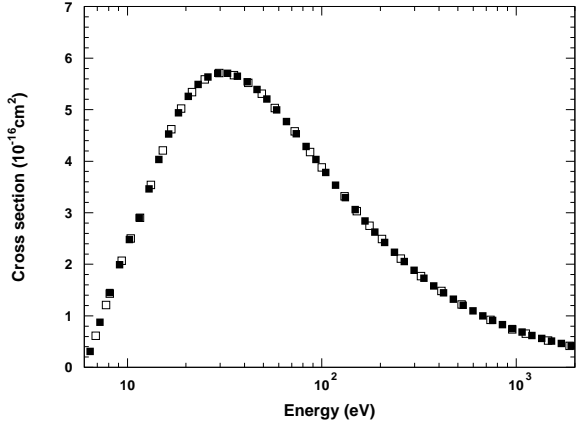}}
\caption{BEB cross section for electron impact on gallium calculated by the
software (white squares) and original values (black squares), as a function of
energy. }
\label{fig_beb_veriorig_31}
\end{figure}

The correctness of the BEB model implementation was verified by comparing
quantitatively cross sections calculated with the same atomic parameters
(binding energies, ionization potentials and electron kinetic energies) used by
the original authors with reference values published in the literature.
The results of the software implementation are consistent with the original
references; an example is illustrated in Fig. \ref{fig_beb_veriorig_31}.

\begin{figure}
\centerline{\includegraphics[angle=0,width=8cm]{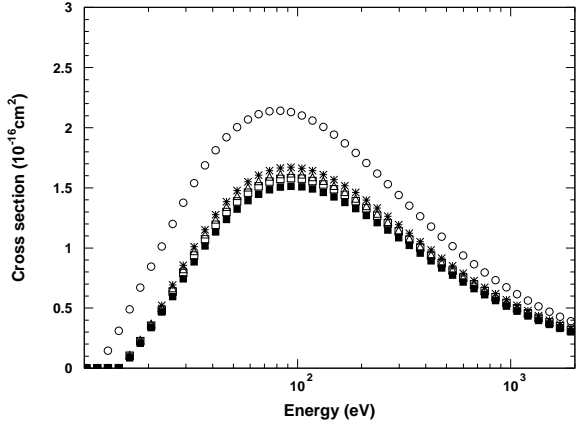}}
\caption{Effect of atomic parameters on BEB cross section for electron impact on
nitrogen: cross sections calculated by the implemented software (white squares)
with parameters as in section \ref{sec_implbeb}, original values (black
squares), calculations as in section \ref{sec_implbeb} except for atomic binding
energies taken from Lotz's compilation as in the DM model (asterisks) and for
average electron kinetic energies as in the original formulation (white
triangles), and with all atomic parameters taken from EADL (white circles). }
\label{fig_beb_veri_7}
\end{figure}

Nevertheless, as discussed in section \ref{sec_implbeb}, the software
implementation uses different values of the the atomic parameters involved in
(\ref{eq_beb}), since the original ones are documented only for a small
number of elements.
The resulting cross section values are sensitive to this modification, as one
can observe in Fig. \ref{fig_beb_veri_7}; the extent of alteration with respect
to the original values depends on the element.

An extensive investigation of the effect of atomic electron binding energies on
various physics quantities relevant to Monte Carlo particle transport, including
the cross sections calculated by the BEB model, is reported in
\cite{tns_binding}.
The effect of atomic parameters on the accuracy of the BEB model at reproducing
experimental data is discussed in section \ref{sec_results}.

% ------------------------------------------------------------------------

\section{The Deutsch-M\"ark model}
\label{sec_dm}

\subsection{Theoretical background}

The Deutsch-M\"ark formalism was originally developed for the calculation of
atomic ionization cross sections \cite{dmDeutsch1987};
it has been subject to evolution \cite{dmMarg1994, dmDeutsch2000,dmDeutsch2004,
dmDeutsch2005,dmDeutsch2005err} since its first formulation.

The DM model has its origin in a classical binary encounter
approximation derived by Thomson \cite{thomson} and its improved form by
Gryzinski \cite{gryzinski}.
%Deutsch and M\”ark introduced the radius of maximum radial density to replace Bohr 
%radius in the model of Gryzinski. 

The current expression of the DM formula calculates
the atomic cross section $\sigma_{DM}$ for single ionization as the sum over all partial ionization
cross sections corresponding to the removal of a single electron from a given
atomic subshell, characterized by quantum numbers $n$ and $l$ as:
\begin{equation}
\sigma_{DM} = 
\sum_{n,l}g_{nl}\pi r^2_{nl}\xi _{nl} b^{(q)}_{nl} (u) \left( \frac{\text{ln}(c_{nl} u)}{u} \right)
\label{eq_dms}
\end{equation}

In this formula $r_{nl}$ is the radius of maximum radial density of the atomic
subshell with quantum numbers $n$ and $l$ and $\xi_{nl}$ is the 
electron occupation number in that subshell; $g_{nl}$ are weighting factors, which were
determined by the original authors from a fitting procedure
\cite{dmDeutsch1987,dmMarg1994} using experimental cross section data.
The quantity $u$ represents the reduced energy $E/E_{nl}$, where $E$ is the
energy of the incident electron and$E_{nl}$ is the ionization energy of the subshell
identified by $n$ and $l$ quantum numbers.
In the original authors' calculations the values of $r_{nl}$ were taken from
Desclaux's compilation \cite{desclaux} and ionization energies form Lotz's
compilation \cite{lotz} of atomic binding energies.
The $c_{nl}$ constant is close to one except for electrons in the $d$ orbital.
The sum extends over all the subshells of the target atom.

The energy-dependent function has the form:

\begin{equation}
 b^{(q)}_{nl} (u) = 
\frac{A_1 - A_2}{1+(u/A_3)^p} + A_2
%\frac{A_1 - A_2}{\left[ 1+(u/A_3)^p \right]} + A_2
\label{eq_dmb}
\end{equation}
where $A_1$, $A_2$, $A_3$ and $p$ are constants, that were determined from
measured cross sections for the various values of $n$ and $l$
\cite{dmDeutsch2005,dmDeutsch2005err}.
The superscript $q$ refers to the number of electrons in the subshell identified
by $n$ and $l$.

% ------------------------------------------------------------------------

\subsection{Implementation of the DM model}
\label{sec_impldm}

The DM cross section model is implemented according to (\ref{eq_dms}).
Most of the parameters in the implementation 
are taken from the sources documented by the original authors.

The values of the radius of maximal radial density derive from the review by
Desclaux \cite{desclaux} as in the original model.
The parameters of the energy dependent function are those reported in an
original reference \cite{dmDeutsch2005}.

Atomic electron binding energies derive from the compilation by Lotz
\cite{lotz}, as in the original calculations, with the exception of the binding
energies for the valence electron, which are taken from NIST collection of
ionization energies \cite{nist_ionipot} for consistency with the BEB
implementation: however, Lotz's and NIST ionization energies are equivalent with
0.05 significance \cite{tns_binding}.
Occupation numbers are also taken from NIST Physics Reference Data, while their
source is not explicitly documented in the original formulation.

Weighting factors in the software implementation are taken from original
publications \cite{dmMarg1994} and\cite{dmDeutsch2000} (the former limited to
the 7s orbital); although more recent values have been determined by the
original authors\cite{dm_privcomm}, they could not be utilized in the
current implementation, as they are not publicly documented.

\subsection{Verification of the DM model implementation}
\label{sec_dm_veri}

The formulation of the DM model was revised in 2004; therefore, only cross
section values published since then
\cite{dmDeutsch2004,dmDeutsch2005,dmDeutsch2008,dmDeutsch2008L} were considered
as a reference in the software verification process.
%In a few cases some differences were observed between the values calculated by
%the software and published references.

\begin{figure}
\centerline{\includegraphics[angle=0,width=8cm]{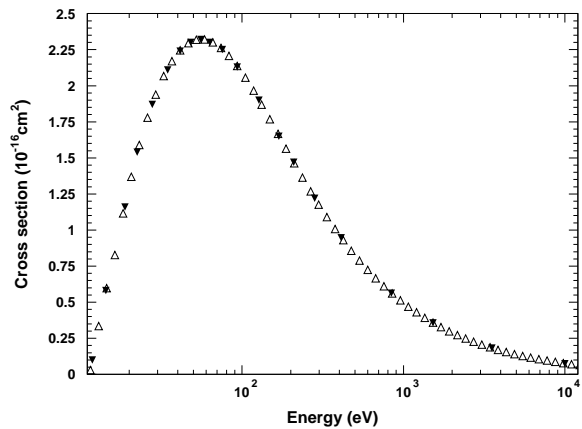}}
\caption{DM cross section for electron impact on carbon calculated by the
software (white triangles) and original values (black triangles), as a function of
energy. }
\label{fig_dmorig_6}
\end{figure}

\begin{figure}
\centerline{\includegraphics[angle=0,width=8cm]{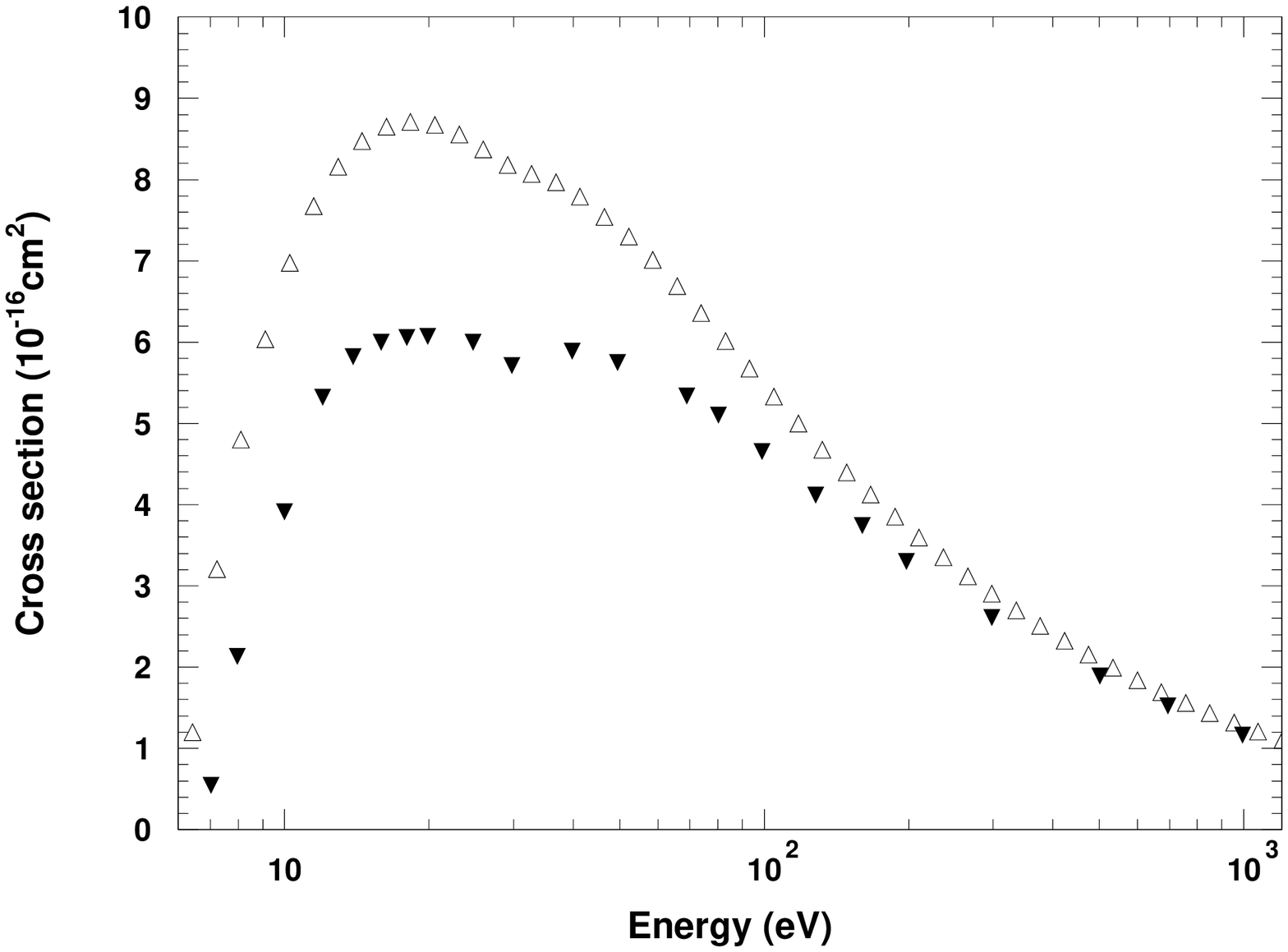}}
\caption{DM cross section for electron impact on gadolinium calculated by the
software (white triangles) and original values (black triangles), as a function of
energy. }
\label{fig_dmorig_64}
\end{figure}

\begin{figure}
\centerline{\includegraphics[angle=0,width=8cm]{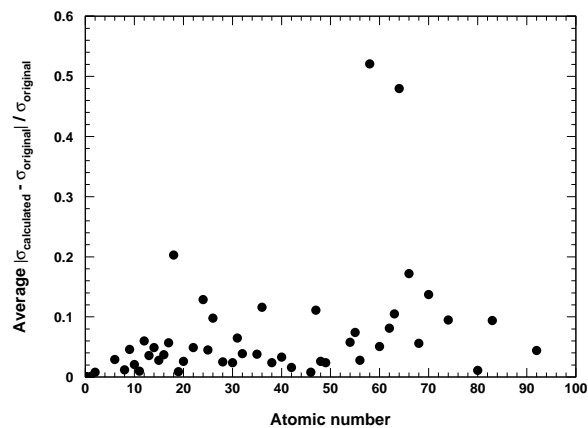}}
\caption{Average relative difference between DM cross sections
calculated by the software and original values, as a function of the atomic
number Z. }
\label{fig_dmorig_diff}
\end{figure}

Original cross section values concerning 48 atoms were retrieved from the
literature and compared to the corresponding values calculated by the software
for the purpose of verification.
Two examples of these comparisons are illustrated in Fig. \ref{fig_dmorig_6} and
\ref{fig_dmorig_64}.
As shown in Fig. \ref{fig_dmorig_diff}, in more than 2/3 of the test cases the
average difference between original and calculated values is smaller than 5\%;
nevertheless, for a few target elements (namely argon, cerium and gadolinium)
it is greater than 20\%.
Goodness-of-fit tests comparing the distributions of original and calculated
cross sections confirm the rejection of the null hypothesis of compatibility
with 0.05 significance in these cases exhibiting large discrepancies.

%Out of them, for 28 atoms the results of the software calculations reproduced
%the reference values within better than 1\%; for 13 atoms \textit{small}
%differences were observed, whereas for 7 atoms (Ar, Cr, Fe, Ge, Kr, Ce and Gd)
%the software calculations exhibited relatively \textit{large} difference with
%respect to the original Deutsch-M\"ark cross section values.
%\textit{(Hee Seo: please document quantitatively what small and large mean; if
%you give me the original DM values for a few atoms showing large differences, I
%can make a GoF test between them and our values and conclude if our values are
%significantly different).}

The most probable source of the observed discrepancies between original and
calculated values is the different set of weighting factors used in the software
implementation and in recent Deutsch-M\"ark calculations, as discussed in
section \ref{sec_impldm}.
This assumption was tested by empirically adjusting the weighting factors values
in the software implementation; as a result of this operation, cross section
values compatible within 1\% with recently published reference ones could be
obtained for the elements exhibiting large discrepancies in Fig.
\ref{fig_dmorig_diff}.
Nevertheless, for better traceability of the results of the software, the
published weighting factors of \cite{dmMarg1994} and\cite{dmDeutsch2000}  were
retained in the implementation.

The role of these discrepancies on the capability of the DM model implementation
to reproduce experimental data is discussed in section \ref{sec_results}.

\section{The Evaluated Electron Data Library}
\label{sec_eedl}

The Evaluated Electron Data Library includes tabulations of ionization
cross sections resulting from theoretical calculations.

For close collisions, the calculations use Seltzer's modification \cite{seltzer1988} of
M{\o}ller's binary collision cross section \cite{moller}, which takes into
consideration the binding of atomic electrons in a given subshell. 
For distant collisions, they use Seltzer's modification \cite{seltzer1988} 
of Weizs\"acker-Williams' method \cite{weizsacker,williams};
this approach is similar to the BED model in that it requires knowledge of the
dipole oscillator strengths of the target, but, being primarily designed for
high energy incident electrons, it may lead to unrealistic results below
a few hundred eV.

Calculations by Scofield \cite{scofield1978} were used to take into account
the density effect; this correction is significant for inner shell cross
sections for incident electron energies above a few hundred MeV, but this effect
sets in at lower energies for outer shells.
%relativistic cross section of Scofield

% ------------------------------------------------------------------------

%components of the model formulation available in the
%literature at the time of writing this paper.

% ------------------------------------------------------------------------
\section{Validation of electron cross section models}
\label{sec_validation}

The electron ionization cross sections calculated by the BEB and DM models, and
those tabulated in EEDL are validated through comparison with experimental data.

Cross sections for the ionization of atoms based on the BEB and DM models have
been previously subject to comparison with experimental measurements (e.g.
\cite{bebKim2001,bebKim2002,bebKim2007,bebAli2008,dmMarg1994,dmDeutsch2005,
dmDeutsch2008,dmDeutsch2008L}); 
these comparisons involve a limited number of target elements and experimental
data sets, and rely on qualitative visual appraisal of the compatibility between
models and measurements.
They concern calculations performed by the original authors, which, as discussed
in the previous sections, in some cases cannot be reproduced, as not all the
original parameters in the model formulation are publicly documented.

To the best of the authors' knowledge, the accuracy of the EEDL ionization cross
section calculations has not yet been quantitatively documented in the
literature.

The validation process described in this paper concerns the cross sections
calculated by the software implementation, which, being intended for open source
release along with the publication of this paper, are reproducible.
It involves a wider collection of experimental data than previously published
comparisons and concerns a larger number of elements; moreover, the
compatibility between the models and experimental measurements is estimated
quantitatively, based on statistical analysis methods.

%In addition, this paper reports for the first time a quantitative comparison of
%the accuracy of the three cross section models, also based on rigorous
%statistical methods.

% ------------------------------------------------------------------------
\subsection{Experimental measurements}
\label{sec_exp}

The validation of the three electron ionization cross section models
is based on a large set of experimental data \cite{expHshah1987} -\cite{exp92}
collected in the literature.
The experiments were performed with different techniques and measured a variety
of physical observables, which are not always exhaustively documented in the
related publications.
Some papers do not report the uncertainties of the measurements.
%Some characteristics of the experimental data could bias the outcome of the
%validation process.

The reference data include measurements of single ionization, i.e. the emission
of a single electron as a result of the primary electron's impact, as well as
experiments that did not distinguish whether more than one electron had been
emitted from the target atom.
When multiple ionization is involved, a further possible source of ambiguity
depends on whether the measurements concerned the so-called ``total counting'' cross
section, which accounts for the number of ions produced,
\begin{equation}
\sigma_{\text{counting}} = \sum{\sigma^{n+}}
\label{sigma_counting}
\end{equation}
or the so-called ``total gross'' cross section, which is determined by measuring the
total ion current,
\begin{equation}
\sigma_{\text{gross}} = \sum{n\sigma^{n+}}
\label{sigma_gross}
\end{equation}
where $n$ represents the number of ionizations.
Multiple ionization is generally small with respect to single ionization: for
instance, for several elements cross sections for double ionization amount to a
few percent of those for single ionization, and cross sections for triple
ionization are approximately an order of magnitude smaller than for double
ionization \cite{expFreund}.
Nevertheless, the contribution of double ionization may be significant for some
elements: for instance, it represents more than 20\% of single ionization for
lead \cite{expFreund}.
% or to other sources of experimental uncertainties.

%Possible sources of systematic effects in the validation process, which could
%originate from
%
%gross ionization
%in some cases the physical characteristics of the
%measurements (e.g. multiple or single ionization, counting or gross cross section) are not documented.

Some experiments measured absolute cross sections; some report relative values
with respect to other references, which are either experimental or theoretical
calculations.
% (e.g. Bray?).
Both techniques have drawbacks: the intrinsic difficulty of making accurate
absolute cross section measurements and the possibility of introducing a
systematic bias in relative measurements.

Other features likely to be associated with systematic effects can be identified
in contradictory measurements of single and total (counting or gross)
ionization cross sections: in some cases (for instance, as reported in 
\cite{expFreund}  and \cite{golovach,shimonAlInTl})
the experimental cross section for single ionization appears larger than
measurements of total gross or counting cross section, of which single
ionization should be a component.

Large discrepancies are evident in some of the experimental data.
Some data sets pertaining to the same target element are patently inconsistent;
systematic effects are likely present in some cases, where the Wald-Wolfowitz
test \cite{wald} detects sequences of positive or negative differences between
experimental data sets, which are incompatible with randomness.

The wide heterogeneity of the experimental data complicates the validation
process; it suggests caution in the interpretation of results of agreement, or
disagreement, of the theoretical models with individual measurements, and induce
to privilege a statistical analysis over a wide experimental sample as an
indicator of the reliability of the theoretical models for use in particle
transport.

\begin{table*}
\begin{center}
\caption{Percentage of test cases in which cross section 
models are compatible with experimental data}
\label{tab_eff}
\begin{tabular}{|l|l|c|c|c|c|}
%\begin{tabular}{|p{2cm}|p{2cm}|p{2cm}|p{2cm}|p{2cm}|p{2cm}|p{2cm}|}
\hline
\multicolumn{2}{|c|}{}	& {\bf All}	& {\bf Single}	& {\bf Absolute}    & {\bf Single} \\
\multicolumn{2}{|c|}{}	&	      	& {\bf ionization} & {\bf measurement} & {\bf Absolute} \\
\hline
\multirow{4}{*}{$<$20 eV} 
& No. data sets	&	 107	& 75	& 73	& 44 \\
& 	{\bf BEB}	& 74$\pm$4	& 71$\pm$5	& 81$\pm$5	& 80$\pm$6 \\
& 	{\bf DM}	& 75$\pm$4	& 75$\pm$5	& 81$\pm$5	& 84$\pm$6 \\
& 	{\bf EEDL}	& 36$\pm$5	& 39$\pm$6	& 38$\pm$6	& 43$\pm$7 \\
\hline		  	  	  	
\multirow{4}{*}{20-50 eV} 	  	  	  	  	  
& No. data sets		& 129	& 90	& 83	& 49 \\
& 	{\bf BEB}	& 63$\pm$5	& 67$\pm$5	& 53$\pm$6	& 53$\pm$8 \\
& 	{\bf DM}	& 71$\pm$4	& 76$\pm$5	& 64$\pm$6	& 69$\pm$7 \\
& 	{\bf EEDL}	& 17$\pm$4	& 21$\pm$5	& 13$\pm$4	& 16$\pm$6 \\
\hline		  	  	  	
\multirow{4}{*}{50-100 eV}	  	  	  	  
& No. data sets		& 124	& 91	& 81	& 50 \\
& 	{\bf BEB}	& 40$\pm$5	& 42$\pm$6	& 37$\pm$6	& 38$\pm$7 \\
& 	{\bf DM}	& 66$\pm$5	& 70$\pm$5	& 64$\pm$6	& 70$\pm$7 \\
& 	{\bf EEDL}	& 18$\pm$4	& 24$\pm$5	& 9$\pm$3	& 14$\pm$5 \\
\hline		  	  	  	
\multirow{4}{*}{100-250 eV} 	  
& No. data sets		& 127	& 93	& 81	& 51 \\
& 	{\bf BEB}	& 44$\pm$5	& 51$\pm$6	& 38$\pm$6	& 45$\pm$8 \\
& 	{\bf DM}	& 70$\pm$4	& 74$\pm$5	& 67$\pm$6	& 73$\pm$7 \\
& 	{\bf EEDL}	& 39$\pm$5	& 49$\pm$6	& 28$\pm$5	& 41$\pm$7 \\
\hline		  	  	  		  	  	  	  	
\multirow{4}{*}{250 eV-1 keV}  	
& No. data sets		& 79	& 58	& 43	& 26 \\
& 	{\bf BEB}	& 62$\pm$5	& 76$\pm$5	& 58$\pm$6	& 81$\pm$6 \\
& 	{\bf DM}	& 78$\pm$4	& 86$\pm$4	& 72$\pm$5	& 85$\pm$5 \\
& 	{\bf EEDL}	& 67$\pm$5	& 79$\pm$5	& 65$\pm$6	& 88$\pm$5 \\
\hline		  	  	  		  	  	  	  	
\multirow{4}{*}{$>$1 keV} 	  		  	  	  	  
& No. data sets		& 25	& 22	& 12	& 11 \\
& 	{\bf BEB}	& 56$\pm$5	& 64$\pm$6	& 75$\pm$5	& 82$\pm$6 \\
& 	{\bf DM}	& 88$\pm$3	& 91$\pm$3	& 92$\pm$3	& 91$\pm$4 \\
& 	{\bf EEDL}	& 72$\pm$4	& 73$\pm$5	& 100$-$12	& 100$-$15 \\
\hline
\end{tabular}
\end{center}
\end{table*}

\subsection{Analysis method}
\label{sec_analysis}

The analysis is articulated over two stages: the first one estimates the
compatibility between cross section models and experimental data;
the second one evaluates whether the three models exhibit any significant
differences in their compatibility with experiment.

Cross sections are compared by means of statistical methods: goodness-of-fit
testing to evaluate the compatibility of the simulation models with experimental
measurements for each element, and categorical analysis based on contingency
tables to evaluate the overall differences in compatibility with experiment
across the models.

The null hypothesis in the goodness-of-fit tests is defined as the equivalence
of the simulated and experimental data distributions subject to comparison.
Unless differently specified, the significance level of the tests, defined as
p-value determining the region of rejection of the null hypothesis, is 0.05.

Goodness of fit tests are performed on pairs of cross section distributions;
for this purpose theoretical BEB and DM cross sections are calculated 
at the same energies as the experimental data, and EEDL cross sections 
corresponding to these energies are obtained through interpolation from 
tabulated values.

Two types of goodness of fit tests, implemented in the Statistical Toolkit
\cite{gof1,gof2}, are exploited in the validation process: the $\chi^2$ \cite{bock} 
test and three tests for unbinned distributions based on the empirical distribution 
function.
Their complementary characteristics address some peculiarities of the
experimental sample, like the lack of documentation of some experimental
uncertainties or their questionable estimate, and mitigate the risk of possible
systematic effects in the validation results related to the mathematical
formulation of a single algorithm.

Among non-parametric goodness-of-fit tests, the $\chi^2$ test takes
into account experimental uncertainties explicitly.
It is applied in this analysis whenever experimental errors are reported in the
literature and the experimental sample subject to test encompasses at least five
data (i.e. the $\chi^2$ test is considered applicable according to statistics
practice).
The $\chi^2$ test statistic is affected by the correct appraisal of the
experimental errors: their unrealistic estimation may lead to incorrect
conclusions regarding the rejection of the null hypothesis.

The Kolmogorov-Smirnov\cite{kolmogorov1933,smirnov1939}, Anderson-Darling
\cite{anderson1952,anderson1954} and Cramer-von Mises
\cite{cramer1928,vonmises1931} tests are applied to all comparisons.
They are the only means of comparing distributions when experimental errors are
unknown, or the sample size is too small for the $\chi^2$ statistic to be
meaningful; they provide complementary information about the compatibility of
the compared distributions in the cases where the $\chi^2$ test is applicable,
but the experimental errors might have not been estimated realistically.
%These tests produce consistent results regarding the rejection of the null
%hypothesis in the majority of the test cases; their different outcome is limited
%to a few cases where the p-value is close to the critical region.
%Some discrepancies in the proximity of the critical region are expected, since 
%the three tests correspond to different mathematical formulations of the distance
%between the distributions subject to comparison.

A criterion is defined to combine the results of the different tests: the null
hypothesis is not rejected if either the p-value of the $\chi^2$ test or the
p-values of at least two out of three unbinned goodness-of-fit tests are larger
than the significance level.
The combined criterion privileges the outcome of the $\chi^2$ test, which takes
into account experimental uncertainties explicitly, and requires some evidence
of consistency from unbinned tests to accept the hypothesis of compatibility
with experiment in cases where experimental errors are unknown, or might have
been underestimated.

The cross section model exhibiting the largest number of test cases where the
null hypothesis is not rejected (i.e. appearing as the most accurate at
reproducing experiment) is taken as a reference in the categorical analysis; the
other models are compared to it by means of contingency tables, to determine
whether they exhibit any statistically significant difference of compatibility
with measurements.

Contingency tables are built on the basis of the results of goodness of fit
tests on individual data samples, which are classified respectively as ``fail'' or ``pass''
according to whether the hypothesis of compatibility of experimental and
calculated data is rejected or not according to the combined criterion.

%to determine whether the examined theoretical
%models exhibit equivalent, or different behavior regarding the compatibility
%with experimental data.

%the analysis estimates whether the
%number of test cases for which a model is compatible with experimental data is
%independent of the model, in other words whether both models under test describe
%the data equally well in statistical terms.

%to whether
%the corresponding p-value is consistent with the defined significance level.
The null hypothesis in the analysis of a contingency table consists of assuming
the equivalence of the two categories of models it compares, regarding their
compatibility with experiment.

Contingency tables are analyzed with Fisher's exact test \cite{fisher} and
with the $\chi^2$ test applying Yates continuity correction \cite{yates}; the
latter ensures meaningful results even with small number of entries in the table.
Pearson $\chi^2$ test \cite{pearson} is also performed on contingency tables,
when their content is consistent with its applicability.
The use of different tests in the analysis of contingency tables contributes to
the robustness of the results, as it mitigates the risk of introducing systematic
effects, which could be due to the peculiar mathematical features of a 
single test.

A 0.05 significance level is set to determine the rejection of the null
hypothesis in the analysis of contingency tables, unless specified differently.

The statistical analysis is articulated in energy ranges relevant to the 
problem domain.
The higher end above 1~keV is covered by all general-purpose Monte Carlo codes
for particle transport; it is considered a conventional r\'egime of calculation
of electron-photon interactions with matter.
The energy range between 250 eV and 1 keV is relevant to simulation applications
using the Geant4 low energy electromagnetic package, regarding the validation of
the current models and the comparison with other specialized cross section
models not yet available in Geant4.
The lower energy end, up to a few tens of eV, pertains to microdosimetry or
nanodosimetry: the validation results assess how the implemented models, which
specifically address this domain, would extend Geant4 simulation capabilities
for applications not yet covered by the toolkit.
The assessment in the intermediate range quantitatively investigates the
possibility of extending the applicability of existing Geant4 models below the
current nominal limit of 250~eV, or the need of new models, such as those
studied in this paper, to fill the gap between the domain of microdosimetry
simulation and conventional particle transport codes.

\section{Results}
\label{sec_results}

\begin{figure}
\centerline{\includegraphics[angle=0,width=8cm]{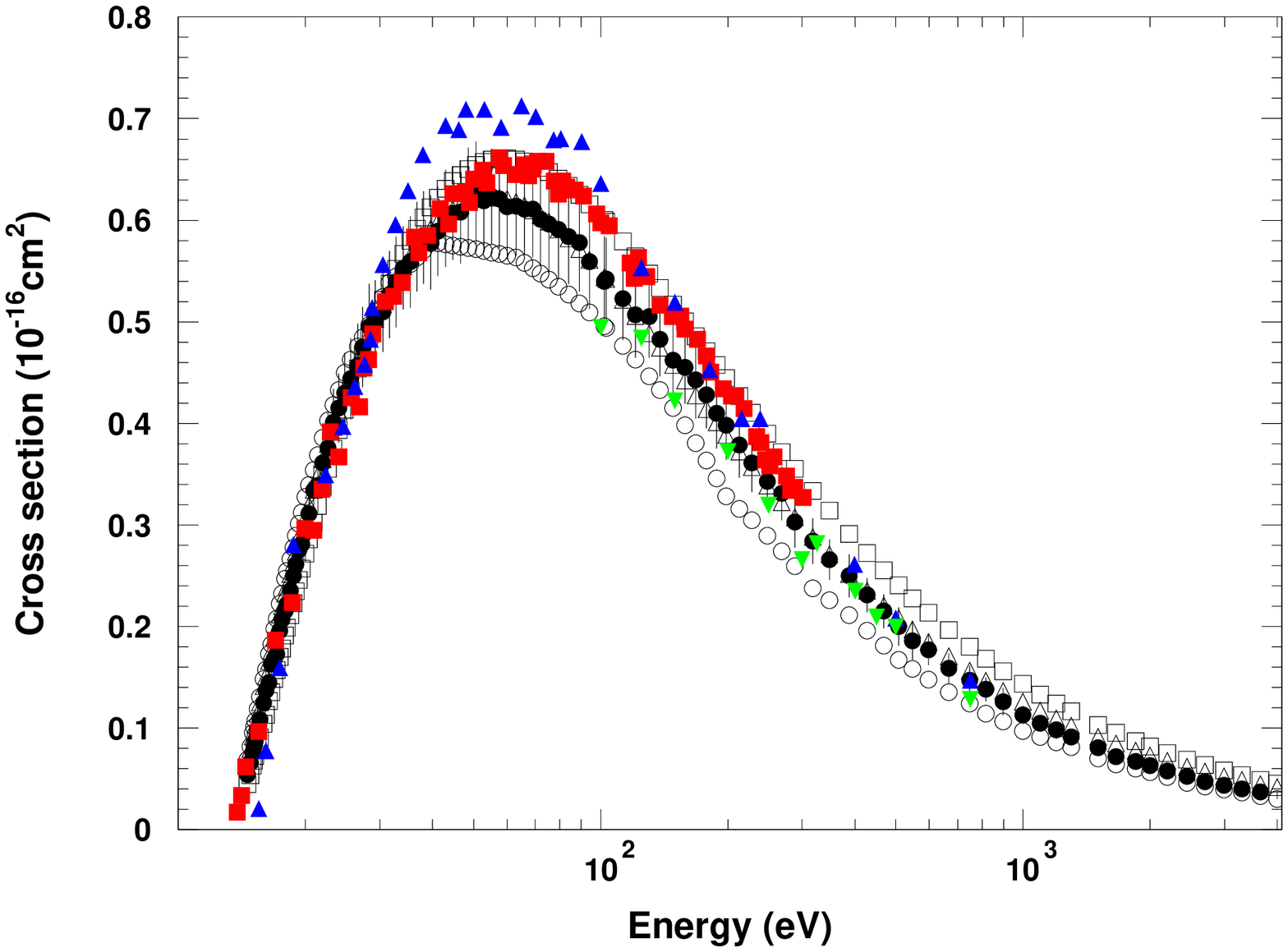}}
\caption{Cross section, Z=1: 
EEDL (empty circles), BEB model (empty squares), 
DM model (empty triangles) 
and experimental data from 
\cite{expHshah1987} (black circles), 
Boksenberg (reported in \cite{expKieffer} (red squares), 
\cite{fite} (blue triangles) and 
\cite{rothe} (green upside-down triangles). }
\label{fig_beb1}
\end{figure}

\begin{figure}
\centerline{\includegraphics[angle=0,width=8cm]{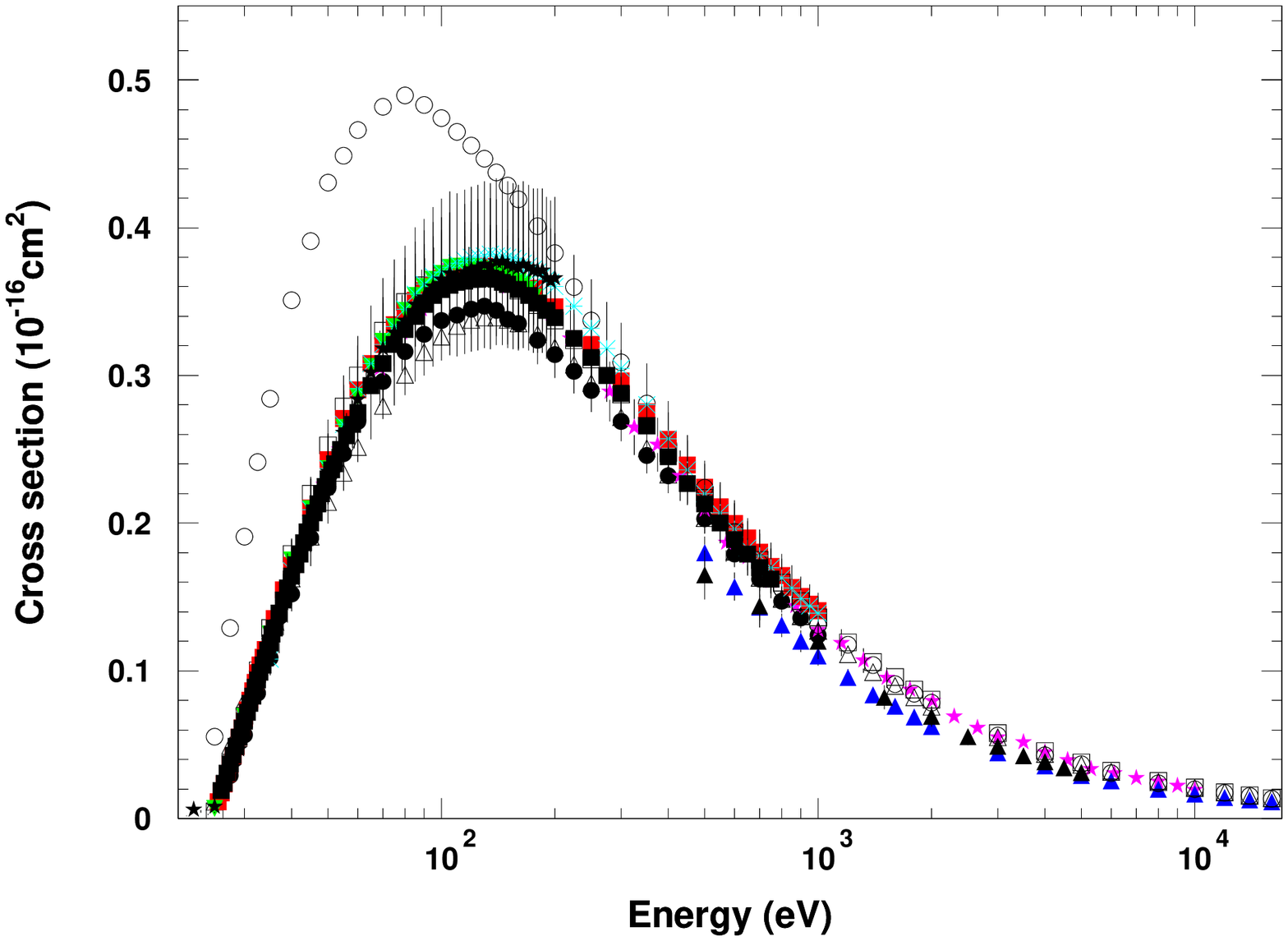}}
\caption{Cross section, Z=2: 
EEDL (empty circles), BEB model (empty squares), 
DM model (empty triangles)
and experimental data from 
\cite{expRejoub2002} (black circles), 
\cite{exp2S} (pink stars),
\cite{rapp} (red squares),
\cite{expHeNeSchram} (blue triangles),
\cite{stephan} (green upside-down triangles),
\cite{krishnakumar} (turquoise asterisks),
\cite{expMontague1984} (black squares),
\cite{expNagy1980} (black triangles) and
\cite{wetzel} (black stars).
}
\label{fig_beb2}
\end{figure}

\begin{figure}
\centerline{\includegraphics[angle=0,width=8cm]{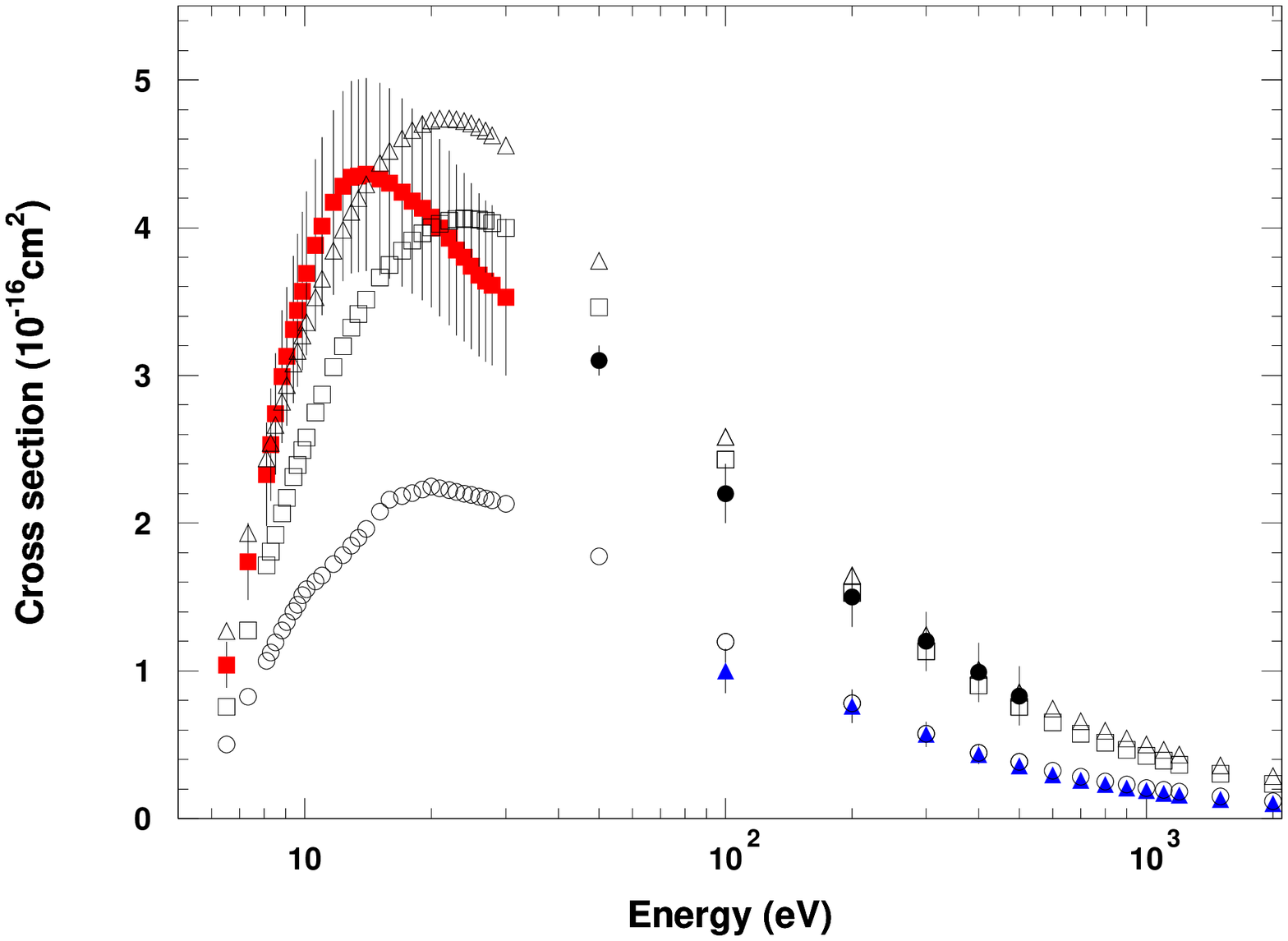}}
\caption{Cross section, Z=3: 
EEDL (empty circles), BEB model (empty squares), 
DM model (empty triangles)
and experimental data from 
\cite{expM1965} (black circles), 
\cite{expZ1969} (red squares) and 
\cite{jalin} (blue triangles).
}
\label{fig_beb3}
\end{figure}

\begin{figure}
\centerline{\includegraphics[angle=0,width=8cm]{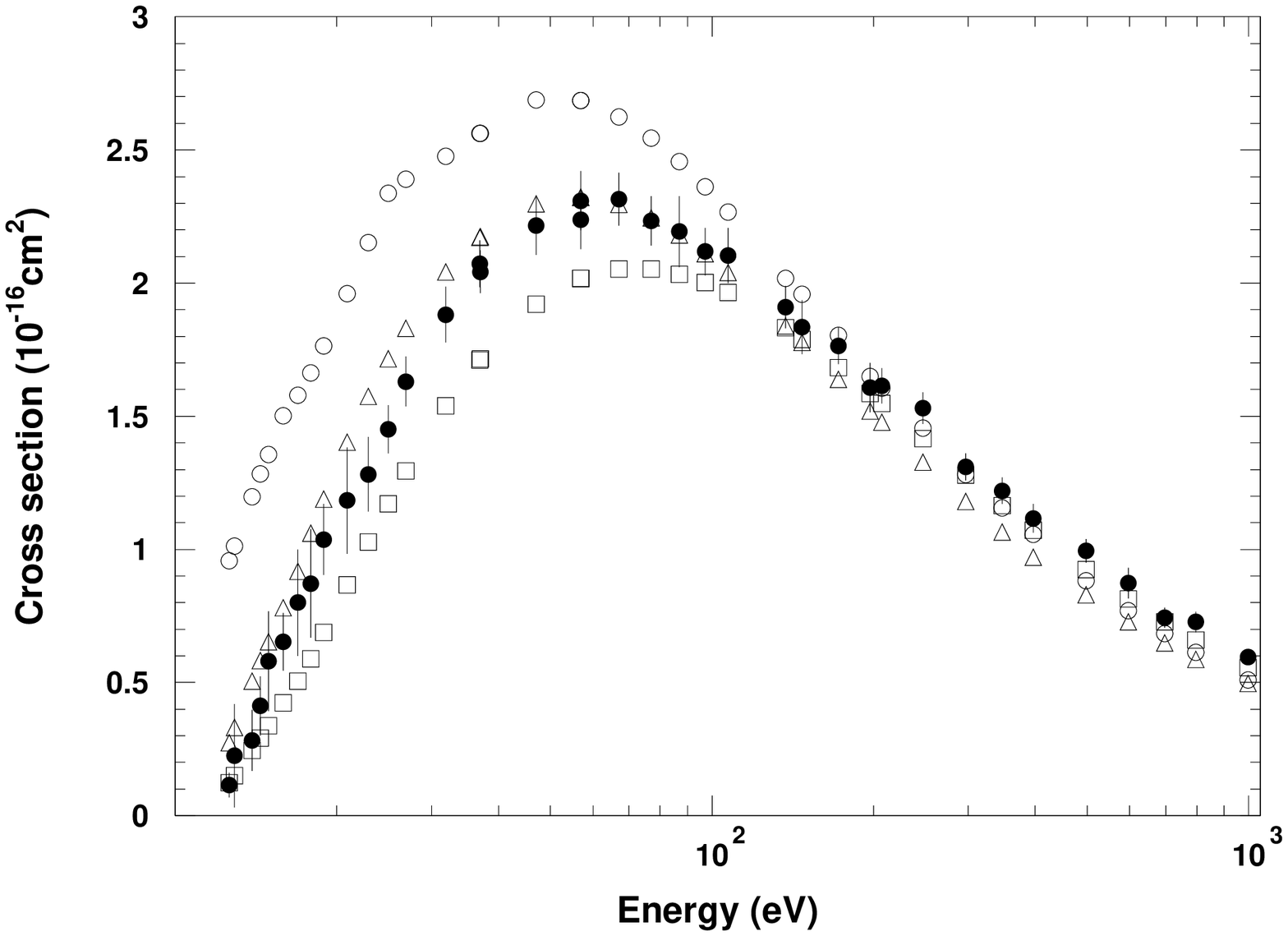}}
\caption{Cross section, Z=6: 
EEDL (empty circles), BEB model (empty squares), 
DM model (empty triangles)
and experimental data from \cite{expBrook1978} 
(black circles). }
\label{fig_beb6}
\end{figure}

\begin{figure}
\centerline{\includegraphics[angle=0,width=8cm]{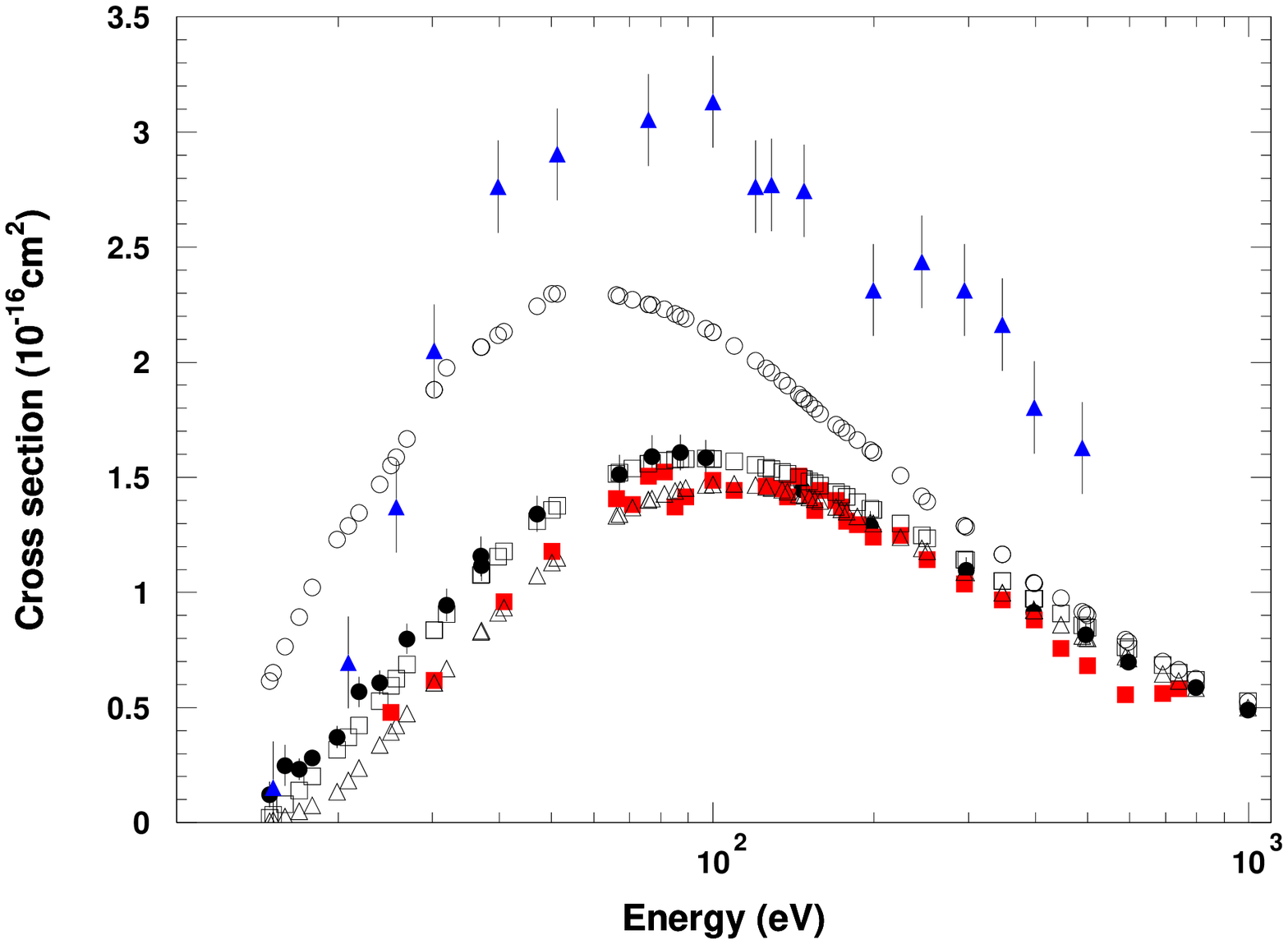}}
\caption{Cross section, Z=7: 
EEDL (empty circles), BEB model (empty squares), 
DM model (empty triangles)
and experimental data from 
\cite{expBrook1978} (black circles),
\cite{smith} (red squares) and
\cite{expPeterson} (blue triangles). }
\label{fig_beb7}
\end{figure}

\begin{figure}
\centerline{\includegraphics[angle=0,width=8cm]{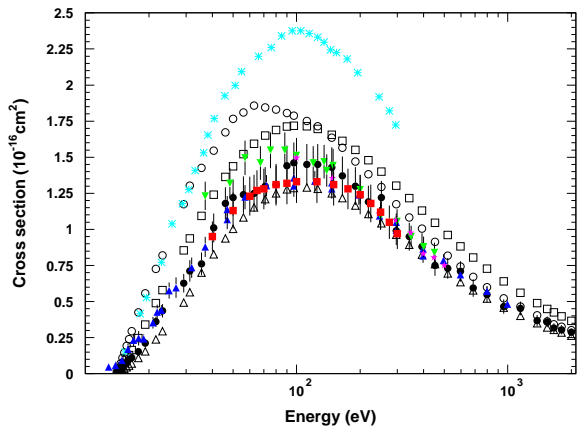}}
\caption{Cross section, Z=8: 
EEDL (empty circles), BEB model (empty squares), 
DM model (empty triangles)
and experimental data from 
\cite{exp8} (black circles),
Boksenberg (reported in \cite{expKieffer} (turquoise asterisks),
\cite{expBrook1978} (blue triangles),
\cite{expOfite1959} (green upside-down triangles),
\cite{rothe} (pink stars) and
\cite{zipf} (red squares). }
\label{fig_beb8}
\end{figure}

\begin{figure}
\centerline{\includegraphics[angle=0,width=8cm]{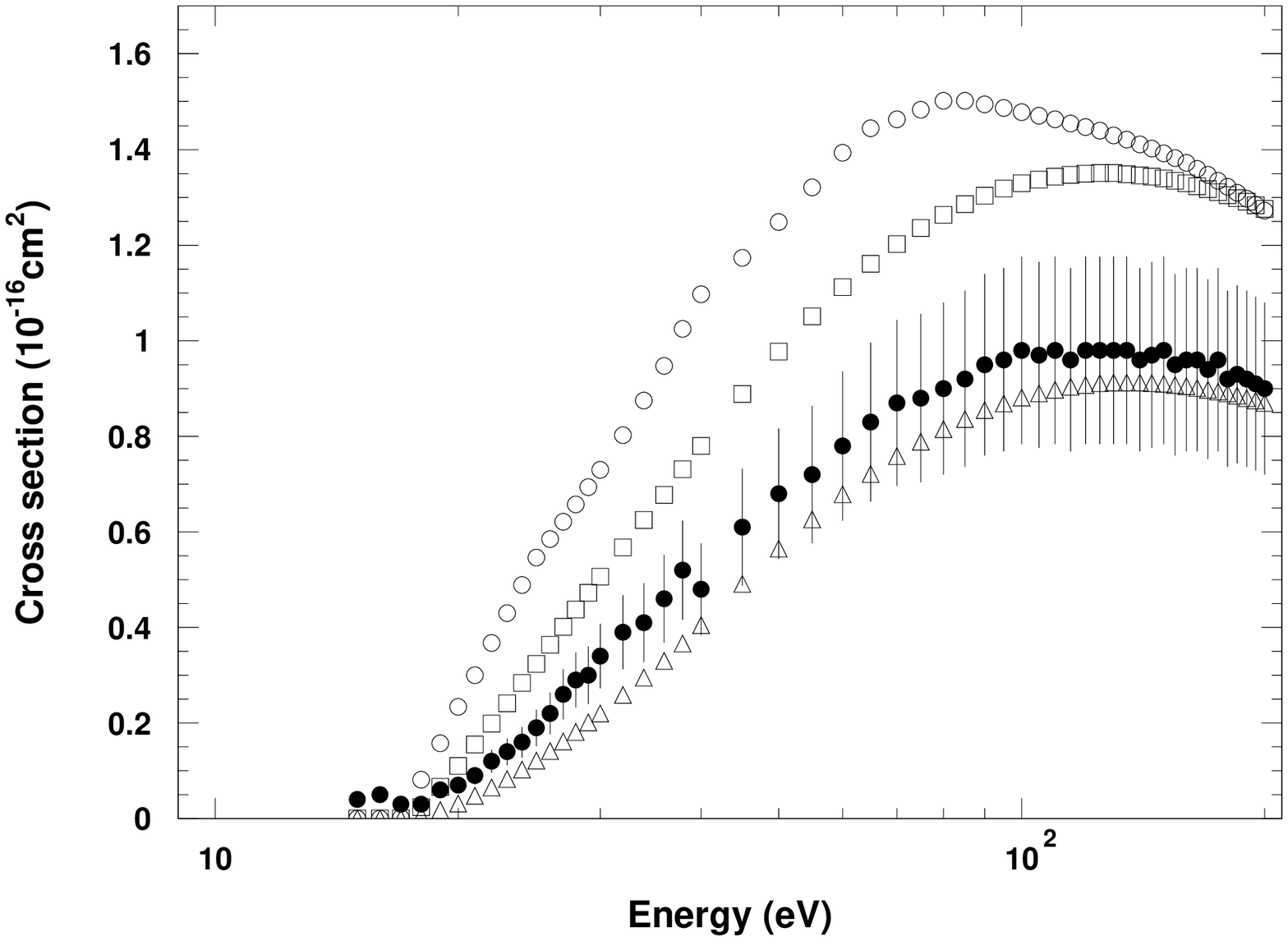}}
\caption{Cross section, Z=9: 
EEDL (empty circles), BEB model (empty squares), 
DM model (empty squares)
and experimental data from 
\cite{expHayes} (black circles). }
\label{fig_beb9}
\end{figure}

\begin{figure}
\centerline{\includegraphics[angle=0,width=8cm]{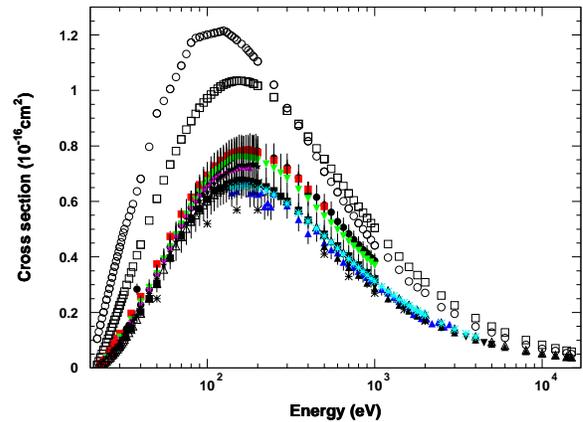}}
\caption{Cross section, Z=10: 
EEDL (empty circles), BEB model (empty squares), 
DM model (empty triangles),
experimental data from 
\cite{rapp} (black circles),
\cite{expRejoub2002} (black squares),
\cite{adamczyk} (black asterisks),
\cite{almeida} (blue triangles),
\cite{fletcher} (red squares),
\cite{krishnakumar} (green upside-down triangles),
\cite{expNagy1980} (black upside-down triangles),
\cite{expHeNeSchram} (black triangles),
\cite{sorokin} (turquoise asterisks),
\cite{stephan} (pink stars) and
\cite{wetzel} (black stars).
}
\label{fig_beb10}
\end{figure}

\begin{figure}
\centerline{\includegraphics[angle=0,width=8cm]{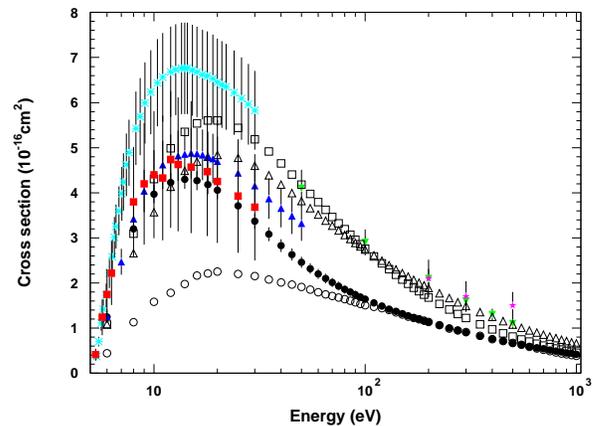}}
\caption{Cross section, Z=11: 
EEDL (empty circles), BEB model (empty squares), 
DM model (empty triangles)
and experimental data from 
\cite{expM1965} (green upside-down triangles),
\cite{expZ1969} (turquoise asterisks),
\cite{brink} (pink stars),
\cite{fujii} (black circles),
\cite{johnston} (blue triangles) and
\cite{tan} (red squares). }
\label{fig_beb11}
\end{figure}

\begin{figure}
\centerline{\includegraphics[angle=0,width=8cm]{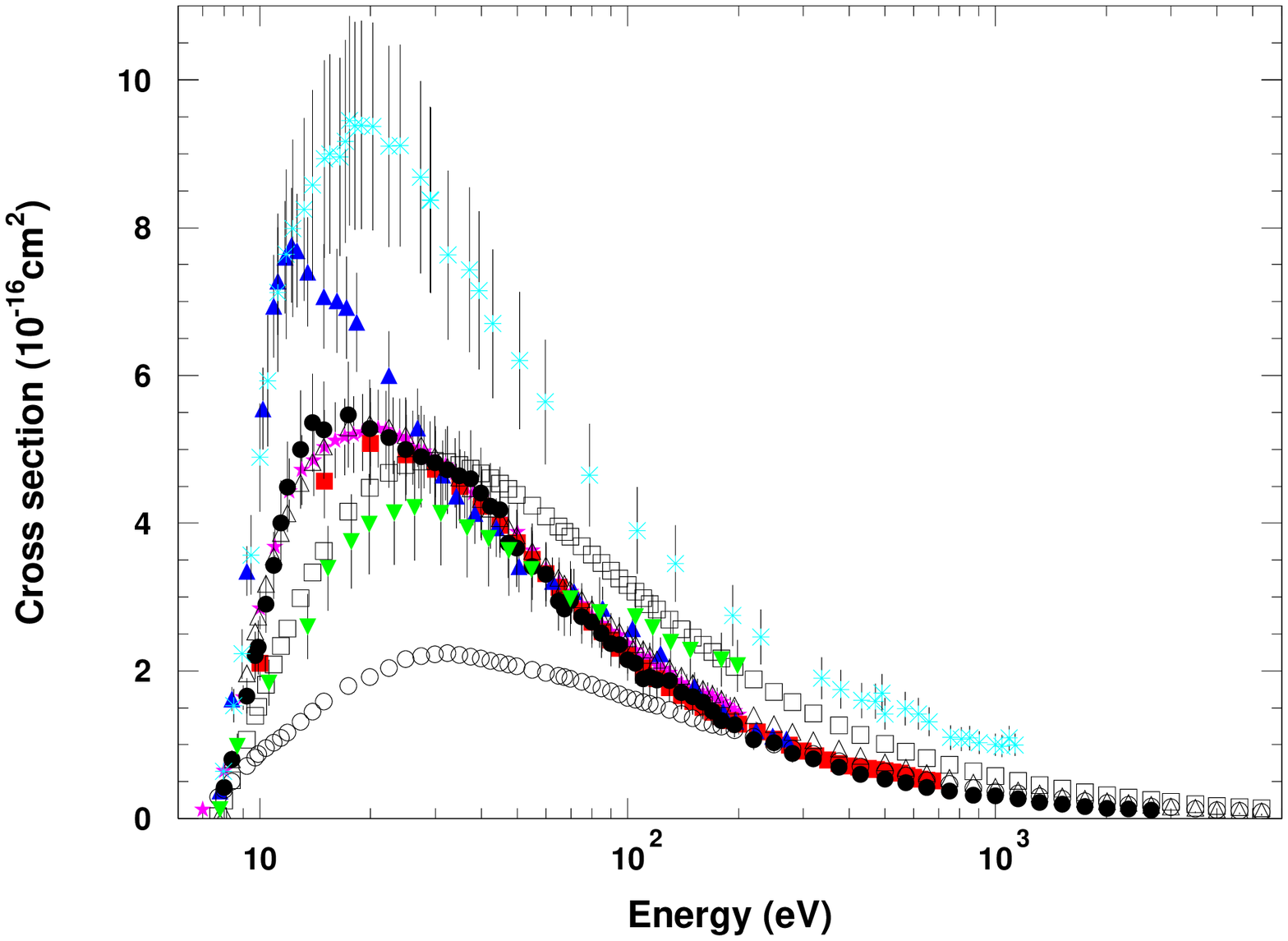}}
\caption{Cross section, Z=12: 
EEDL (empty circles), BEB model (empty squares), 
DM model (empty triangles)
and experimental data from 
\cite{expFreund} (pink stars), 
\cite{boivin} (red squares),
\cite{karstensen} (blue triangles), 
\cite{expMgMcCallion} (black circles),
\cite{expVainsh} (green upside-down triangles) and
\cite{okunoMg} (turquoise asterisks). }
\label{fig_beb12}
\end{figure}

\begin{figure}
\centerline{\includegraphics[angle=0,width=8cm]{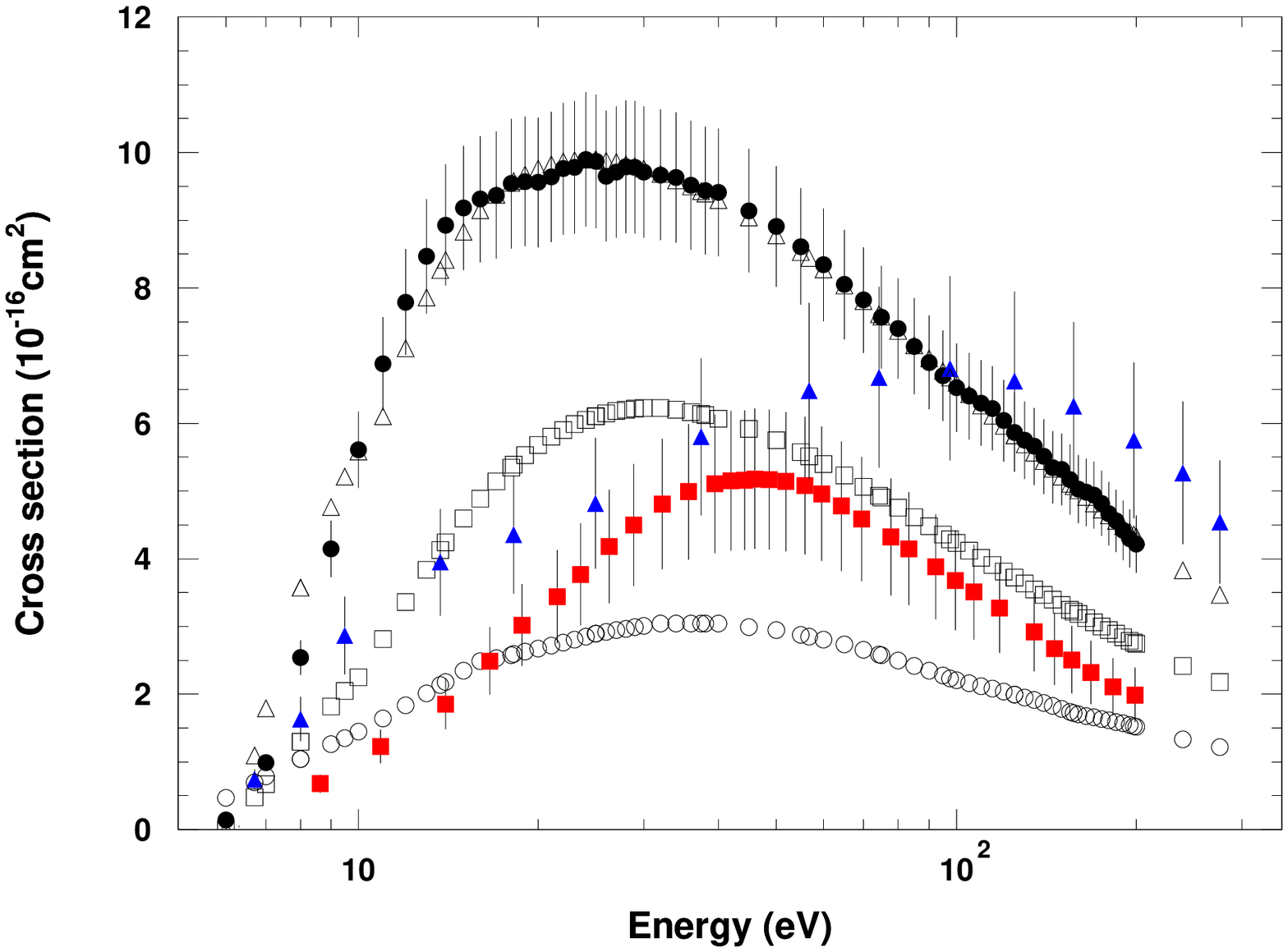}}
\caption{Cross section, Z=13: 
EEDL (empty circles), BEB model (empty squares), 
DM model (empty triangles)
and experimental data from \cite{expFreund} (black circles),
\cite{golovach} (red squares) and
\cite{shimonAlInTl} (blue triangles). }
\label{fig_beb13}
\end{figure}

\begin{figure}
\centerline{\includegraphics[angle=0,width=8cm]{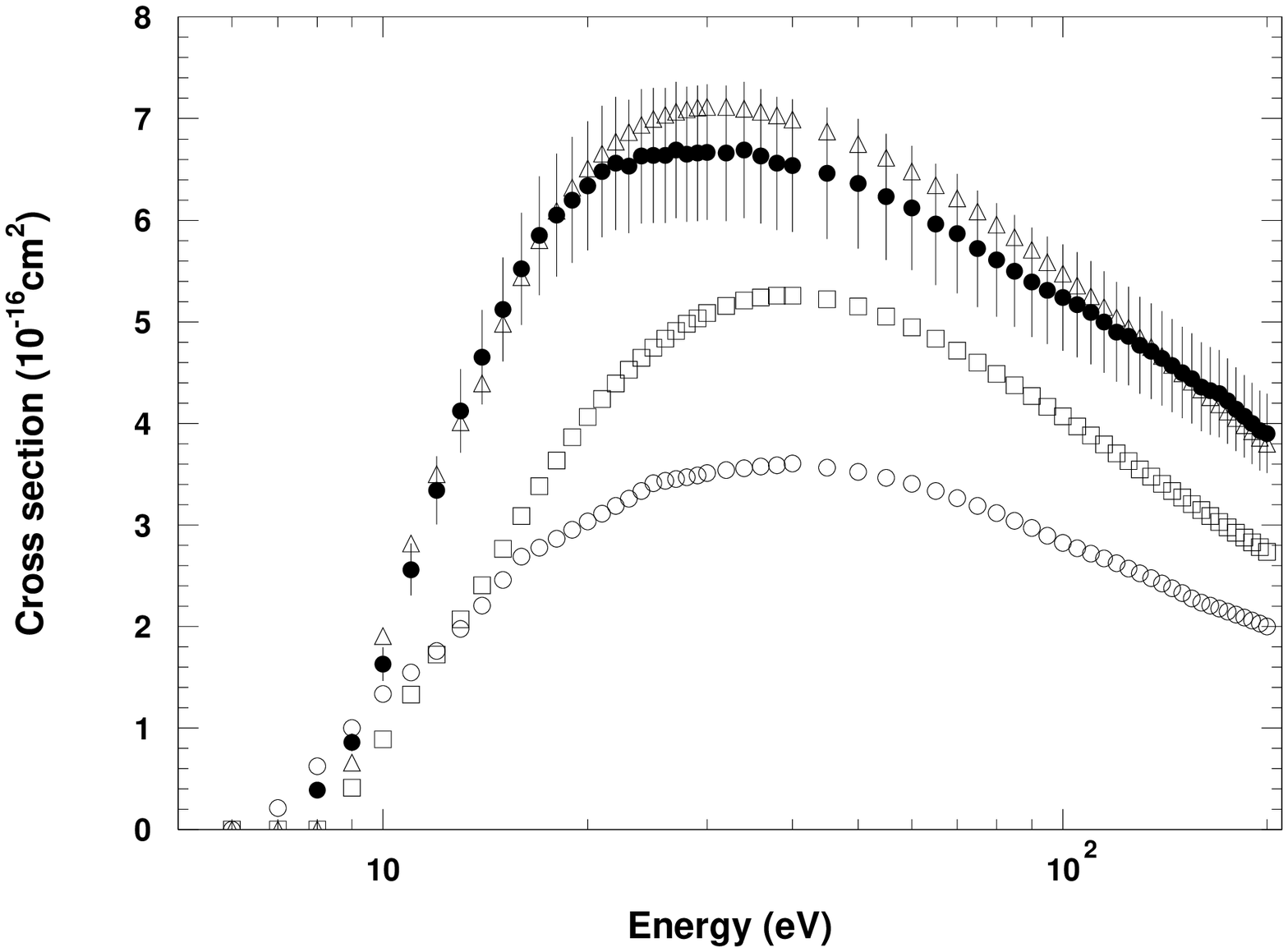}}
\caption{Cross section, Z=14: 
EEDL (empty circles), BEB model (empty squares), 
DM model (empty triangles)
and experimental data from 
\cite{expFreund} (black circles). }
\label{fig_beb14}
\end{figure}

\begin{figure}
\centerline{\includegraphics[angle=0,width=8cm]{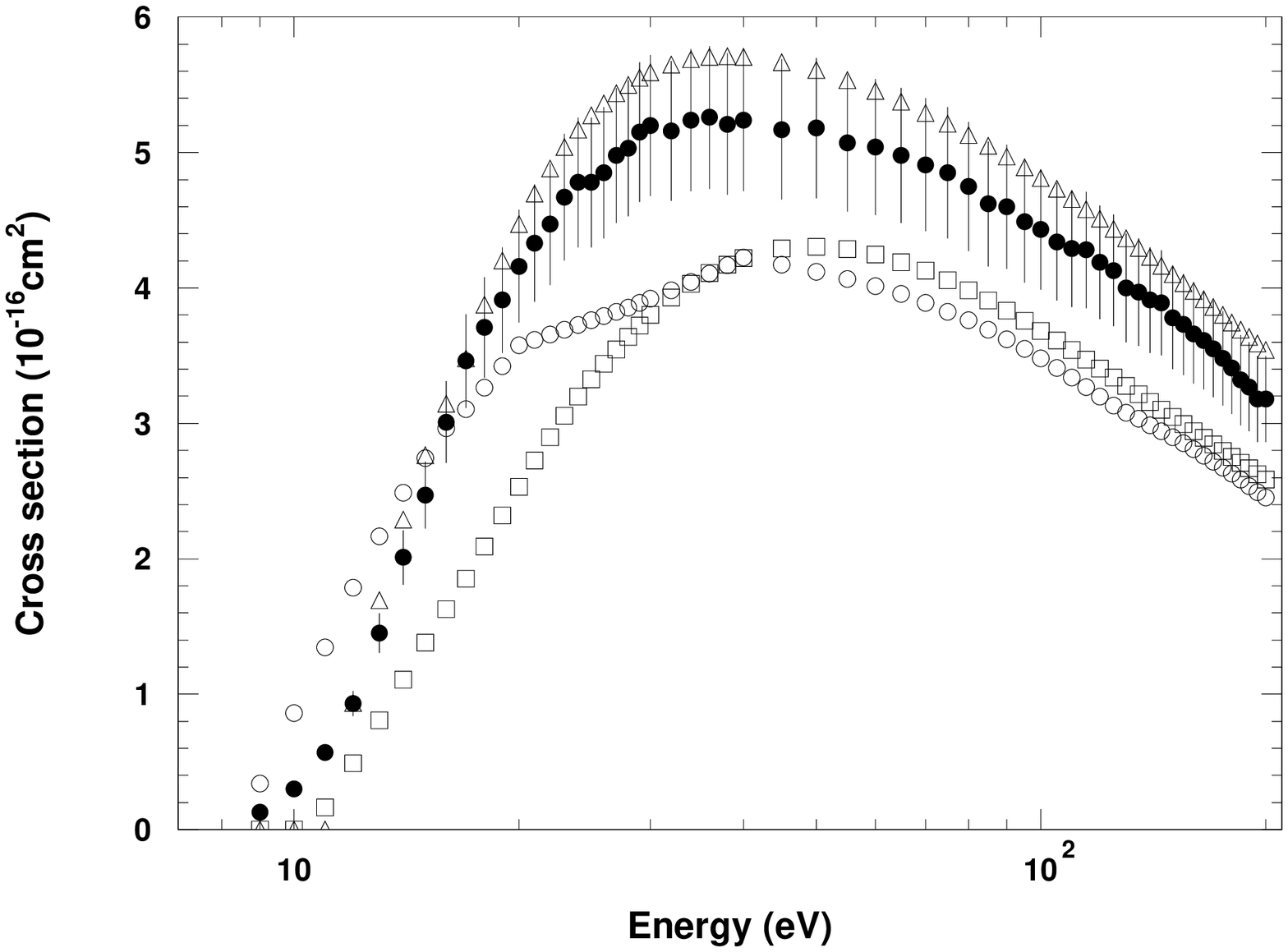}}
\caption{Cross section, Z=15: 
EEDL (empty circles), BEB model (empty squares), 
DM model (empty triangles)
and experimental data from \cite{expFreund} 
(black circles). }
\label{fig_beb15}
\end{figure}

\begin{figure}
\centerline{\includegraphics[angle=0,width=8cm]{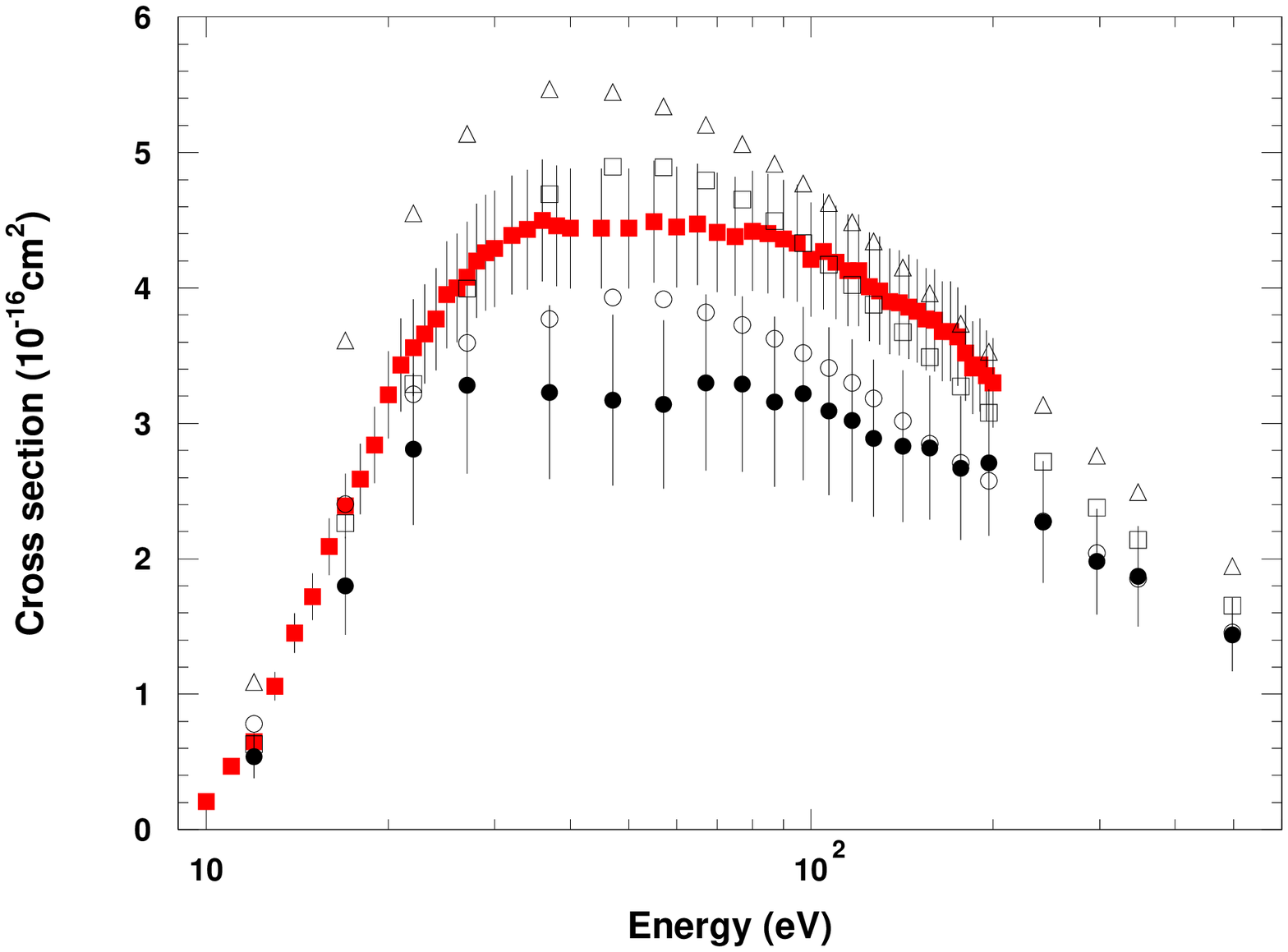}}
\caption{Cross section, Z=16: 
EEDL (empty circles), BEB model (empty squares), 
DM model (empty triangles)
and experimental data from \cite{expFreund} 
(black circles) and \cite{ziegler} (red squares). }
\label{fig_beb16}
\end{figure}

\begin{figure}
\centerline{\includegraphics[angle=0,width=8cm]{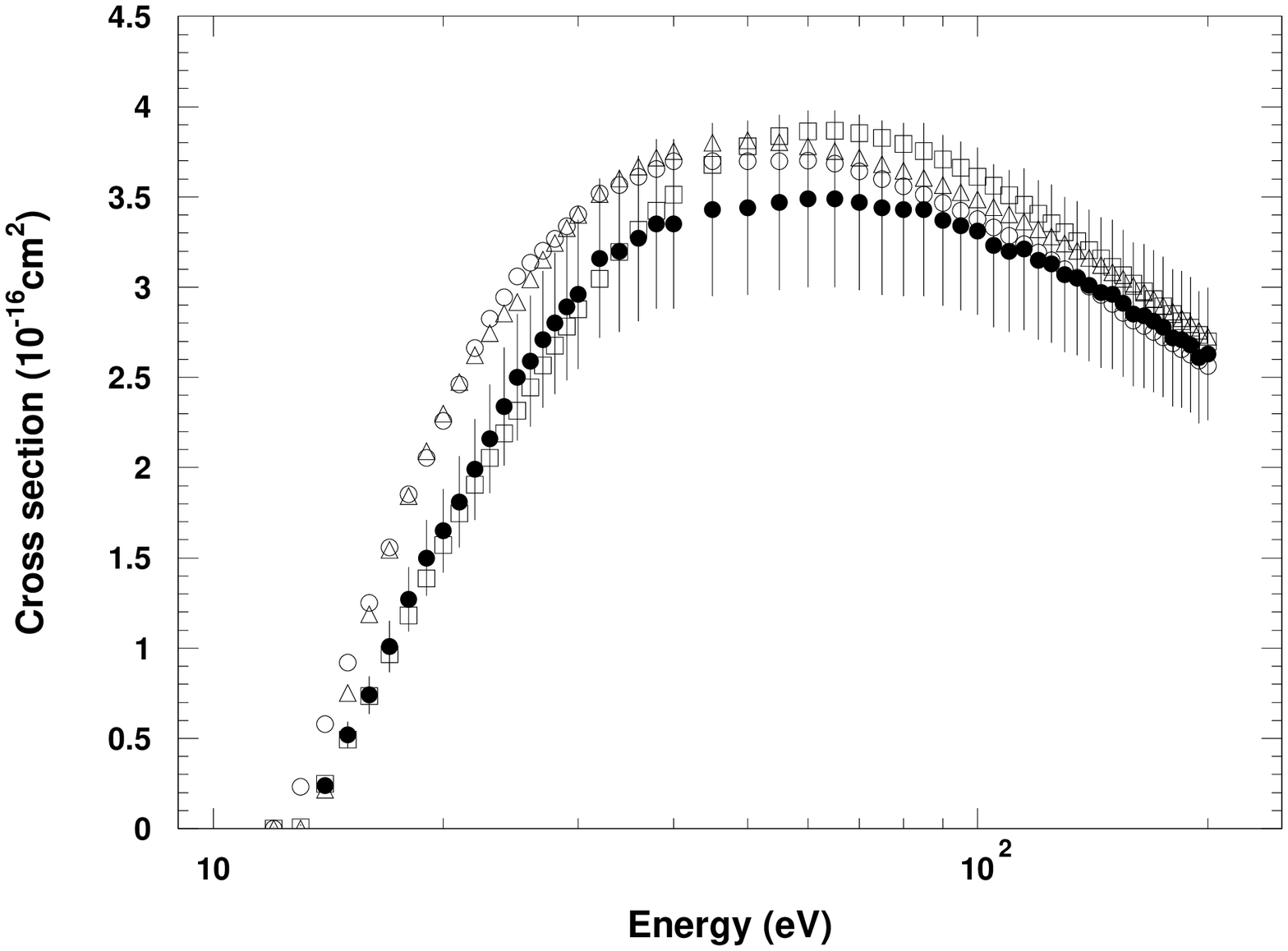}}
\caption{Cross section, Z=17: 
EEDL (empty circles), BEB model (empty squares), 
DM model (empty triangles)
and experimental data from \cite{expHayes} 
(black circles). }
\label{fig_beb17}
\end{figure}

\begin{figure}
\centerline{\includegraphics[angle=0,width=8cm]{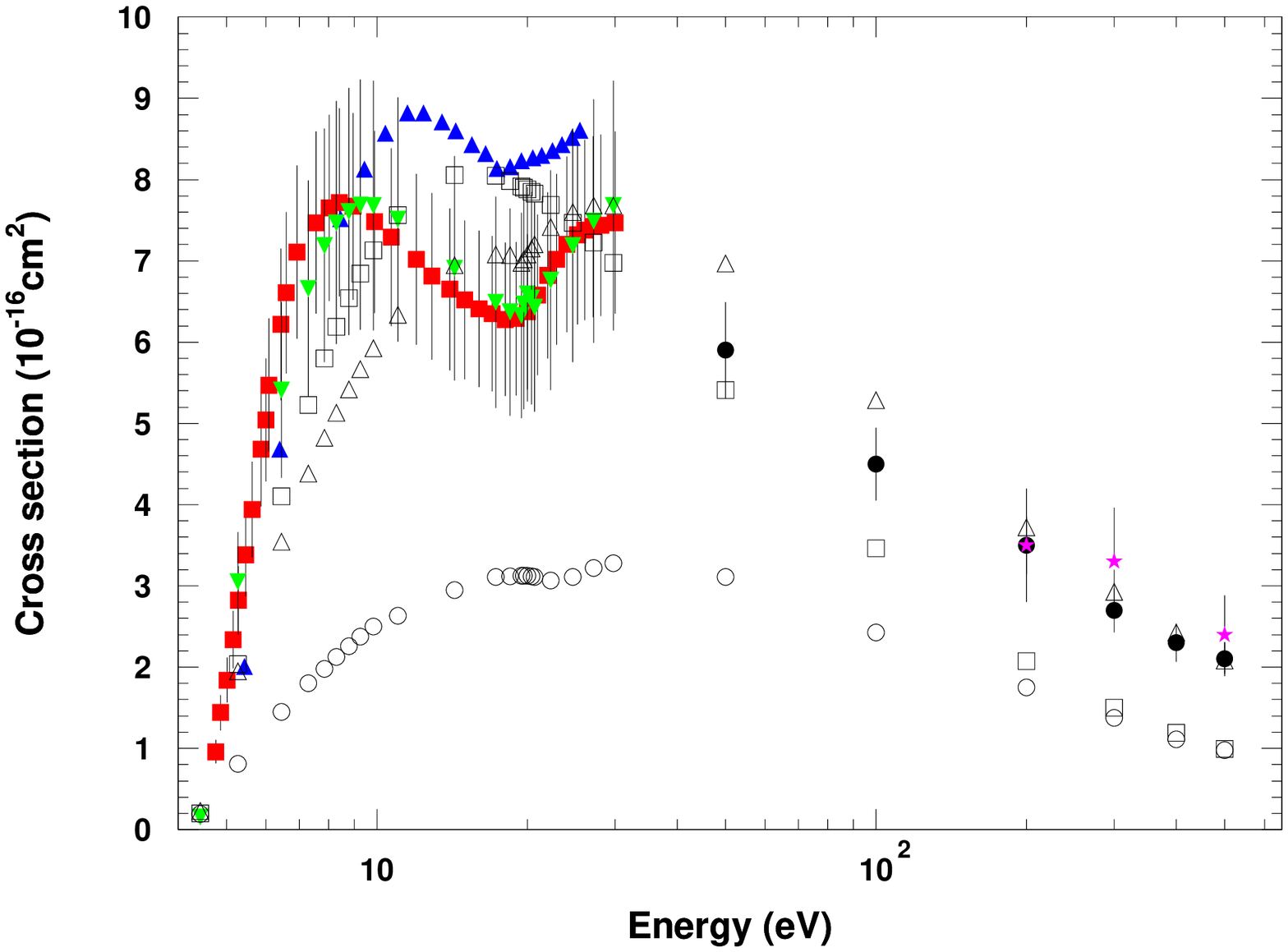}}
\caption{Cross section, Z=19: 
EEDL (empty circles), BEB model (empty squares), 
DM model (empty triangles)
and experimental data from 
\cite{expM1965} (black circles),
\cite{expZ1969} (red squares),
\cite{brink} (pink stars),
\cite{korchevoi} (blue triangles) and 
\cite{expKNygaard} (green upside-down triangles). }
\label{fig_beb19}
\end{figure}

\begin{figure}
\centerline{\includegraphics[angle=0,width=8cm]{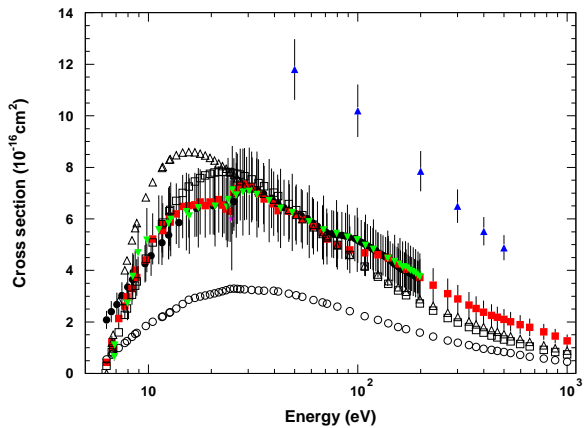}}
\caption{Cross section, Z=20: 
EEDL (empty circles), BEB model (empty squares), 
DM model (empty triangles)
and experimental data from 
\cite{expVainsh} (black circles),
\cite{expM1967}  (blue triangles),
\cite{rakhovskii} (green upside-down triangles),
\cite{okuno} (red squares) and 
\cite{schneider}  (pink stars). }
\label{fig_beb20}
\end{figure}

\begin{figure}
\centerline{\includegraphics[angle=0,width=8cm]{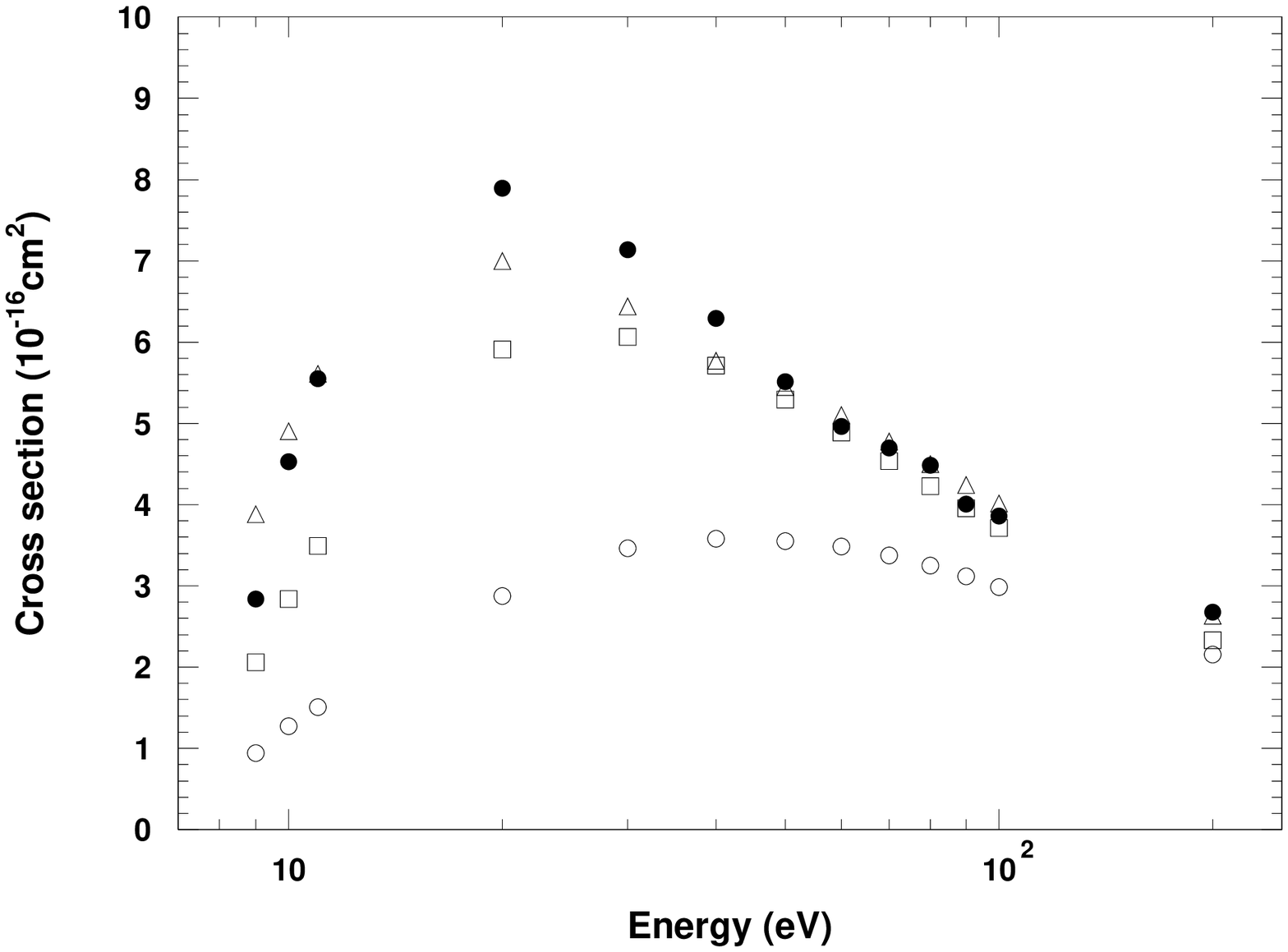}}
\caption{Cross section, Z=22: 
EEDL (empty circles), BEB model (empty squares), 
DM model (empty triangles)
and experimental data from Koparnski reported in \cite{dmMarg1994} 
(black circles). }
\label{fig_beb22}
\end{figure}

\begin{figure}
\centerline{\includegraphics[angle=0,width=8cm]{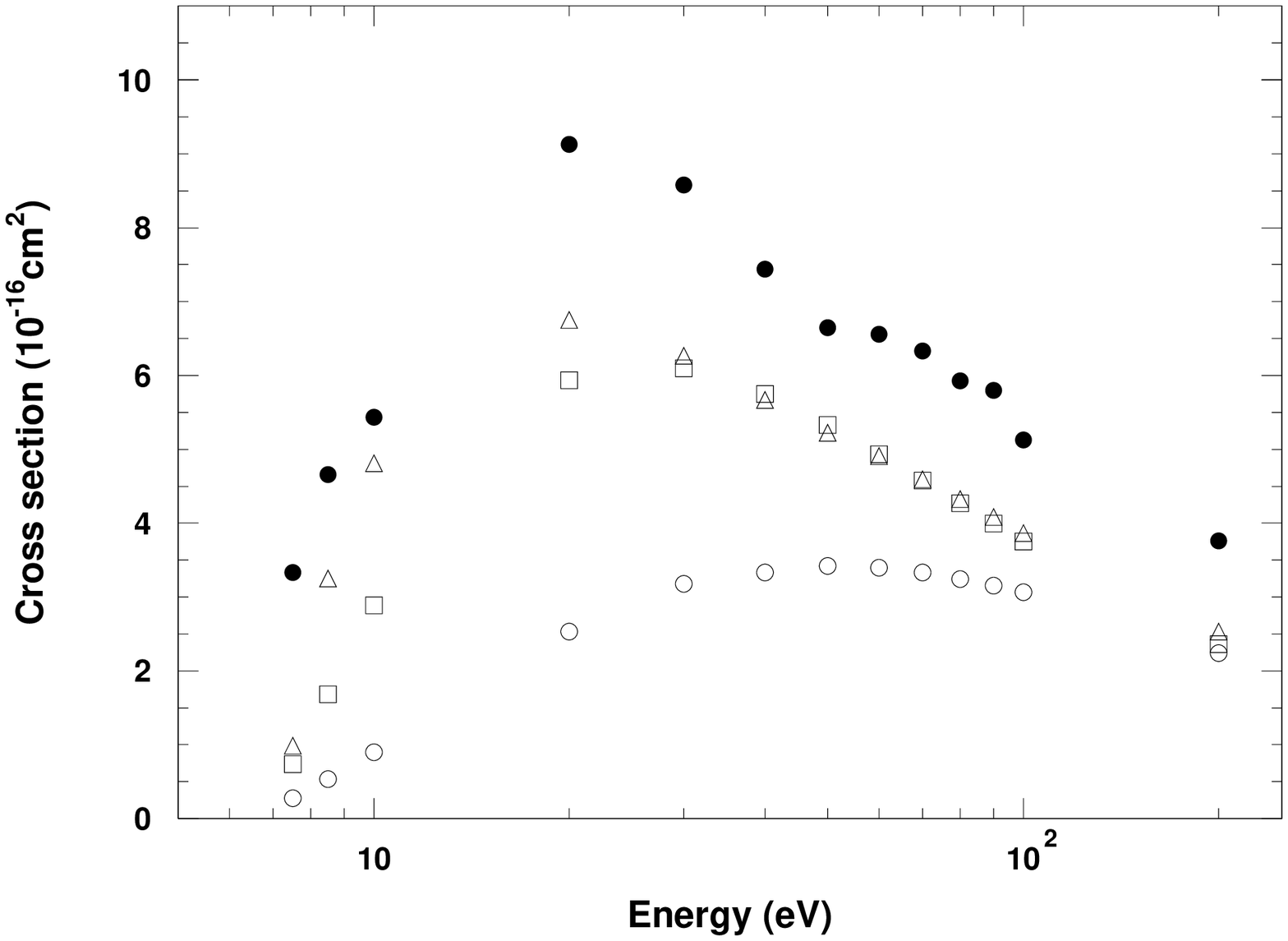}}
\caption{Cross section, Z=23: 
EEDL (empty circles), BEB model (empty squares), 
DM model (empty triangles)
and experimental data from Koparnski reported in \cite{dmMarg1994} 
(black circles). }
\label{fig_beb23}
\end{figure}

\begin{figure}
\centerline{\includegraphics[angle=0,width=8cm]{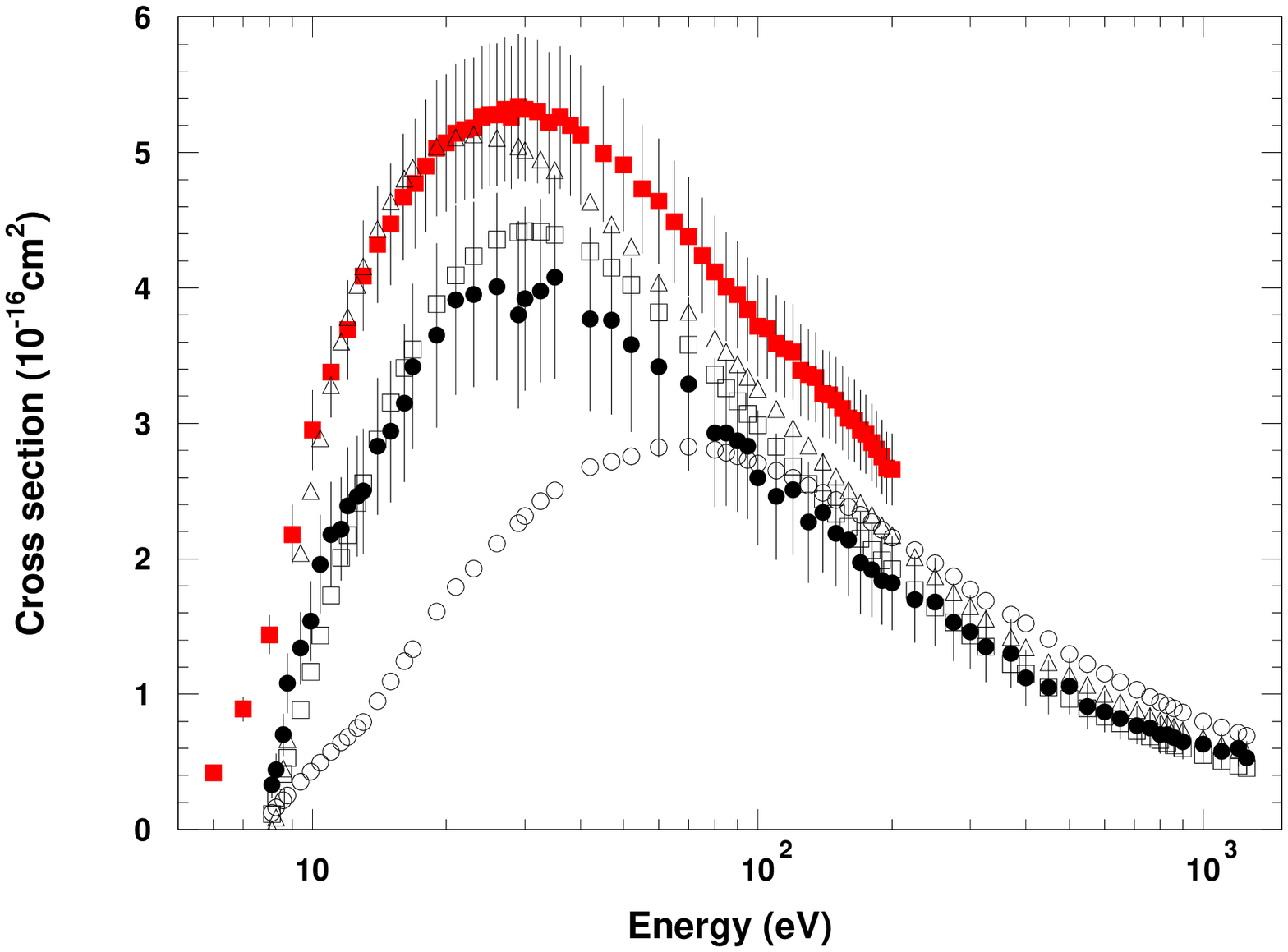}}
\caption{Cross section, Z=26: 
EEDL (empty circles), BEB model (empty squares), 
DM model (empty triangles)
and experimental data from \cite{expFreund} (red squares) and 
\cite{expFeShah1993} (black circles). }
\label{fig_beb26}
\end{figure}

\begin{figure}
\centerline{\includegraphics[angle=0,width=8cm]{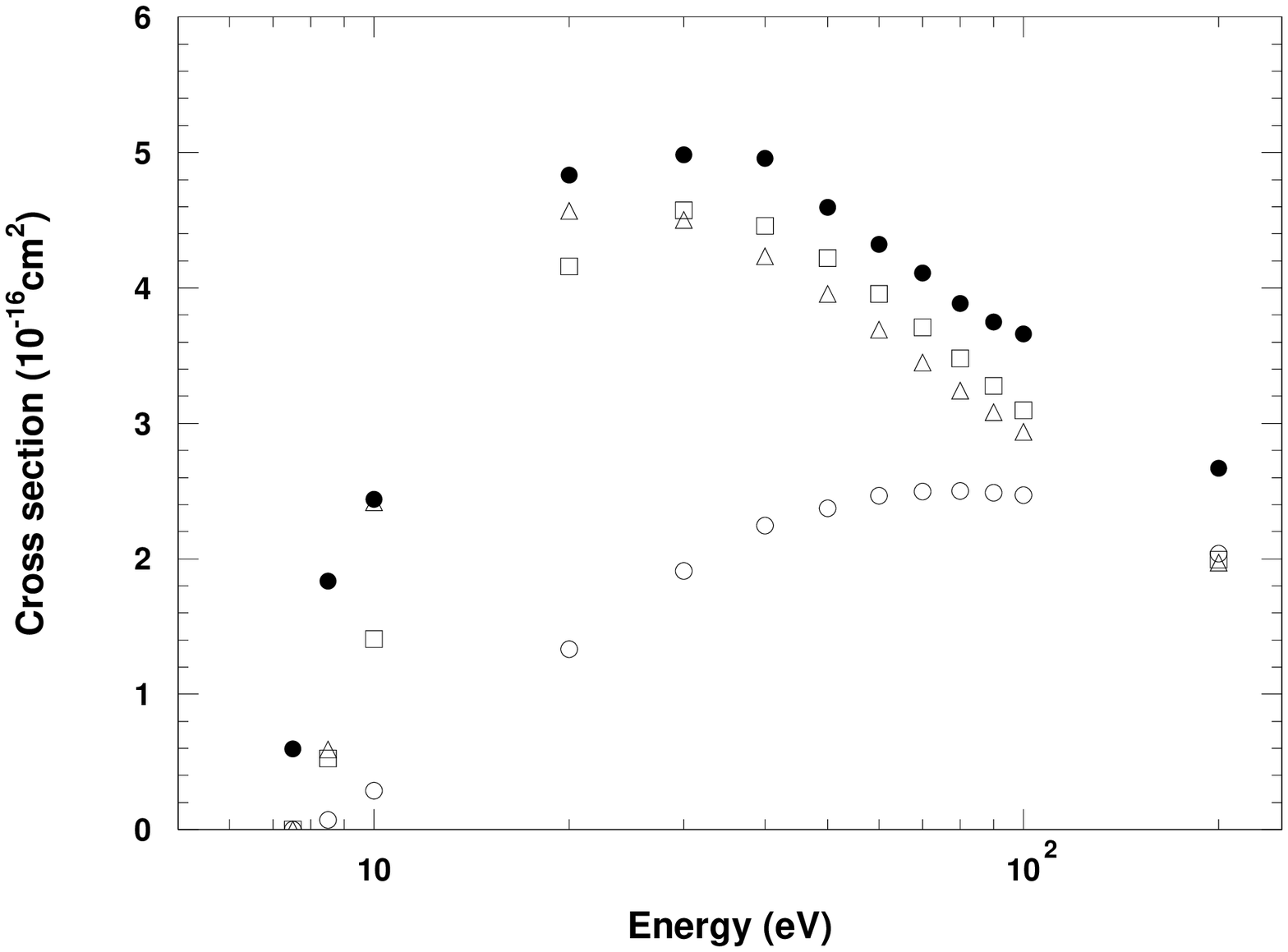}}
\caption{Cross section, Z=28: 
EEDL (empty circles), BEB model (empty squares), 
DM model (empty triangles)
and experimental data from Koparnski reported in \cite{dmMarg1994}  
(black circles). }
\label{fig_beb28}
\end{figure}

\begin{figure}
\centerline{\includegraphics[angle=0,width=8cm]{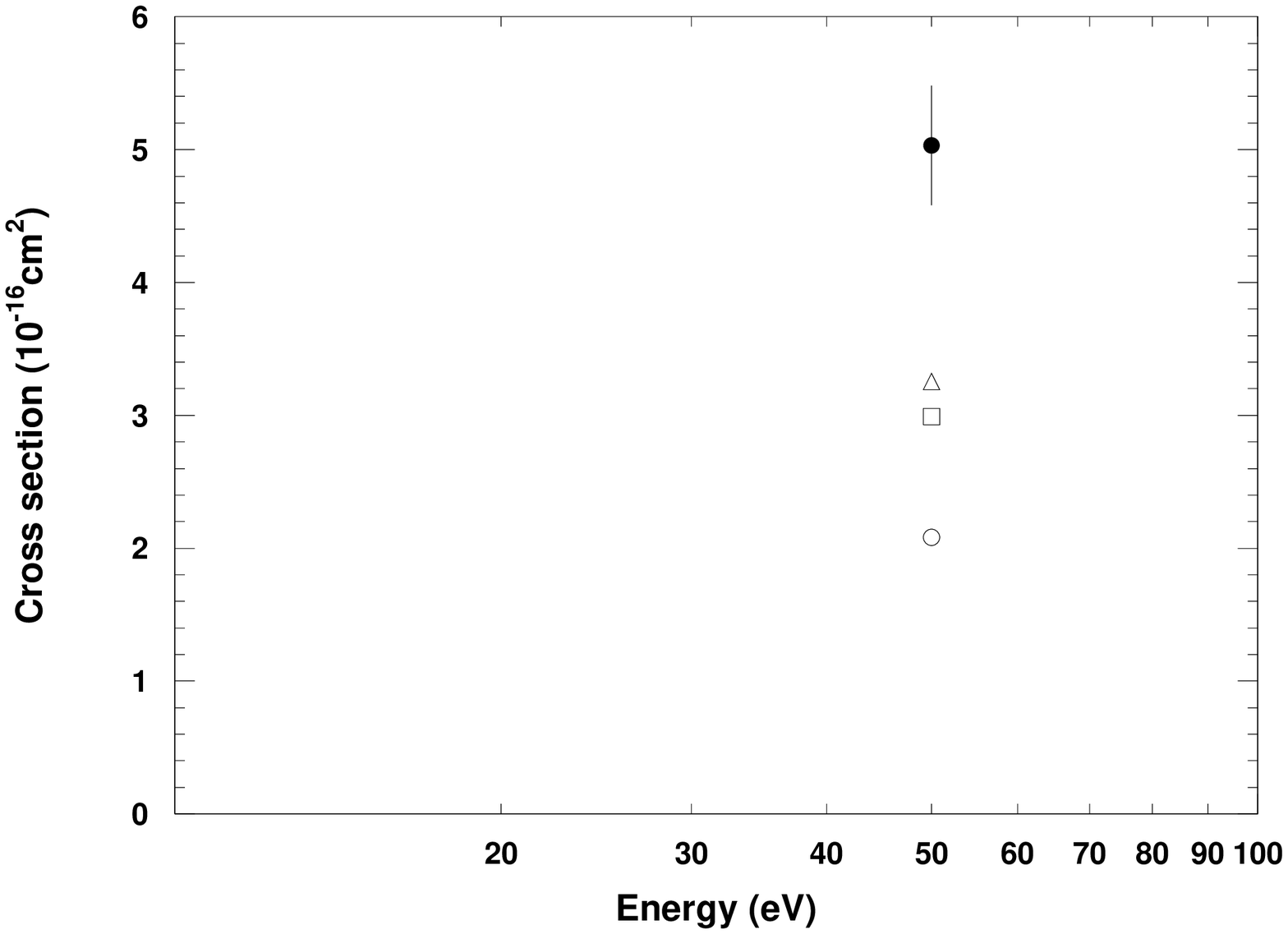}}
\caption{Cross section, Z=30: 
EEDL (empty circles), BEB model (empty squares), 
DM model (empty triangles)
and experimental data from \cite{expZnCd} (black circles). }
\label{fig_beb30}
\end{figure}

\begin{figure}
\centerline{\includegraphics[angle=0,width=8cm]{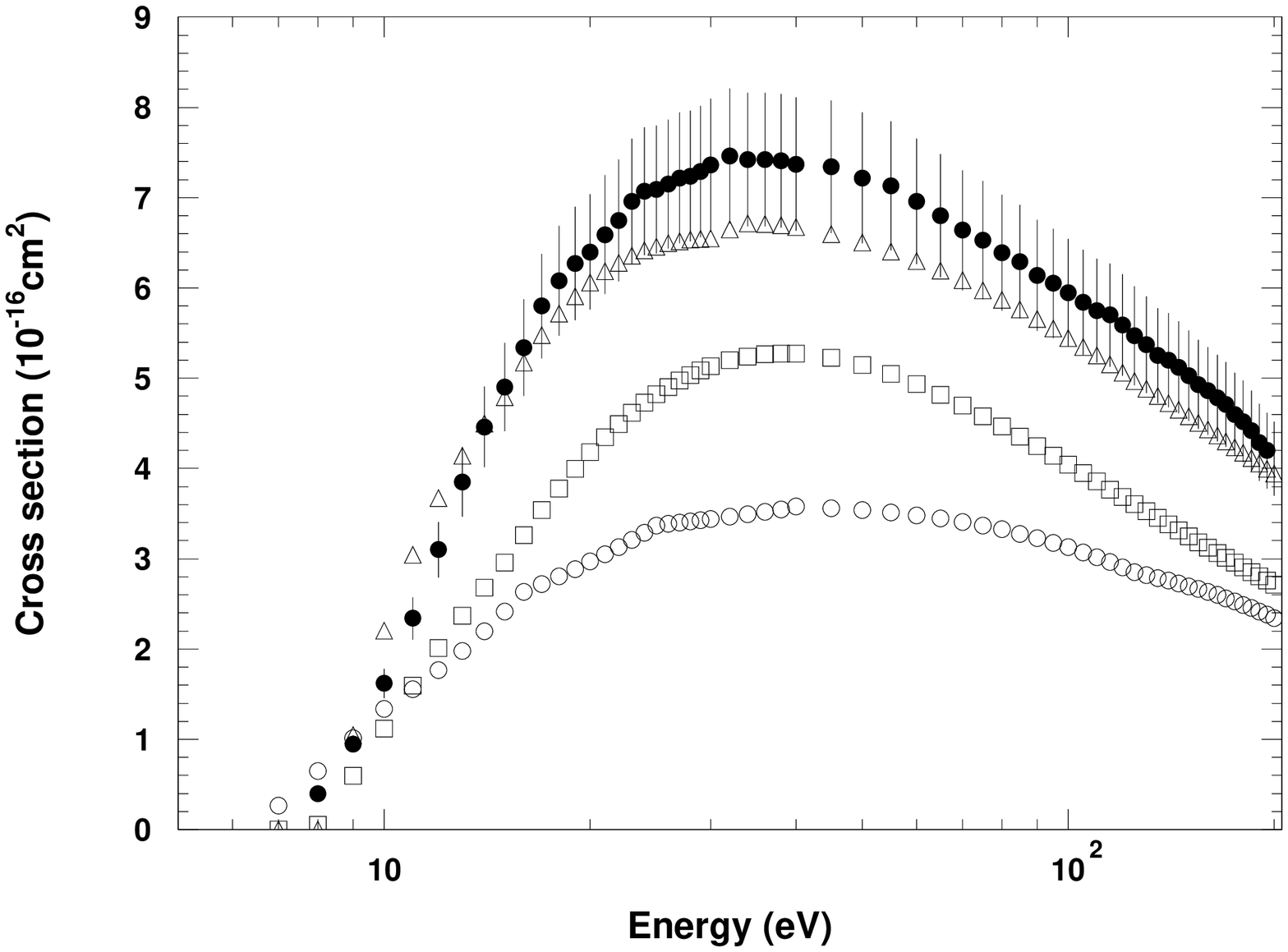}}
\caption{Cross section, Z=32: 
EEDL (empty circles), BEB model (empty squares), 
DM model (empty triangles)
and experimental data from \cite{expFreund} 
(black circles). }
\label{fig_beb32}
\end{figure}

\begin{figure}
\centerline{\includegraphics[angle=0,width=8cm]{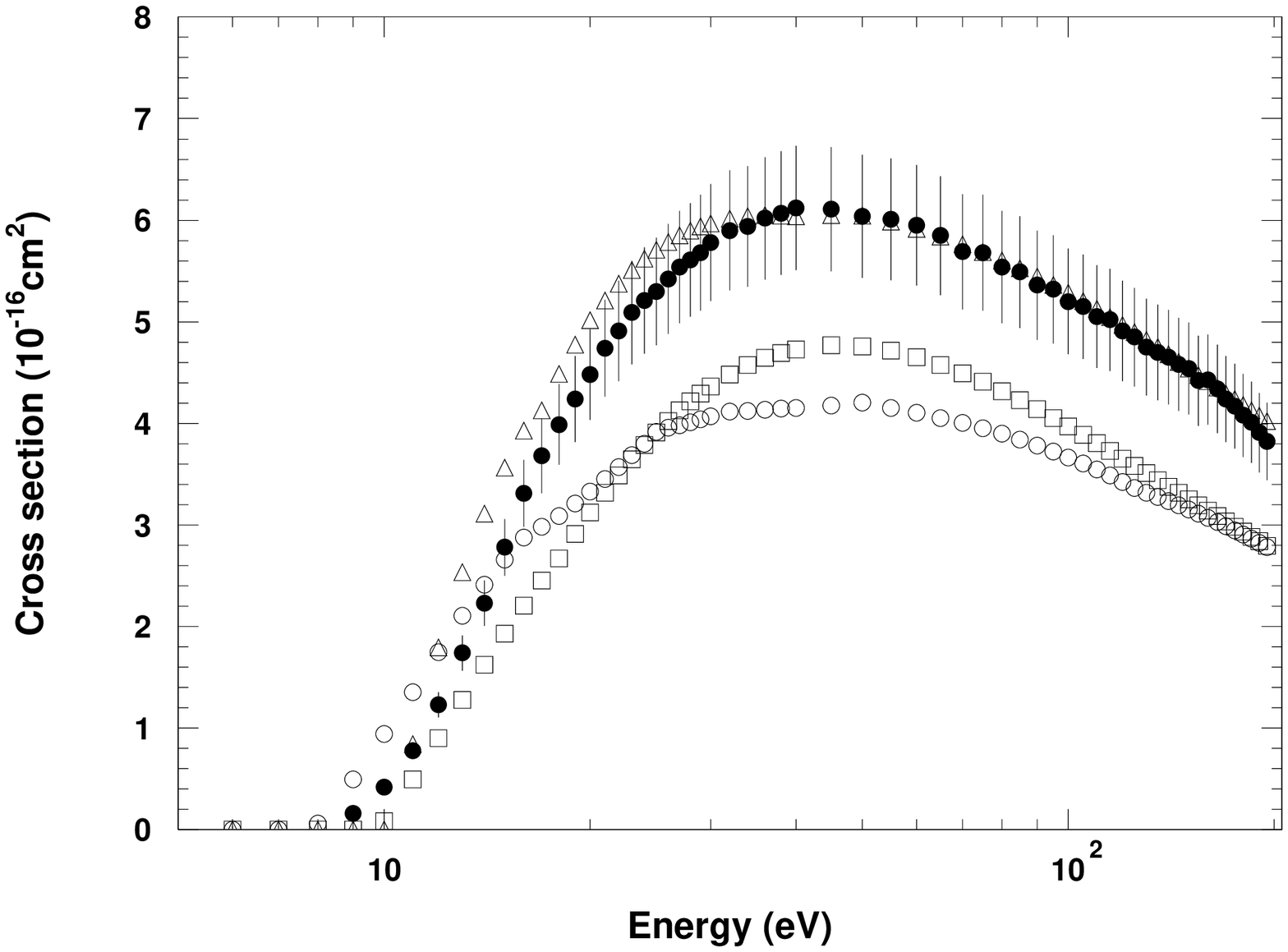}}
\caption{Cross section, Z=33: 
EEDL (empty circles), BEB model (empty squares), 
DM model (empty triangles)
and experimental data from \cite{expFreund} 
(black circles). }
\label{fig_beb33}
\end{figure}

\begin{figure}
\centerline{\includegraphics[angle=0,width=8cm]{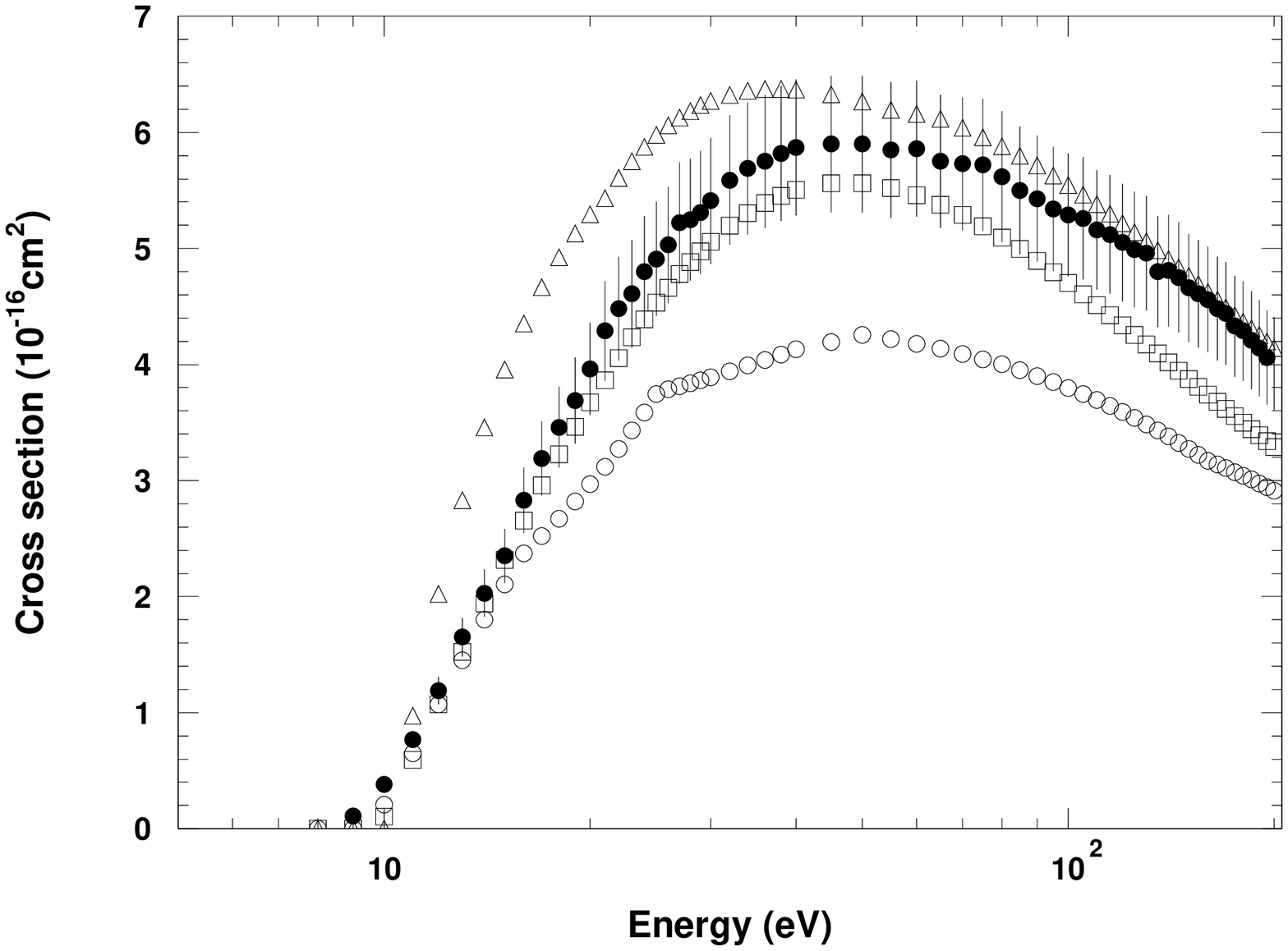}}
\caption{Cross section, Z=34: 
EEDL (empty circles), BEB model (empty squares), 
DM model (empty triangles)
and experimental data from \cite{expFreund} 
(black circles). }
\label{fig_beb34}
\end{figure}

\begin{figure}
\centerline{\includegraphics[angle=0,width=8cm]{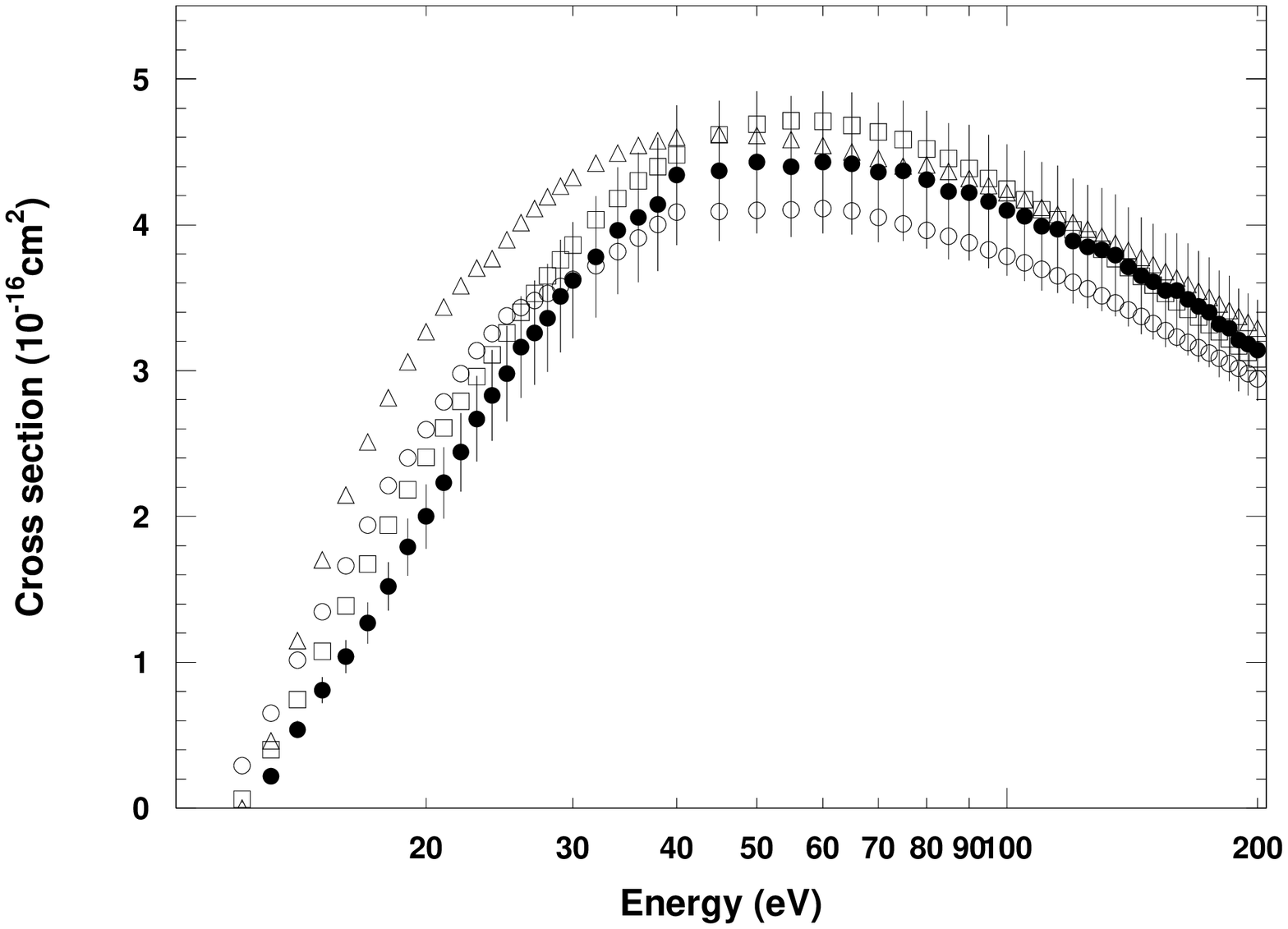}}
\caption{Cross section, Z=35: 
EEDL (empty circles), BEB model (empty squares), 
DM model (empty triangles)
and experimental data from \cite{expHayes} 
(black circles). }
\label{fig_beb35}
\end{figure}

\begin{figure}
\centerline{\includegraphics[angle=0,width=8cm]{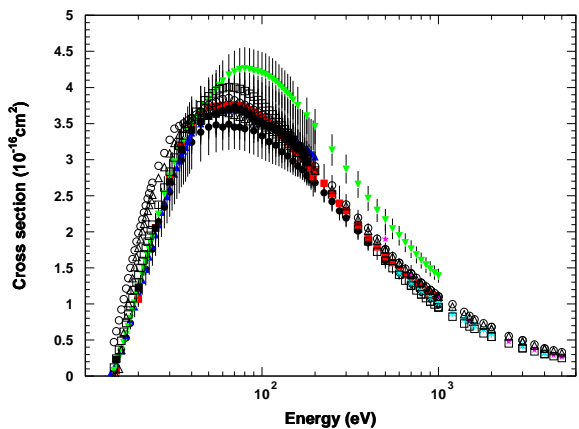}}
\caption{Cross section, Z=36: 
EEDL (empty circles), BEB model (empty squares), 
DM model (empty triangles)
and experimental data from 
\cite{expRejoub2002} (black circles),
\cite{wetzel}  (blue triangles),
\cite{krishnakumar}  (red squares),
\cite{expNagy1980}  (pink stars),
\cite{rapp}  (green upside-down triangles),
\cite{expArKrXeSchram}  (turquoise asterisks) and
\cite{stephan}  (black squares). }
\label{fig_beb36}
\end{figure}

\begin{figure}
\centerline{\includegraphics[angle=0,width=8cm]{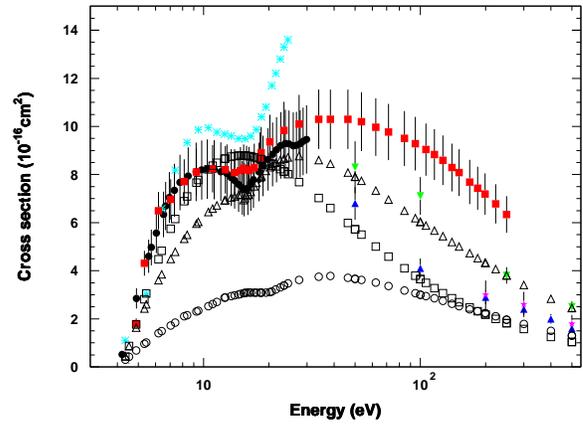}}
\caption{Cross section, Z=37: 
EEDL (empty circles), BEB model (empty squares), 
DM model (empty triangles)
and experimental data from 
\cite{expM1965}  (blue triangles),
\cite{expZ1969} (black circles),
\cite{brink}  (pink stars),
\cite{korchevoi}  (turquoise asterisks),
\cite{expRbNygaard} (red squares) and
\cite{schappe} (green upside-down triangles). }
\label{fig_beb37}
\end{figure}

\begin{figure}
\centerline{\includegraphics[angle=0,width=8cm]{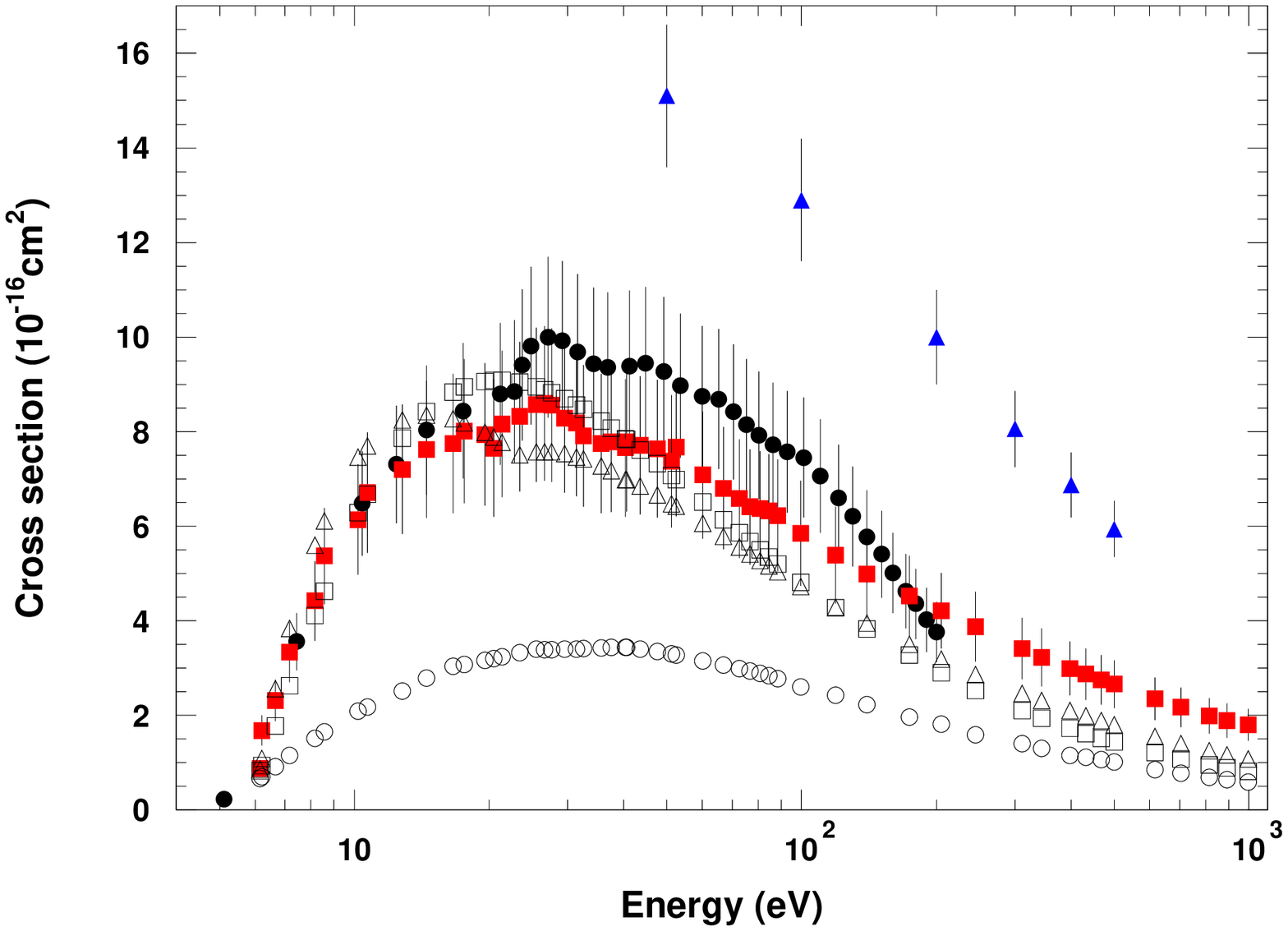}}
\caption{Cross section, Z=38: 
EEDL (empty circles), BEB model (empty squares), 
DM model (empty triangles)
and experimental data from 
\cite{expM1967}  (blue triangles),
\cite{expVainsh} (black circles) and
\cite{okuno} (red squares). }
\label{fig_beb38}
\end{figure}

\begin{figure}
\centerline{\includegraphics[angle=0,width=8cm]{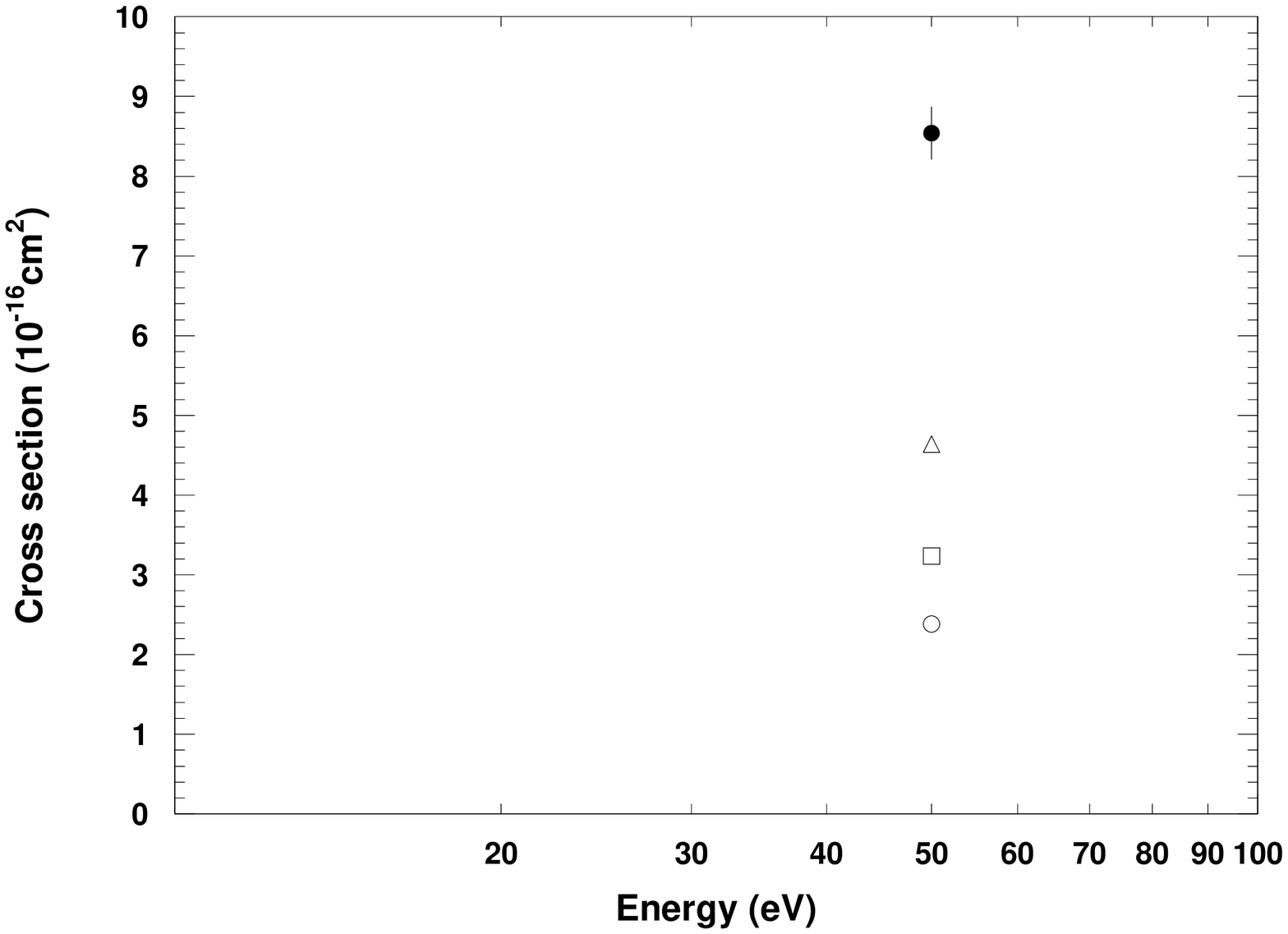}}
\caption{Cross section, Z=48: 
EEDL (empty circles), BEB model (empty squares), 
DM model (empty triangles)
and experimental data from \cite{expZnCd} (black circles). }
\label{fig_beb48}
\end{figure}

\begin{figure}
\centerline{\includegraphics[angle=0,width=8cm]{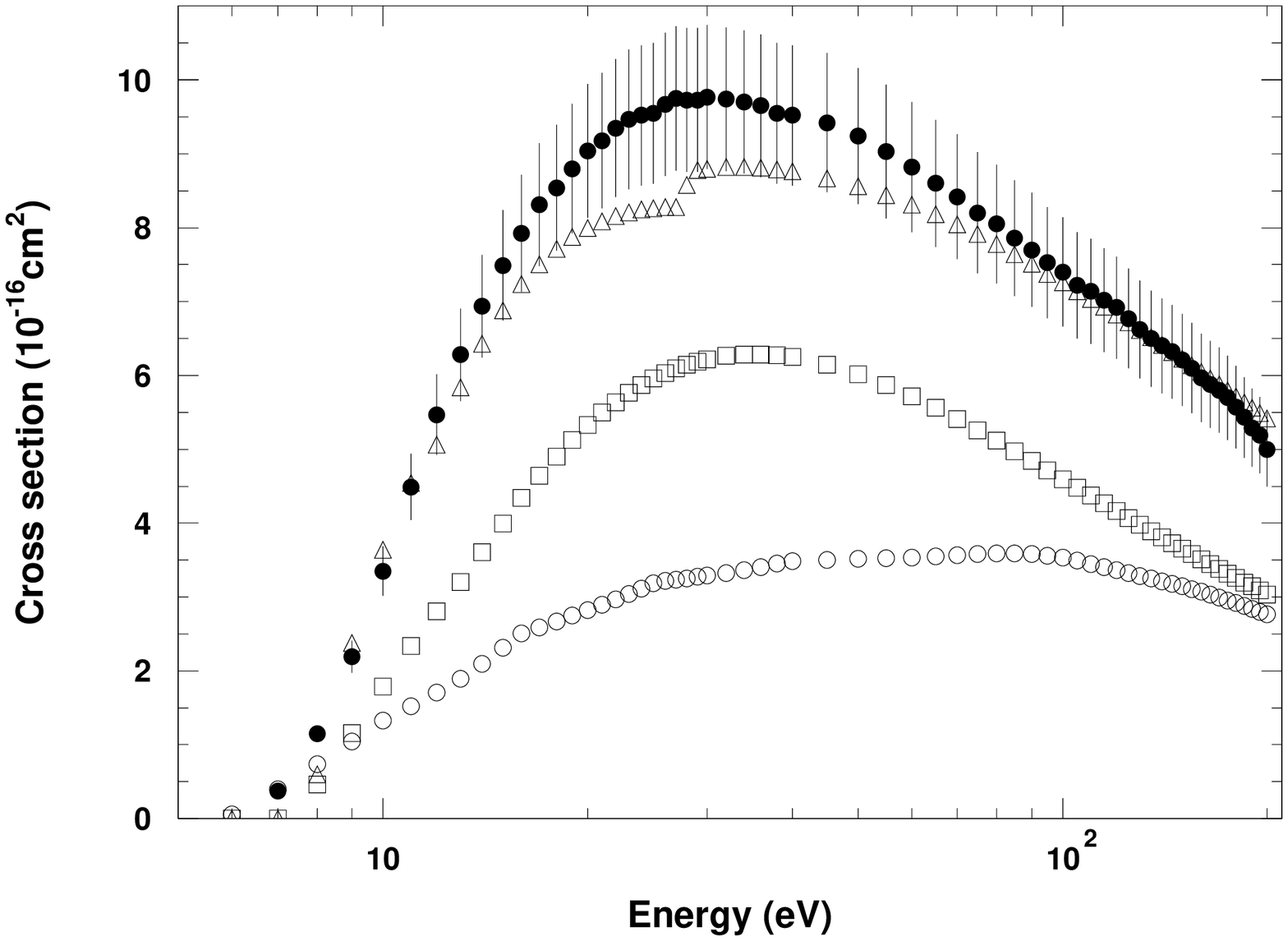}}
\caption{Cross section, Z=50: 
EEDL (empty circles), BEB model (empty squares), 
DM model (empty triangles)
and experimental data from \cite{expFreund} 
(black circles). }
\label{fig_beb50}
\end{figure}

\begin{figure}
\centerline{\includegraphics[angle=0,width=8cm]{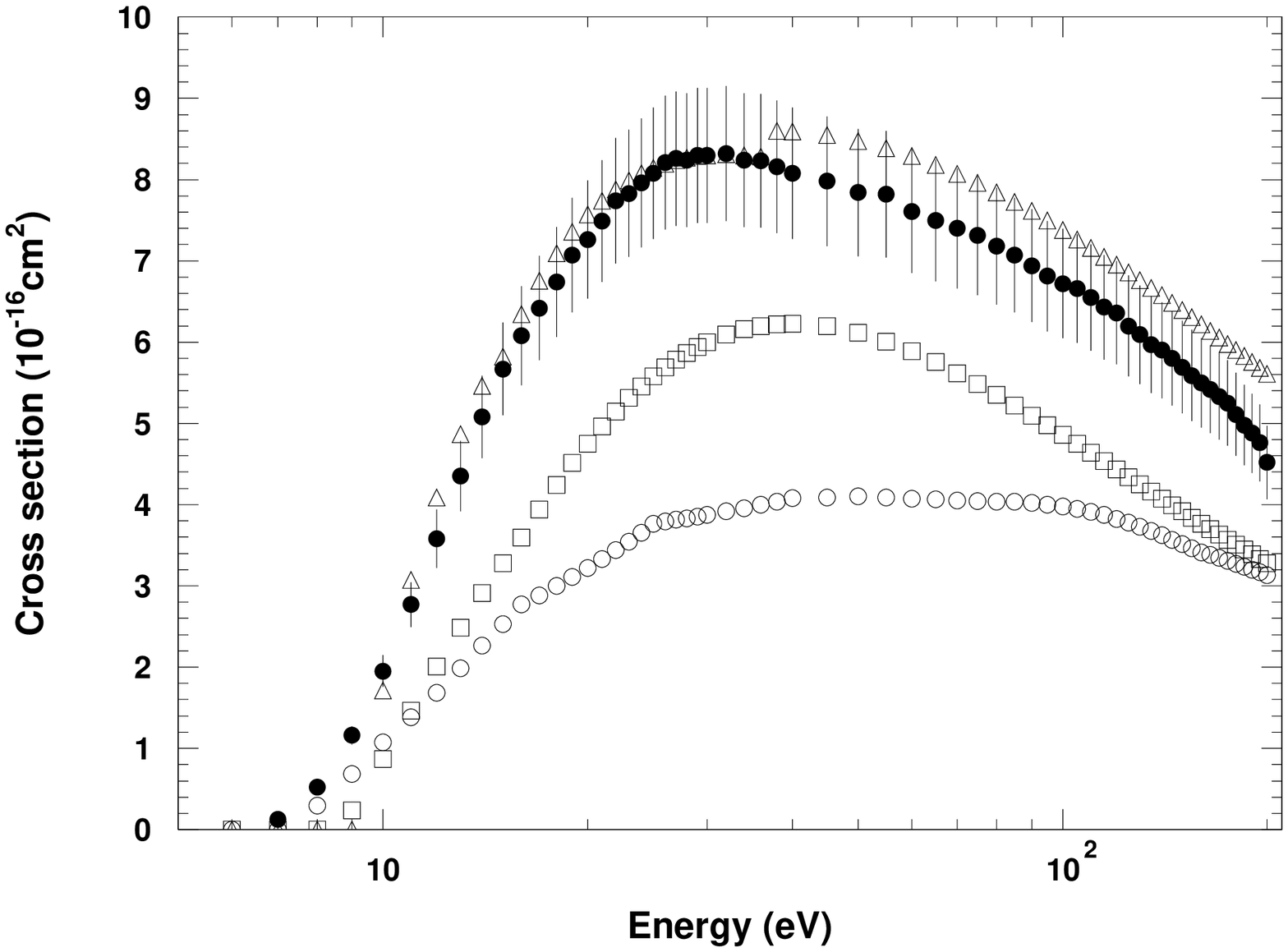}}
\caption{Cross section, Z=51: 
EEDL (empty circles), BEB model (empty squares), 
DM model (empty triangles)
and experimental data from \cite{expFreund} 
(black circles). }
\label{fig_beb51}
\end{figure}

\begin{figure}
\centerline{\includegraphics[angle=0,width=8cm]{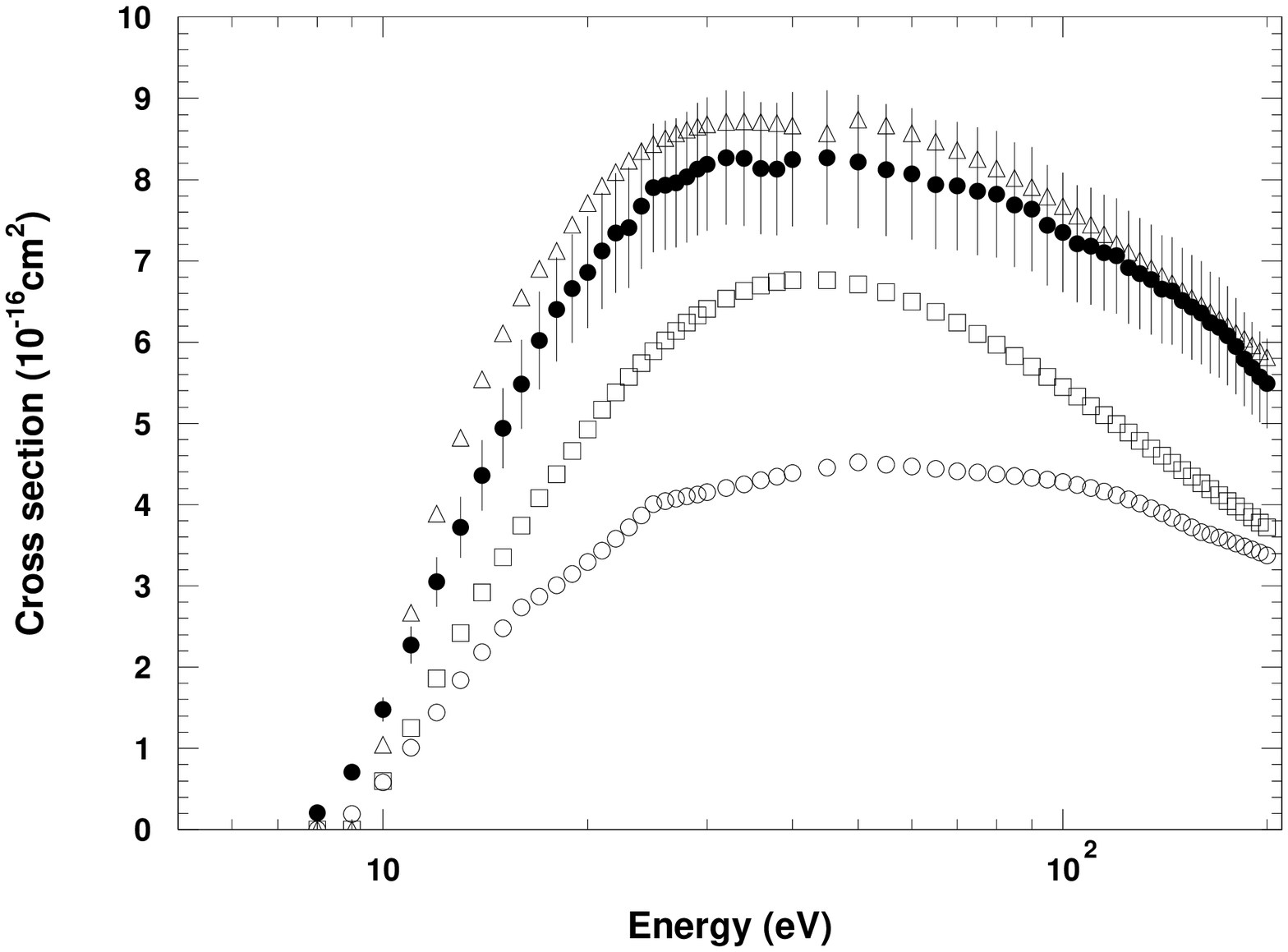}}
\caption{Cross section, Z=52: 
EEDL (empty circles), BEB model (empty squares), 
DM model (empty triangles)
and experimental data from \cite{expFreund} 
(black circles). }
\label{fig_beb52}
\end{figure}

\begin{figure}
\centerline{\includegraphics[angle=0,width=8cm]{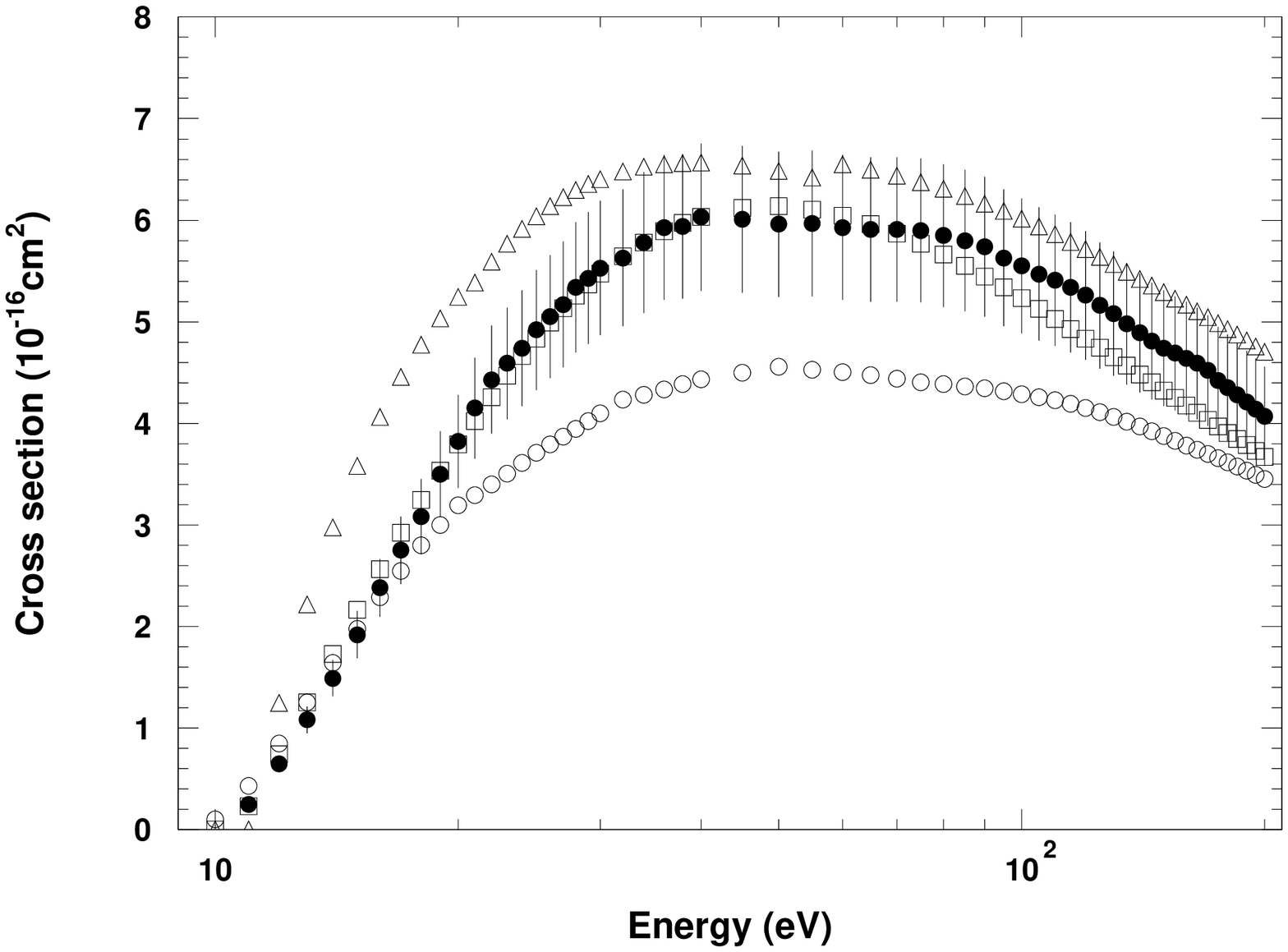}}
\caption{Cross section, Z=53: 
EEDL (empty circles), BEB model (empty squares), 
DM model (empty triangles)
and experimental data from \cite{expHayes} 
(black circles). }
\label{fig_beb53}
\end{figure}

\begin{figure}
\centerline{\includegraphics[angle=0,width=8cm]{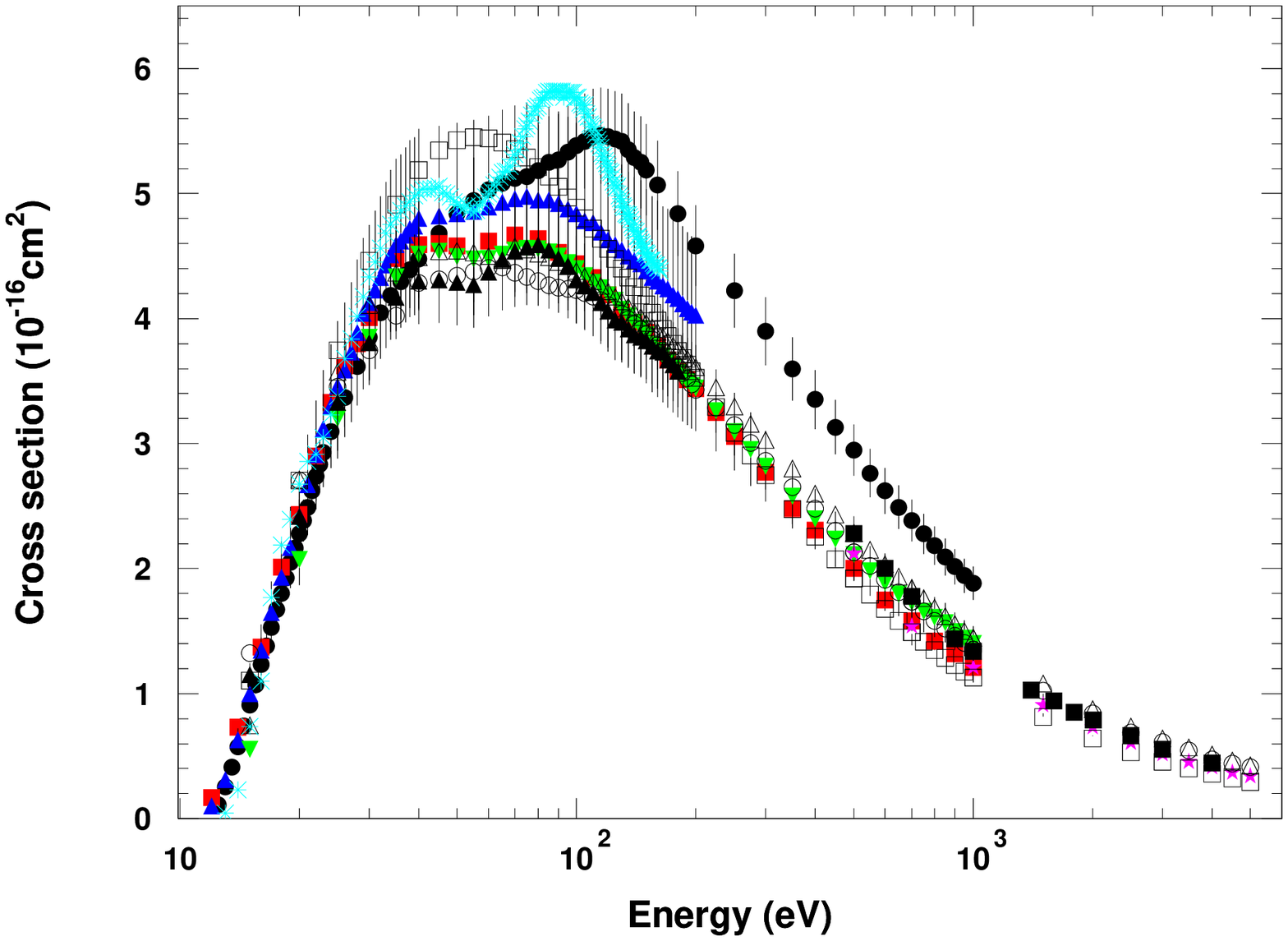}}
\caption{Cross section, Z=54: 
EEDL (empty circles), BEB model (empty squares), 
DM model (empty triangles)
and experimental data from 
\cite{rapp} (black circles),
\cite{expRejoub2002} (red squares),
\cite{wetzel} (blue triangles),
\cite{krishnakumar}  (green upside-down triangles),
\cite{mathur} (turquoise asterisks),
\cite{expNagy1980} (pink stars),
\cite{expArKrXeSchram} (black squares) and 
\cite{expStephan1984} (black triangles). }
\label{fig_beb54}
\end{figure}

\begin{figure}
\centerline{\includegraphics[angle=0,width=8cm]{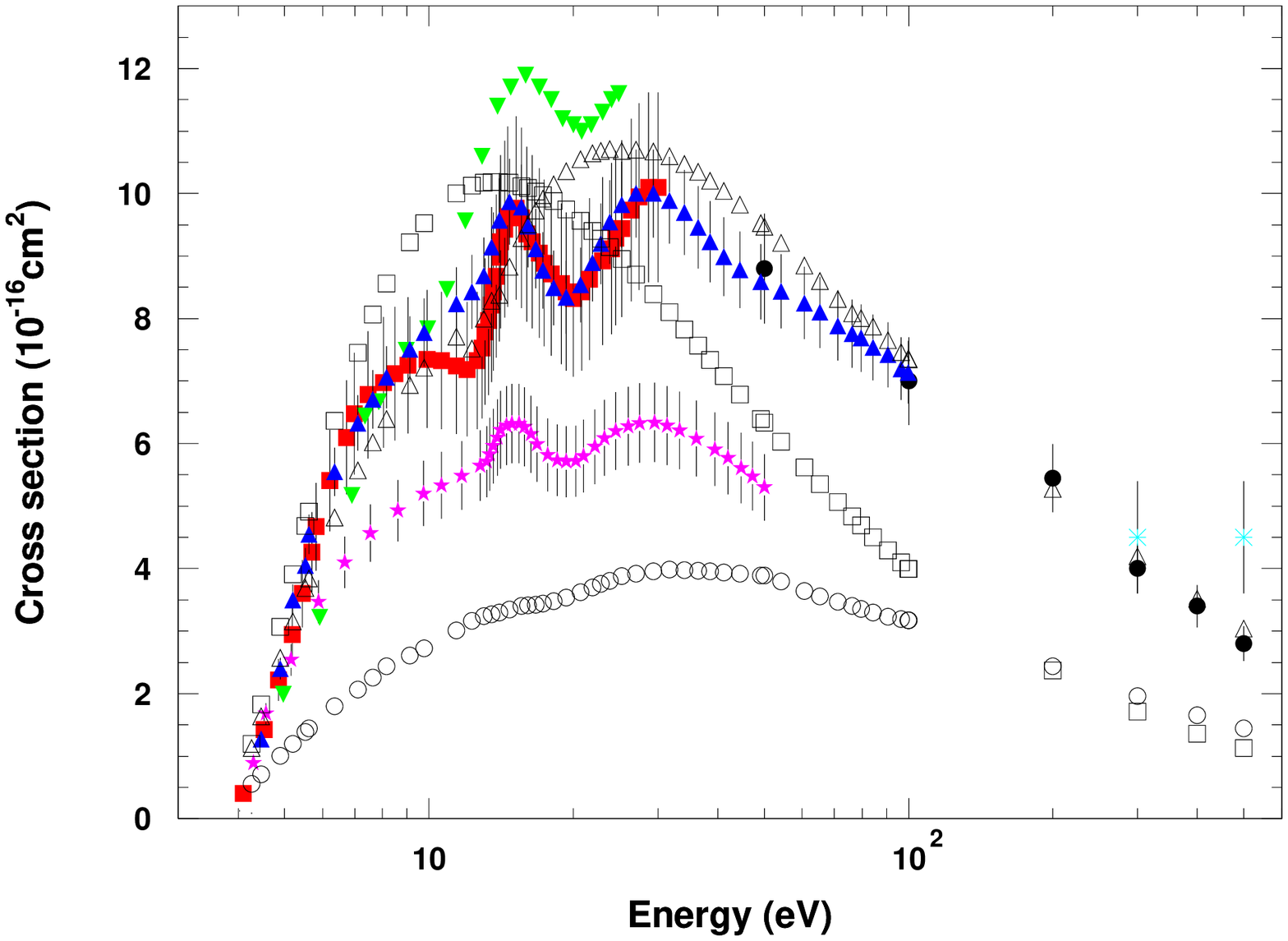}}
\caption{Cross section, Z=55: 
EEDL (empty circles), BEB model (empty squares), 
DM model (empty triangles)
and experimental data from 
\cite{expM1965} (black circles),
\cite{expZ1969}  (red squares),
\cite{brink} (turquoise asterisks),
\cite{heil} (pink stars),
\cite{korchevoi} (green upside-down triangles) and
\cite{expCsNygaard} (blue triangles).}
\label{fig_beb55}
\end{figure}

\begin{figure}
\centerline{\includegraphics[angle=0,width=8cm]{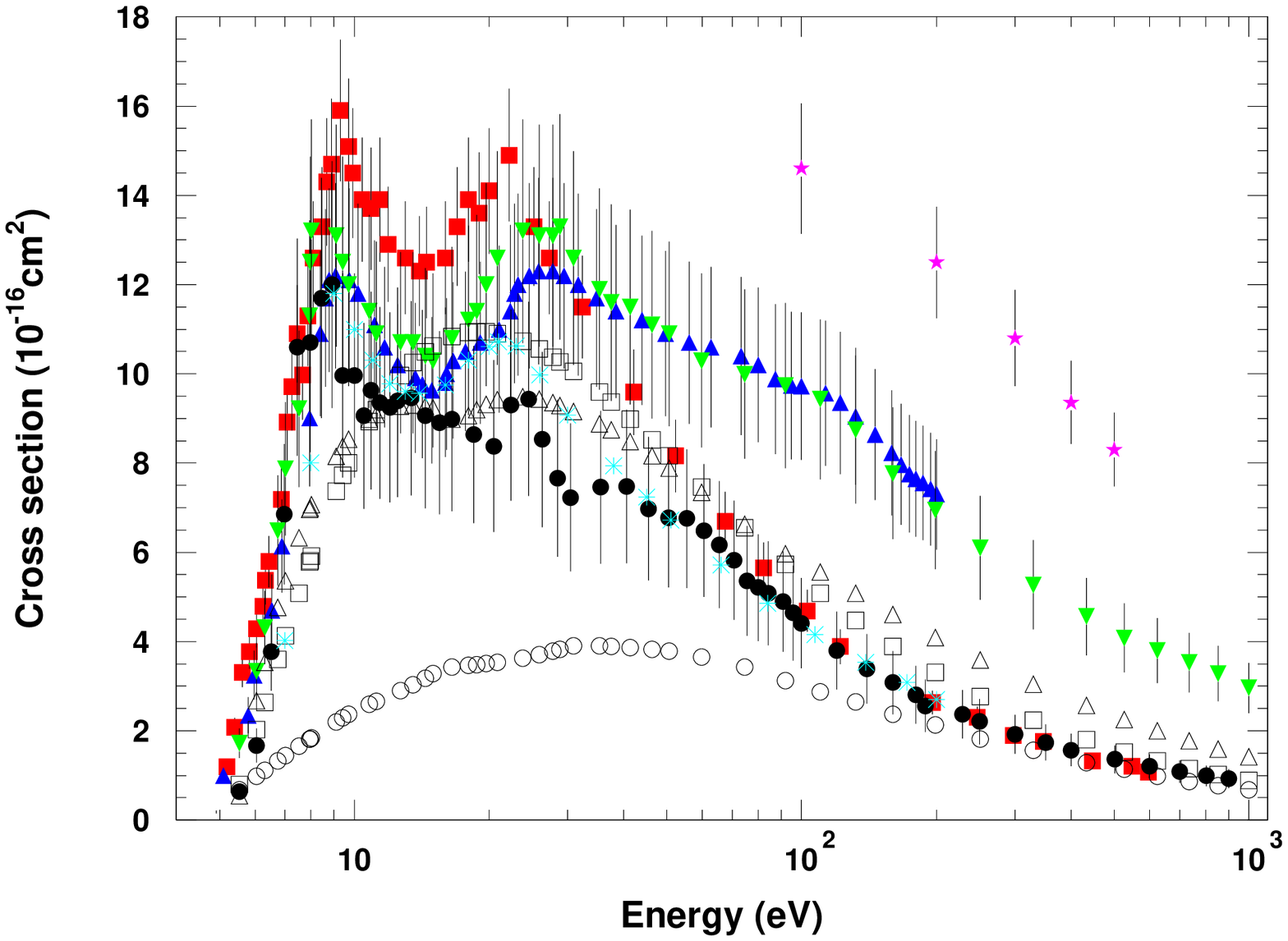}}
\caption{Cross section, Z=56: 
EEDL (empty circles), BEB model (empty squares), 
DM model (empty triangles)
and experimental data from 
\cite{expM1967}  (pink stars),
\cite{dettmann}  (red squares),
\cite{expVainsh}  (blue triangles),
\cite{expYagi2000} (black circles),
\cite{okuno} (green upside-down triangles),
\cite{golovachBaPb} (turquoise asterisks). }
\label{fig_beb56}
\end{figure}

\begin{figure}
\centerline{\includegraphics[angle=0,width=8cm]{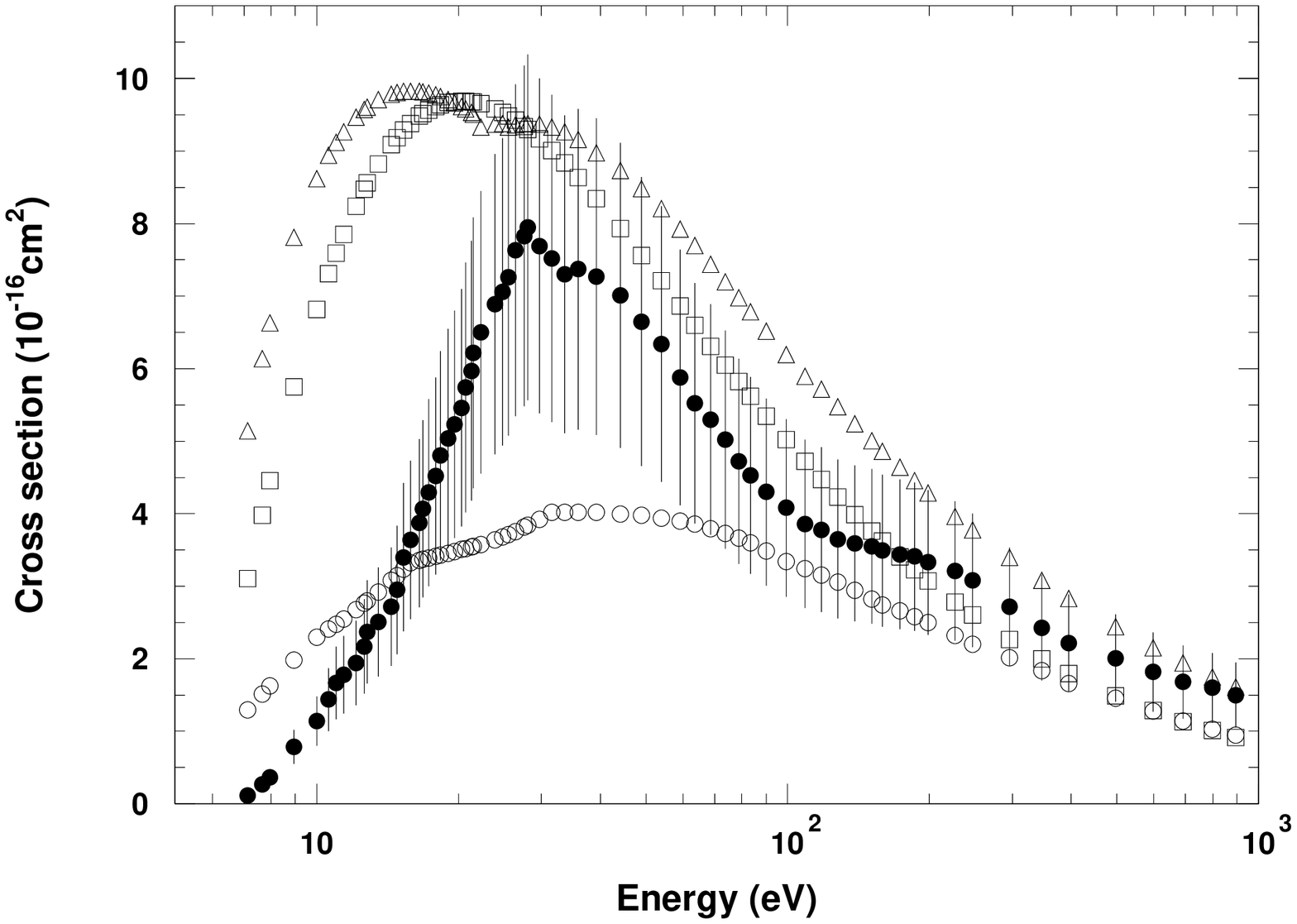}}
\caption{Cross section, Z=58: 
EEDL (empty circles), BEB model (empty squares), 
DM model (empty triangles)
and experimental data from \cite{expYagi2001} 
(black circles). }
\label{fig_beb58}
\end{figure}

\begin{figure}
\centerline{\includegraphics[angle=0,width=8cm]{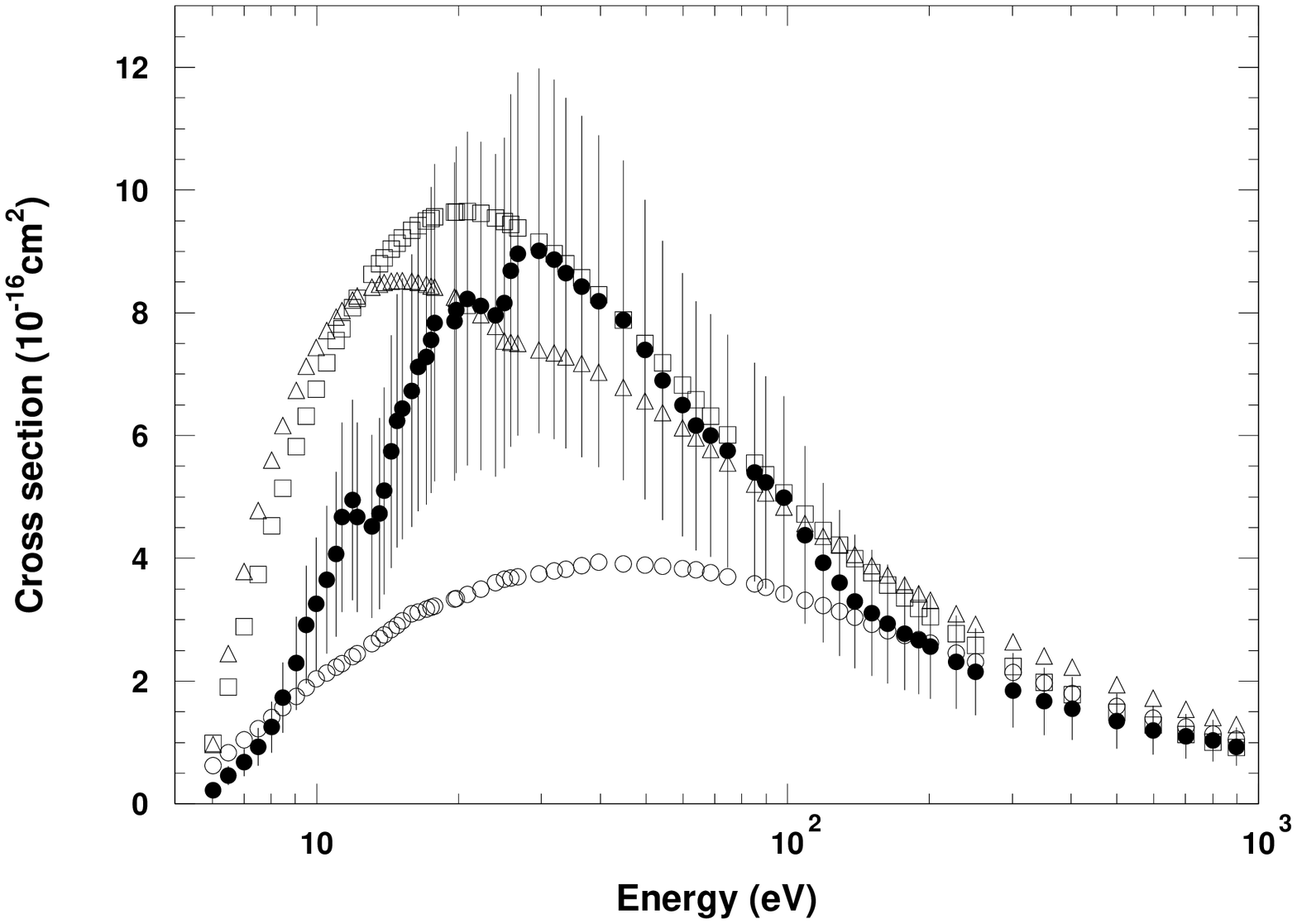}}
\caption{Cross section, Z=60: 
EEDL (empty circles), BEB model (empty squares), 
DM model (empty triangles)
and experimental data from \cite{expYagi2001} 
(black circles). }
\label{fig_beb60}
\end{figure}

\begin{figure}
\centerline{\includegraphics[angle=0,width=8cm]{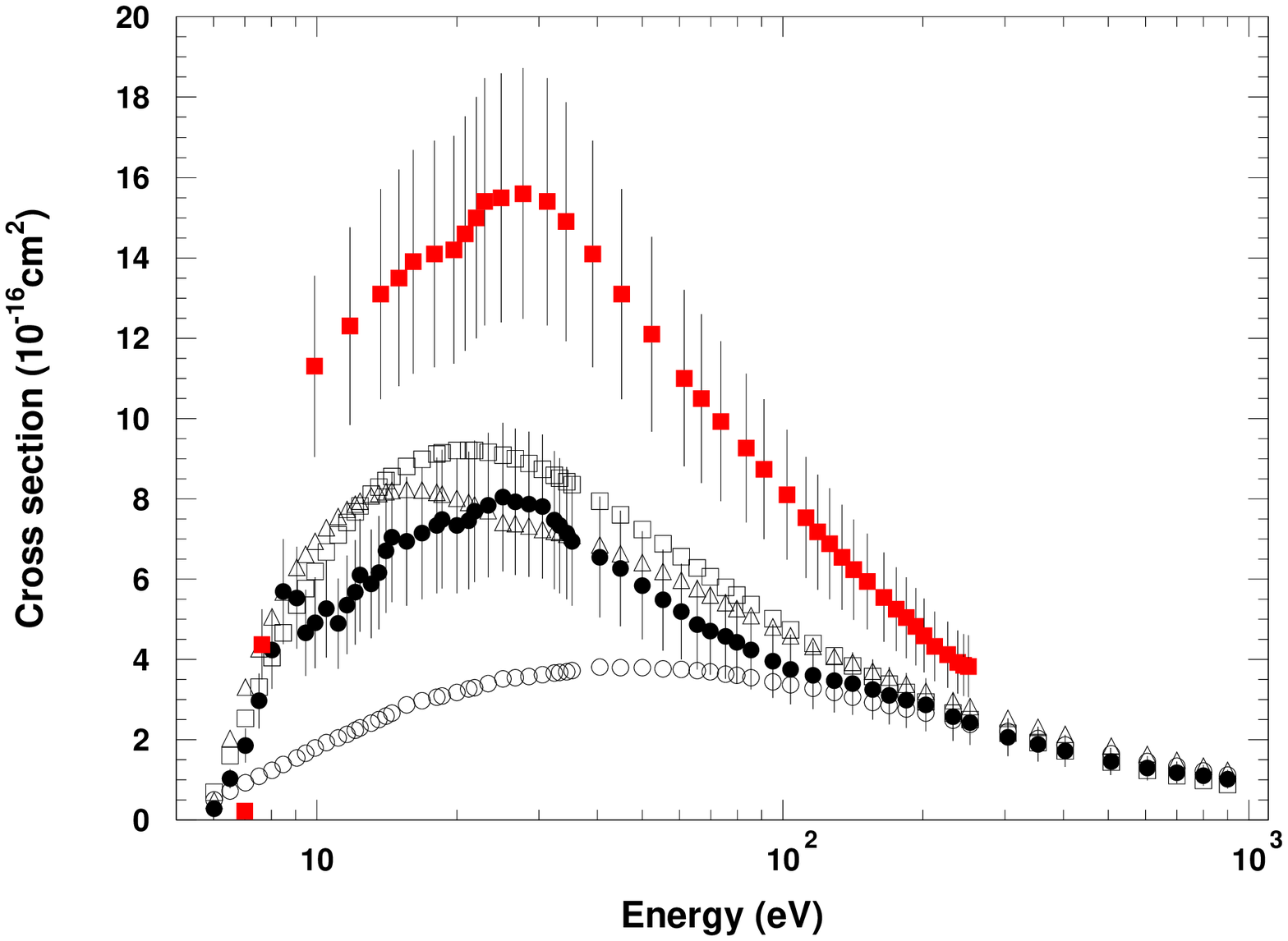}}
\caption{Cross section, Z=62: 
EEDL (empty circles), BEB model (empty squares), 
DM model (empty triangles)
and experimental data from \cite{expYagi2001} (black circles)
and \cite{shimon} (red squares). }
\label{fig_beb62}
\end{figure}

\begin{figure}
\centerline{\includegraphics[angle=0,width=8cm]{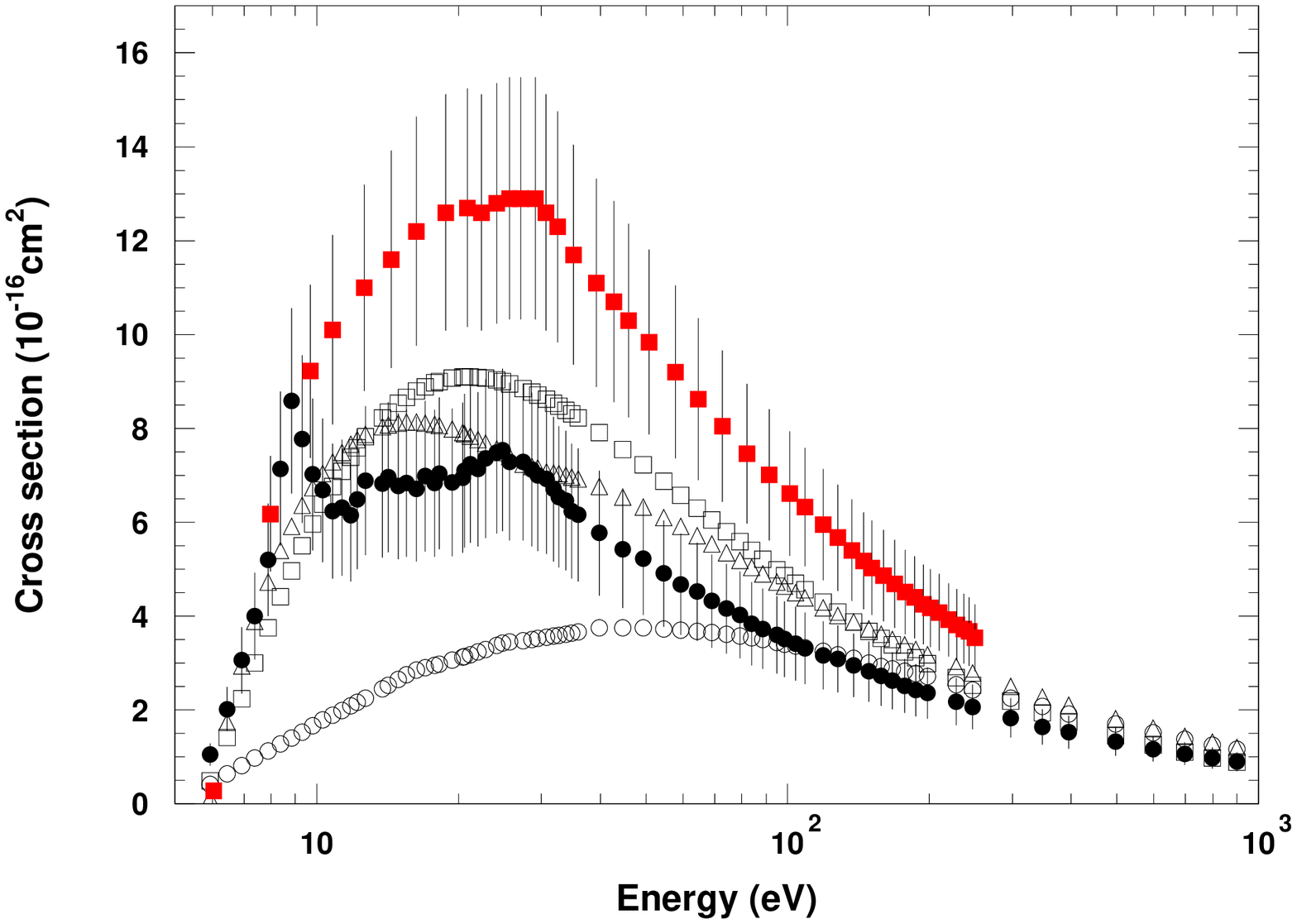}}
\caption{Cross section, Z=63: 
EEDL (empty circles), BEB model (empty squares), 
DM model (empty triangles)
and experimental data from \cite{expYagi2000} 
(black circles) and \cite{shimon} (red squares). }
\label{fig_beb63}
\end{figure}

\begin{figure}
\centerline{\includegraphics[angle=0,width=8cm]{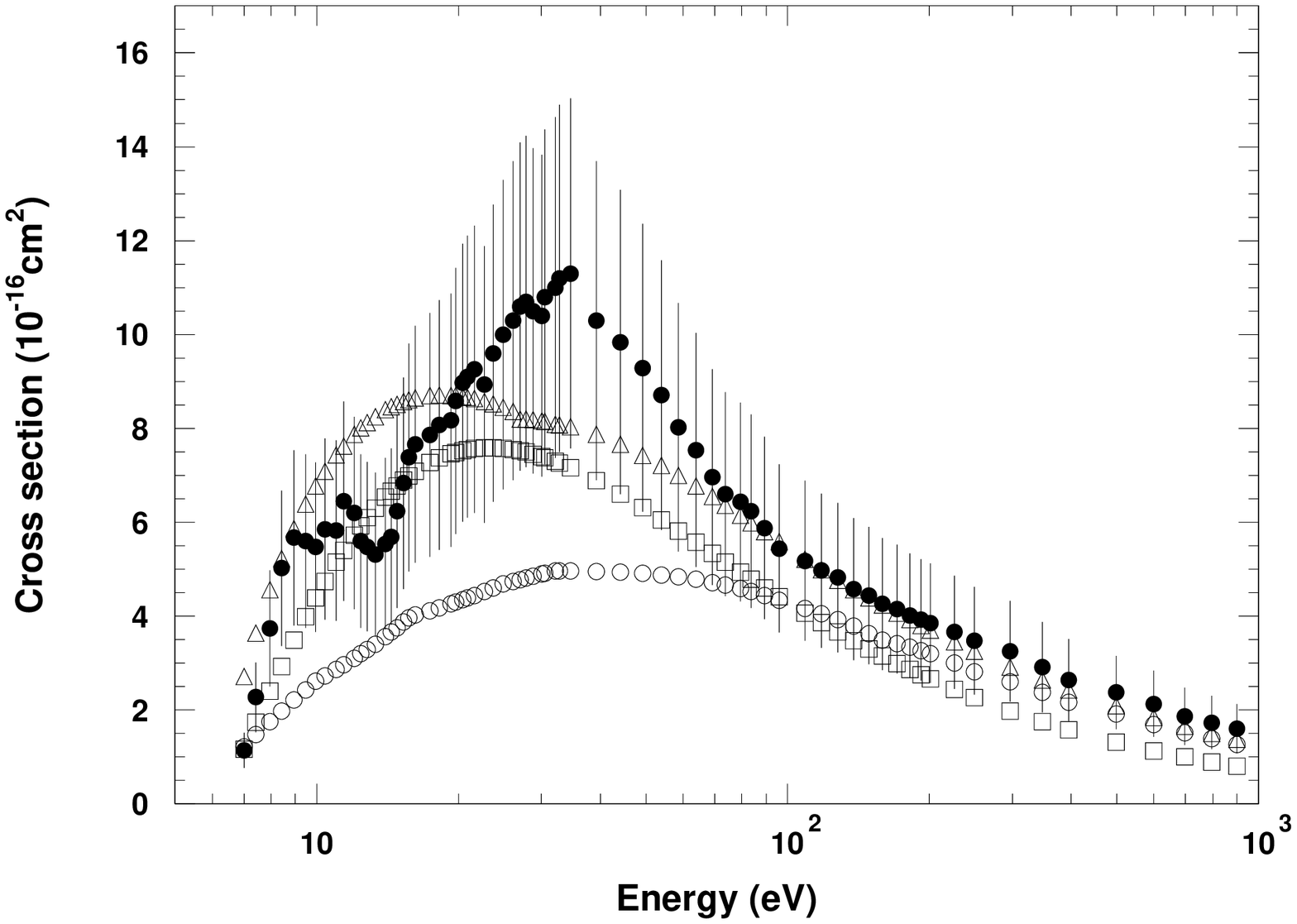}}
\caption{Cross section, Z=64: 
EEDL (empty circles), BEB model (empty squares), 
DM model (empty triangles)
and experimental data from \cite{expYagi2001} 
(black circles). }
\label{fig_beb64}
\end{figure}

\begin{figure}
\centerline{\includegraphics[angle=0,width=8cm]{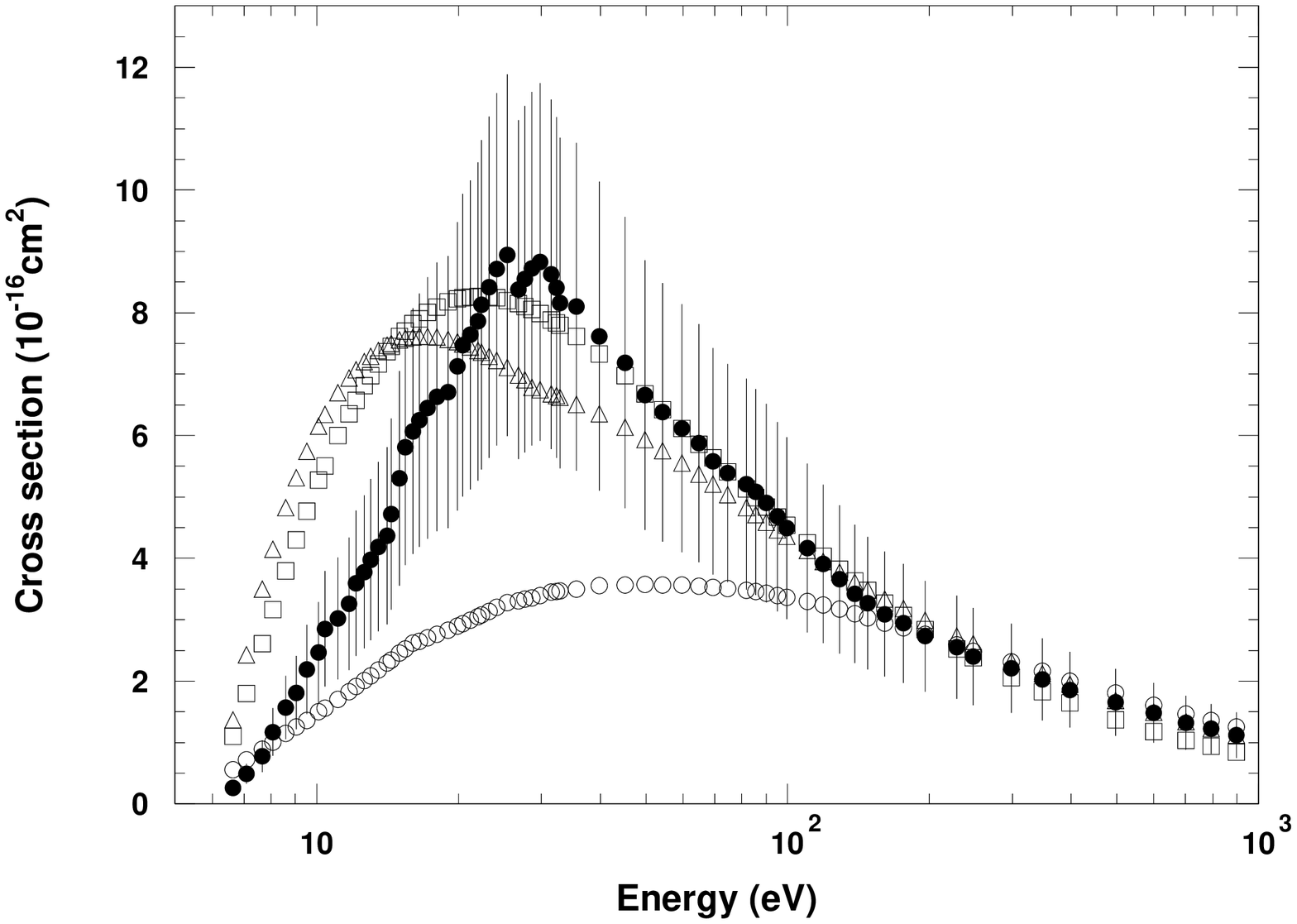}}
\caption{Cross section, Z=66: 
EEDL (empty circles), BEB model (empty squares), 
DM model (empty triangles)
and experimental data from \cite{expYagi2001} 
(black circles). }
\label{fig_beb66}
\end{figure}

\begin{figure}
\centerline{\includegraphics[angle=0,width=8cm]{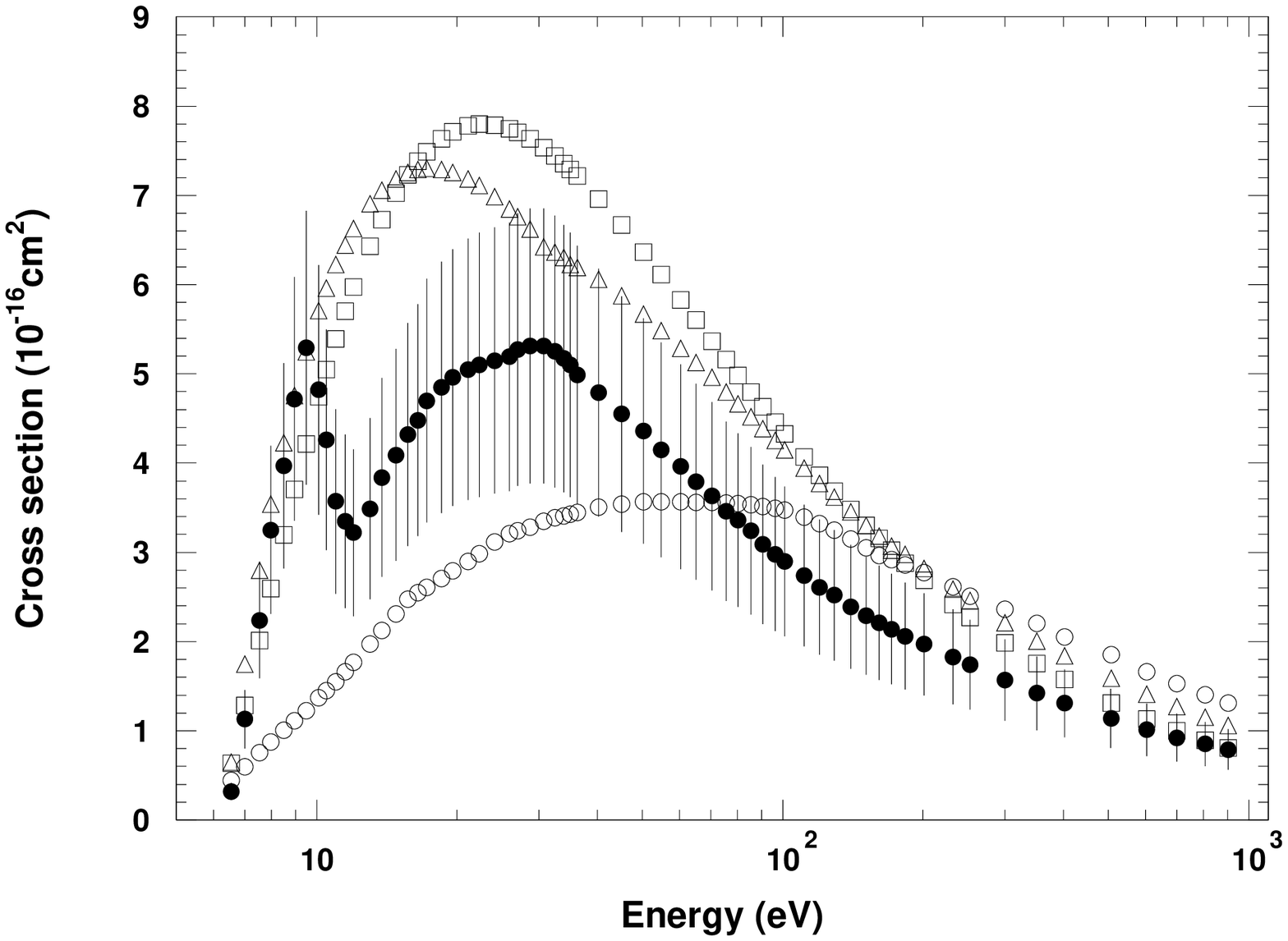}}
\caption{Cross section, Z=68: 
EEDL (empty circles), BEB model (empty squares), 
DM model (empty triangles)
and experimental data from \cite{expYagi2001} 
(black circles). }
\label{fig_beb68}
\end{figure}

\begin{figure}
\centerline{\includegraphics[angle=0,width=8cm]{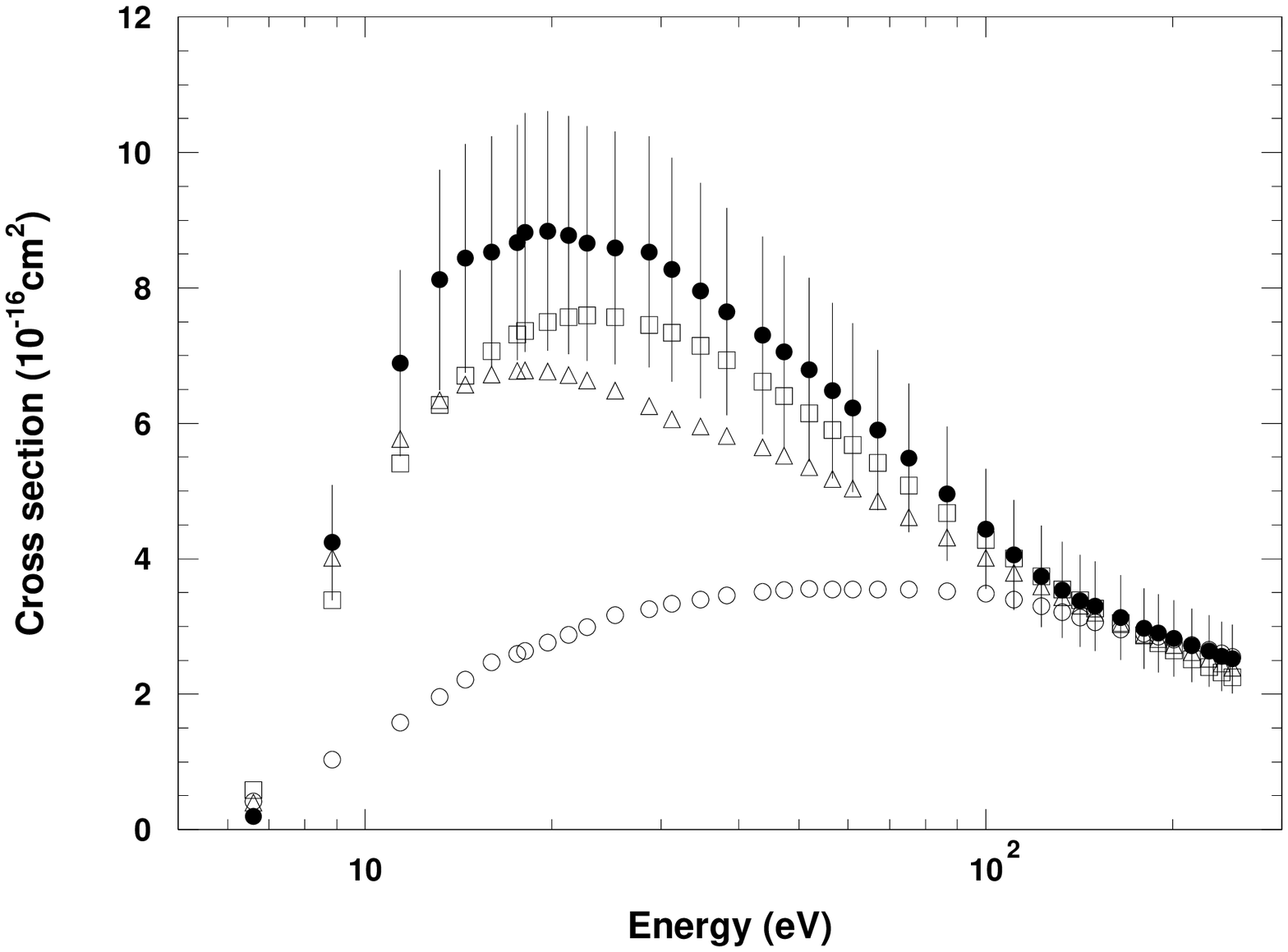}}
\caption{Cross section, Z=69: 
EEDL (empty circles), BEB model (empty squares), 
DM model (empty triangles)
and experimental data from \cite{shimon} 
(black circles). }
\label{fig_beb69}
\end{figure}

\begin{figure}
\centerline{\includegraphics[angle=0,width=8cm]{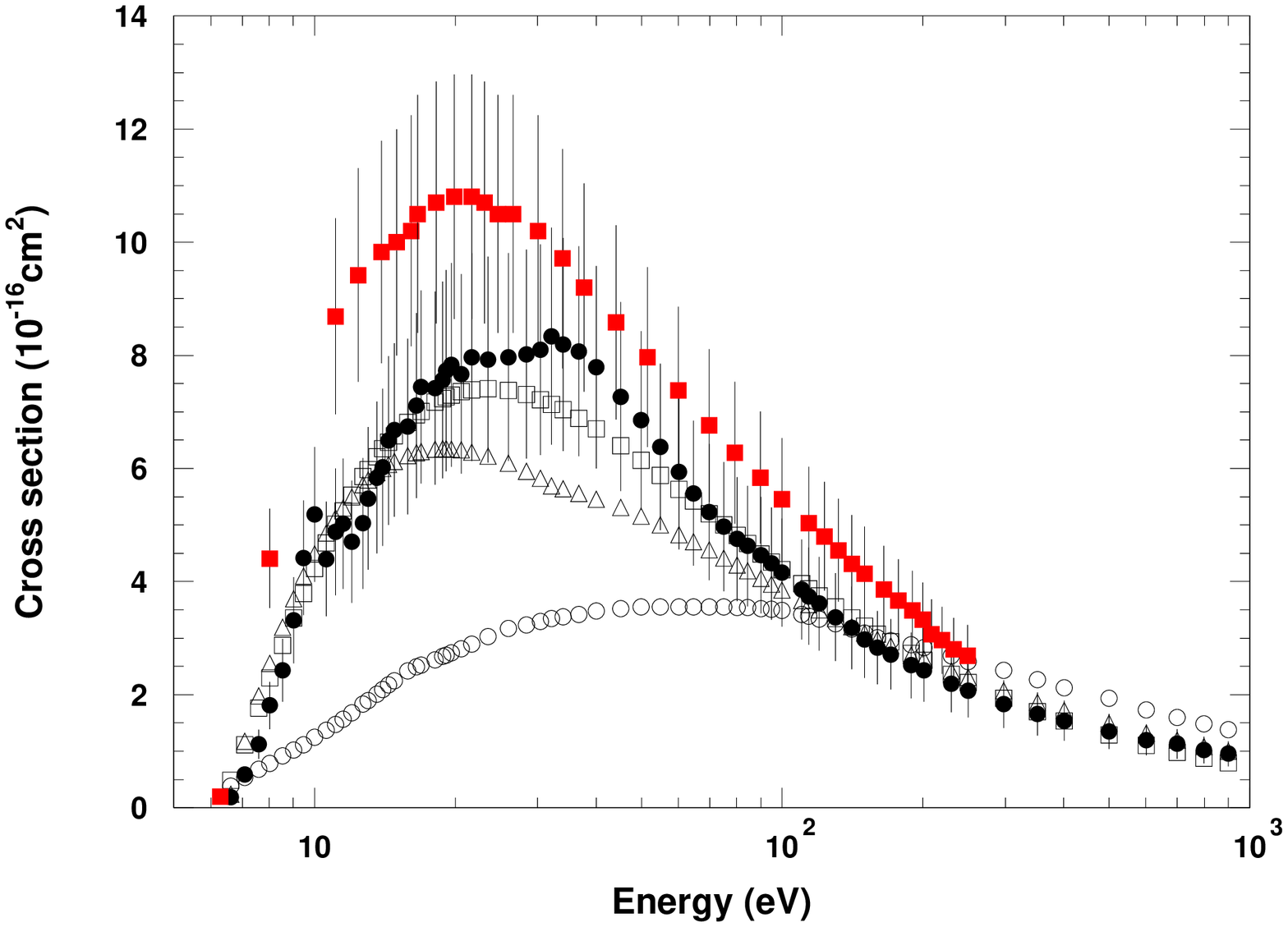}}
\caption{Cross section, Z=70: 
EEDL (empty circles), BEB model (empty squares), 
DM model (empty triangles)
and experimental data from \cite{expYagi2001} 
(black circles) and \cite{shimon} (red squares). }
\label{fig_beb70}
\end{figure}

\begin{figure}
\centerline{\includegraphics[angle=0,width=8cm]{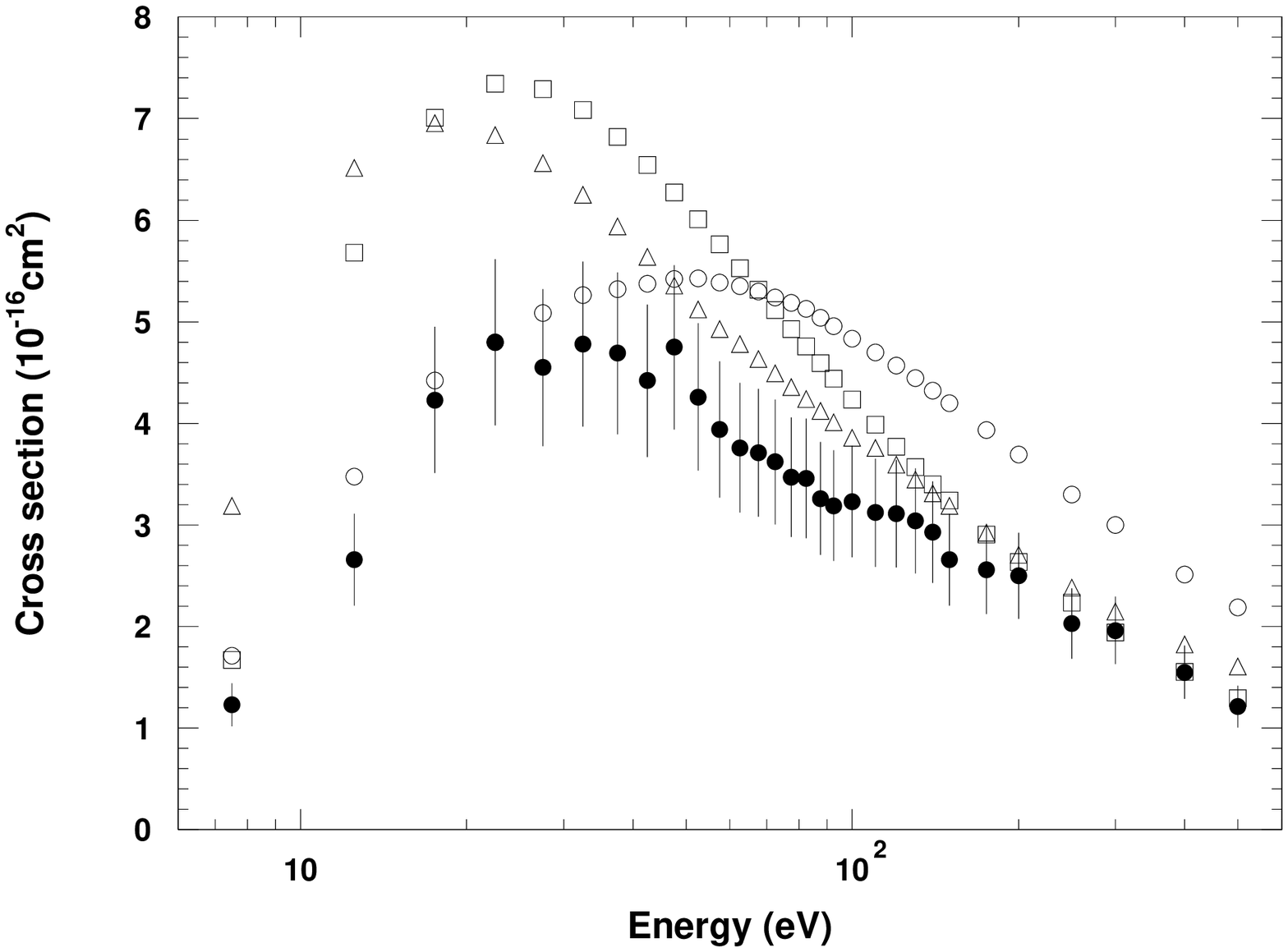}}
\caption{Cross section, Z=92:
EEDL (empty circles), BEB model (empty squares), 
DM model (empty triangles)
and experimental data from \cite{exp92} 
(black circles). }
\label{fig_beb92}
\end{figure}

The cross sections calculated by the BEB, DM and EEDL models are plotted in
Figs. \ref{fig_beb1}-\ref{fig_beb92} along with experimental data.
The results of their quantitative comparisons are detailed in the next section,
while possible sources of systematic effects, which might affect the validation
of the simulation models, are discussed in the following sections.

\subsection{Compatibility with experimental data}

The results of goodness-of-fit tests over all experimental data samples
are summarized in Table \ref{tab_eff}; they report the percentage of test cases 
in which the null hypothesis is not rejected, i.e. the theoretical models
describe the data adequately.
The table lists results for different experimental data types: the whole data
sample, single ionization cross sections, absolute measurements, and absolute
measurements of single ionization.

\begin{figure}
\centerline{\includegraphics[angle=0,width=8.5cm]{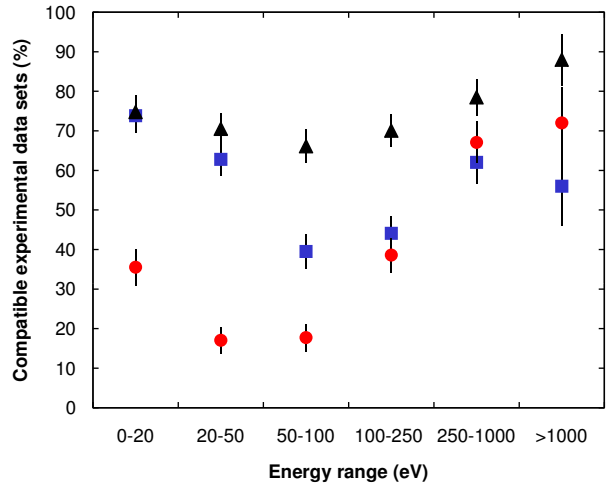}}
\caption{Fraction of test cases in which cross sections calculated by the
implemented models are compatible with experimental data at 0.05 significance
level: BEB model (blue squares), DM model (black triangles) and EEDL (red
circles). The fraction is calculated over the whole collection of data sets.}
\label{fig_models_all}
\end{figure}

The results over the whole data sample are visualized in Fig.
\ref{fig_models_all}, where one can observe that the DM model exhibits the best
overall compatibility with experimental data, while EEDL capability at
reproducing measurements drops significantly below the limit of 250 eV
recommended for its use in Geant4.

\begin{figure}
\centerline{\includegraphics[angle=0,width=8.5cm]{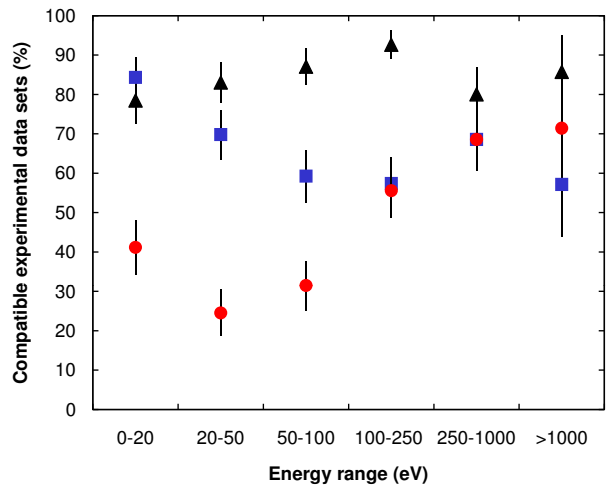}}
\caption{Fraction of elements subject to test for which cross sections
calculated by the implemented models are compatible with at least one
experimental data set at 0.05 significance level: BEB model (blue squares), DM
model (black triangles) and EEDL (red circles). }
\label{fig_models_zall}
\end{figure}

Fig. \ref{fig_models_zall} summarizes the results in relation to elements
rather than individual experimental data sets; it shows the percentage of
elements subject to test for which the null hypothesis is not rejected for at
least one set of experimental measurements.
The trend is quite similar to that observed in Fig. \ref{fig_models_all}, with
the DM model exhibiting in general the best capability at calculating cross
sections compatible with experiment.

\begin{figure}
\centerline{\includegraphics[angle=0,width=8cm]{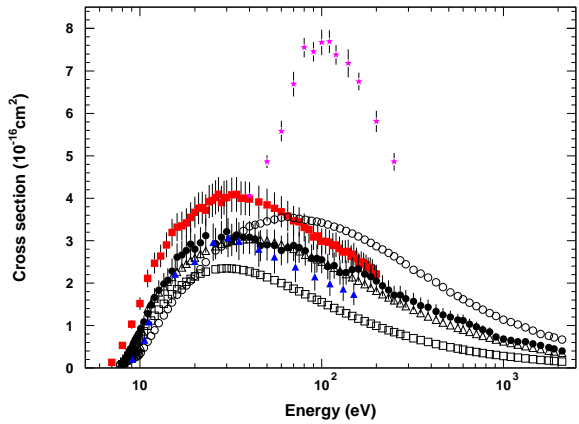}}
\caption{Cross section, Z=29: 
EEDL (empty circles), BEB model (empty squares), 
DM model (empty triangles)
and experimental data from \cite{expFreund} (red squares),
\cite{bolorizadeh} (black circles),
\cite{pavlovCuAg} (blue triangles) and 
\cite{schroeer} (pink stars). }
\label{fig_beb29}
\end{figure}

\begin{figure}
\centerline{\includegraphics[angle=0,width=8cm]{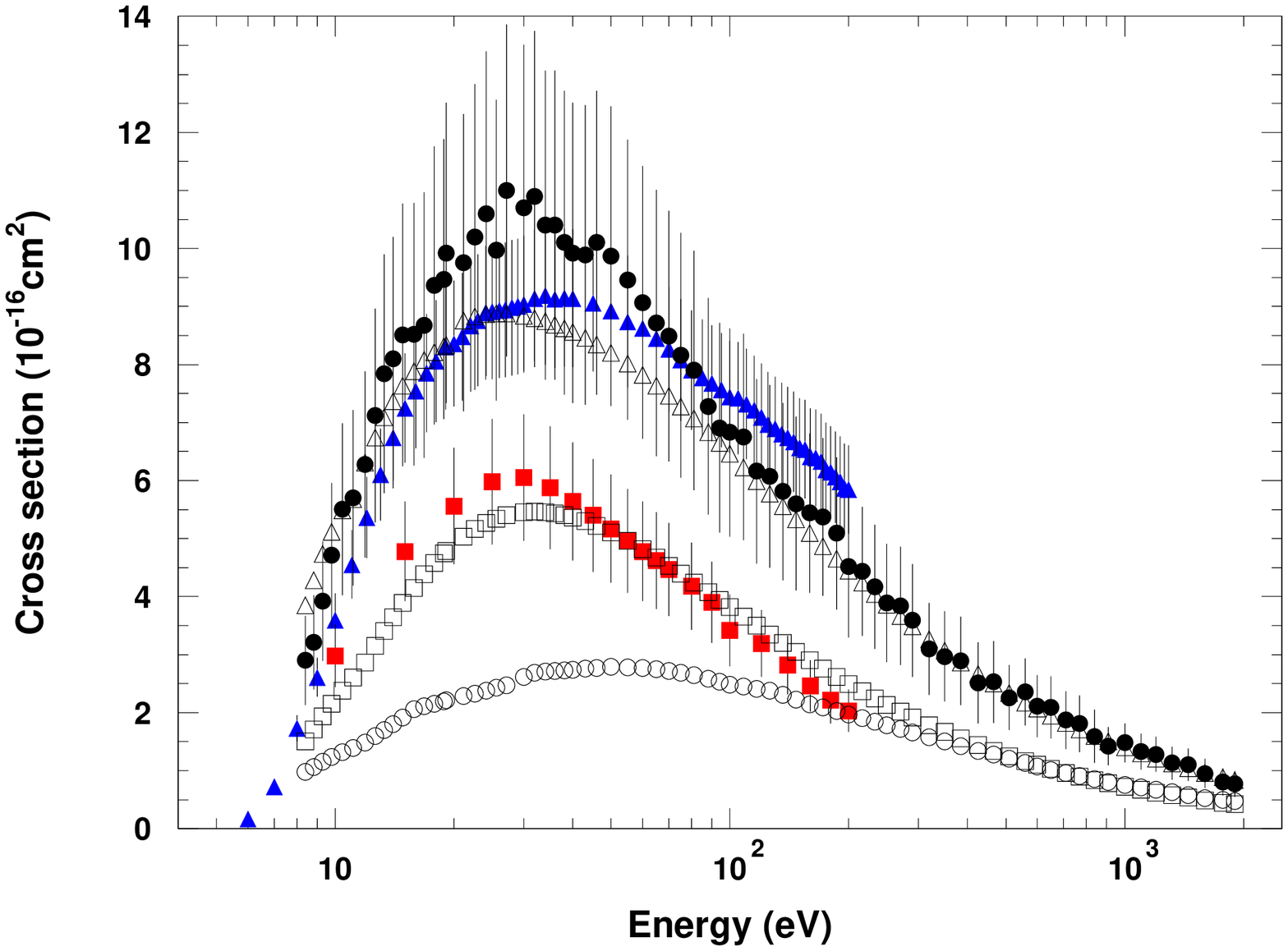}}
\caption{Cross section, Z=31: 
EEDL (empty circles), BEB model (empty squares), 
DM model (empty triangles)
and experimental data from 
\cite{expShul1989} (blue triangles),
\cite{expGaInVainsh} (red squares) and
\cite{patton} (black circles). }
\label{fig_beb31}
\end{figure}

\begin{figure}
\centerline{\includegraphics[angle=0,width=8cm]{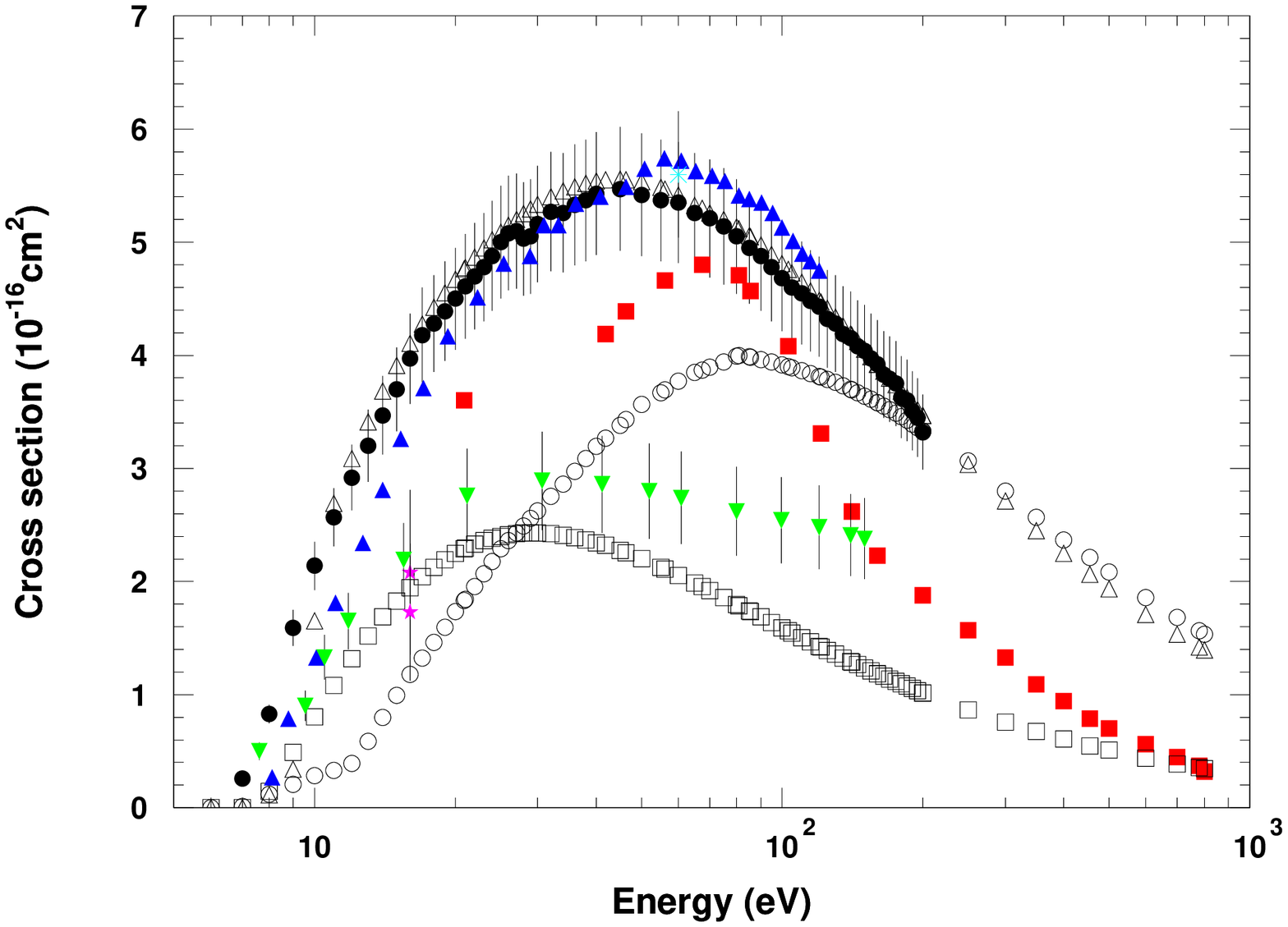}}
\caption{Cross section, Z=47: 
EEDL (empty circles), BEB model (empty squares), 
DM model (empty triangles)
and experimental data from \cite{expFreund} (black circles),
\cite{crawford} (red squares),
\cite{franzreb}  (blue triangles),
\cite{lin}  (turquoise asterisks),
\cite{lyubimov}  (pink stars) and
\cite{pavlovCuAg} (green upside-down triangles). }
\label{fig_beb47}
\end{figure}

\begin{figure}
\centerline{\includegraphics[angle=0,width=8cm]{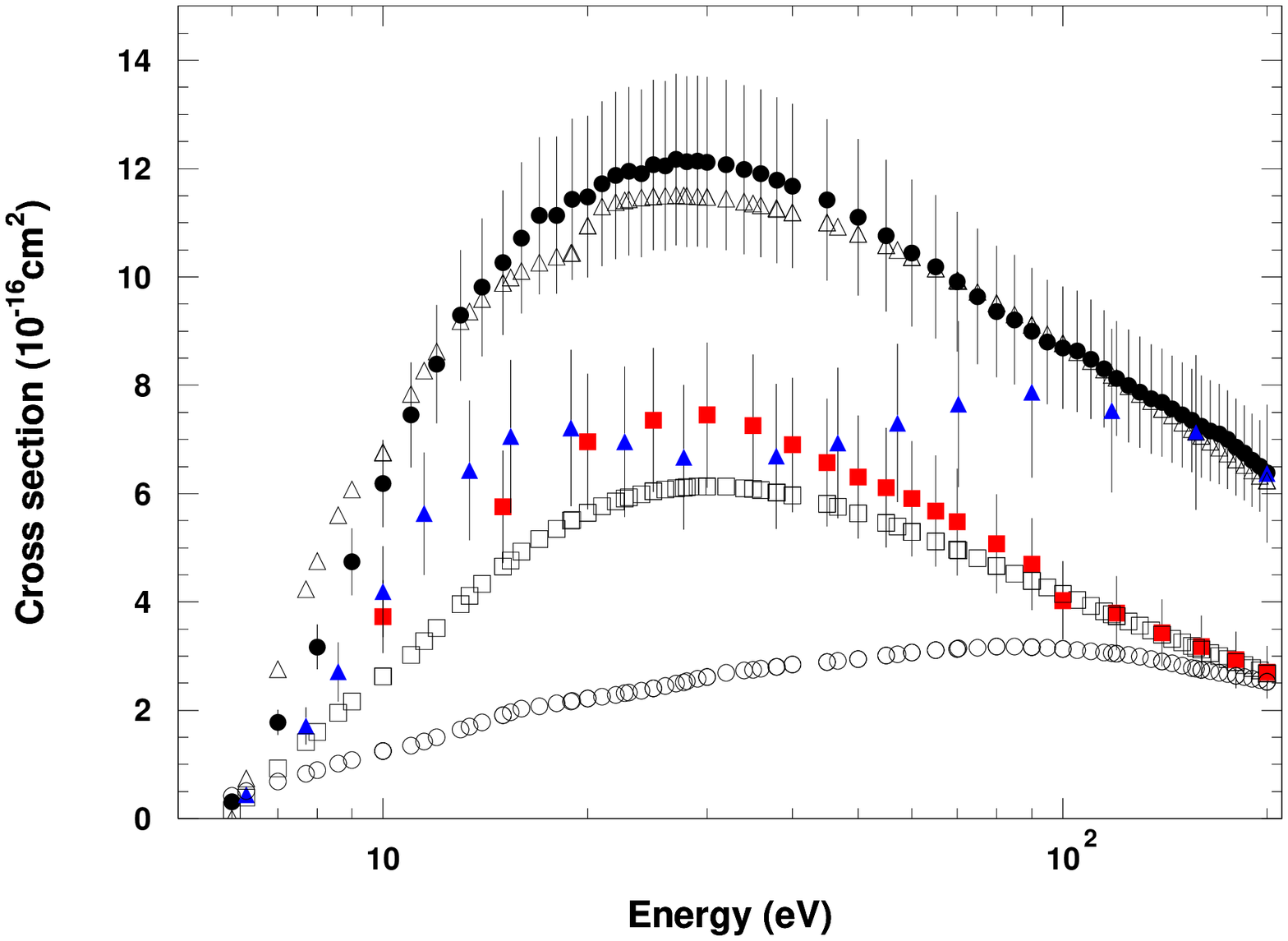}}
\caption{Cross section, Z=49: 
EEDL (empty circles), BEB model (empty squares), 
DM model (empty triangles)
and experimental data from 
\cite{expShul1989} (black circles),
\cite{expGaInVainsh} (red squares) and 
\cite{shimonAlInTl}  (blue triangles). }
\label{fig_beb49}
\end{figure}

Due to the discrepancies of measurements discussed in section \ref{sec_exp}, the
results reported in Table \ref{tab_eff} and Figs.
\ref{fig_models_all}-\ref{fig_models_zall} should not be interpreted
straightforwardly as estimates of the efficiency of the implemented models at
calculating correct cross sections
The presence of experimental data affected by systematic errors contributes to
underestimate the accuracy of theoretical models, which may appear compatible
with only a subset of experimental data samples: Fig. \ref{fig_beb29} and Fig.
\ref{fig_beb47} are an example.
On the other hand, the contradiction of patently discrepant theoretical models
that appear compatible with discrepant experimental data sets contributes to
overestimate the accuracy: examples are Fig. \ref{fig_beb31} and Fig.
\ref{fig_beb49}.

%Considering the discrepancies of measurements discussed in section
%\ref{sec_exp}, caution should be exercised in interpreting the values reported
%in Table \ref{tab_eff} and Figs. \ref{fig_models_all}-\ref{fig_models_zall} as
%estimates of the efficiency of the implemented models at producing cross
%sections compatible with experiment.
%Experimental data affected by systematic errors contribute to
%underestimate the accuracy of theoretical models, which may appear compatible
%with only a subset of experimental data samples: Fig. \ref{fig_beb29} and Fig.
%\ref{fig_beb47} are an example.
%On the other hand, the contradiction of patently discrepant theoretical models
%that appear compatible with discrepant experimental data sets contributes to
%overestimate the accuracy: examples are Fig. \ref{fig_beb31} and Fig.
%\ref{fig_beb49}.

The categorical analysis estimates whether the differences of the models in
compatibility with experiment are statistically significant.
The results are summarized in Tables \ref{tab_contbeb} and \ref{tab_conteedl},
respectively comparing the compatibility of the BEB model and of EEDL over the
whole collection of data samples, and in Table \ref{tab_contz} regarding the
compatibility with at least one experimental sample per element.

The outcome of this statistical analysis supports the qualitative appraisal
of Fig. \ref{fig_models_all} and \ref{fig_models_zall}.
Over the whole collection of data samples, in the low energy range up to 50 eV
the BEB model is equivalent to the DM one, while EEDL is statistically
equivalent to the DM model above 250 eV.
If one considers the compatibility with at least one experimental data set per
element, the BEB model is statistically equivalent to the DM one also above 250
eV.

Some possible sources of systematic effects, which may bias the results of
the validation process, are analyzed in the following sections.

%the fraction of test cases compatible with experiment underestimates the models'
%accuracy, when conflicting experimental measurements for the same target element
%lead to the acceptance of the null hypothesis for for
%, the values in  Table \ref{tab_eff} are prone to
%at 95$\pm$ confidence level: in fact, 

%\begin{table}
%\begin{center}
%\caption{Percentage of test elements for which cross section 
%models are compatible with experimental data}
%\label{tab_effz}
%\begin{tabular}{|l|l|c|}
%\hline
%\multirow{4}{*}{$<$20 eV}	
%& No. data sets		&51    \\
%& 	{\bf BEB}	&84$\pm$  \\
%& 	{\bf DM}	&78\%  \\
%& 	{\bf EEDL}	&41\%  \\
%\hline		  	      
%\multirow{4}{*}{20-50 eV}      	 
%& No. data sets		&53   \\
%& 	{\bf BEB}	&70\%  \\
%& 	{\bf DM}	&83\%  \\
%& 	{\bf EEDL}	&25\%  \\
%\hline		  	      
%\multirow{4}{*}{50-100 eV}  
%& No. data sets		&54   \\
%& 	{\bf BEB}	&59\%  \\
%& 	{\bf DM}	&87\%  \\
%& 	{\bf EEDL}	&31\%  \\
%\hline		  	      
%\multirow{4}{*}{100-250 eV} 
%& No. data sets		&54   \\
%& 	{\bf BEB}	&57\%  \\
%& 	{\bf DM}	&93\%  \\
%& 	{\bf EEDL}	&56\%  \\
%\hline		  	       	  	  	
%\multirow{4}{*}{250 eV - 1 keV}	
%& No. data sets		&35   \\
%& 	{\bf BEB}	&69\%  \\
%& 	{\bf DM}	&80\%  \\
%& 	{\bf EEDL}	&69\%  \\
%\hline		  	        	  		  	  	  	  	
%\multirow{4}{*}{$>$1 keV}      	  		  	  	  	  
%& No. data sets		&14  \\
%& 	{\bf BEB}	&57\% \\
%& 	{\bf DM}	&86\% \\
%& 	{\bf EEDL}	&71\% \\
%\hline
%\end{tabular}
%\end{center}
%\end{table}						

\begin{table*}
\begin{center}
\caption{Contingency tables related to DM and BEB cross section compatibility
with experimental data}
\label{tab_contbeb}
\begin{tabular}{|l|l|cc|cc|cc|cc|}
\hline
Energy 	&Goodness-of-fit test &\multicolumn{2}{|c|}{\bf All}	&\multicolumn{2}{|c|}{\bf Single}      &\multicolumn{2}{|c|}{\bf Absolute}  &\multicolumn{2}{|c|}{\bf Single} \\
& 	& \multicolumn{2}{|c|}{}				&\multicolumn{2}{|c|}{\bf ionization} &\multicolumn{2}{|c|}{\bf measurement} &\multicolumn{2}{|c|}{\bf Absolute} \\
\hline
\multirow{6}{*}{$<$ 20 eV} 
& 				&DM	&BEB			&DM     &BEB  			&DM     &BEB			&DM     &BEB		\\
%\cline{2-12}						        		                 				            
& Pass				&80	&79 			&56	&53   			&59	&59      		&37	&35   	\\
& Fail				&27	&28 			&19	&22   			&14	&14      		&7       &9   	\\
\cline{2-10}
& p-value Fisher test		& \multicolumn{2}{|c|}{1} 	&\multicolumn{2}{|c|}{0.714}  	&\multicolumn{2}{|c|}{1}  	&\multicolumn{2}{|c|}{0.783}   \\
& p-value Pearson $\chi^2$	& \multicolumn{2}{|c|}{0.876} 	&\multicolumn{2}{|c|}{0.583}  	&\multicolumn{2}{|c|}{1}  	&\multicolumn{2}{|c|}{0.580}  \\
& p-value Yates $\chi^2$	& \multicolumn{2}{|c|}{1} 	&\multicolumn{2}{|c|}{0.714}  	&\multicolumn{2}{|c|}{1}  	&\multicolumn{2}{|c|}{0.782}  \\
\hline
\multirow{6}{*}{20-50 eV} 
& 				&DM	&BEB			&DM     &BEB			&DM     &BEB         		&DM     &BEB    \\
%\cline{2-12}						                                                                                
& Pass				&91	&81 			&68       &60        		&53       &44             	&34       &26       \\
& Fail				&38	&48 			&22       &30           	&30       &39             	&15       &23       \\
\cline{2-10}	
& p-value Fisher test		& \multicolumn{2}{|c|}{0.235} 	&\multicolumn{2}{|c|}{0.250}  	&\multicolumn{2}{|c|}{0.208}  	&\multicolumn{2}{|c|}{0.146} \\
& p-value Pearson $\chi^2$	& \multicolumn{2}{|c|}{0.187} 	&\multicolumn{2}{|c|}{0.188}  	&\multicolumn{2}{|c|}{0.156}  	&\multicolumn{2}{|c|}{0.097} \\
& p-value Yates $\chi^2$	& \multicolumn{2}{|c|}{0.235} 	&\multicolumn{2}{|c|}{0.250}  	&\multicolumn{2}{|c|}{0.208}  	&\multicolumn{2}{|c|}{0.147} \\
\hline
\multirow{6}{*}{50-100 eV} 
& 				&DM	&BEB			&DM     &BEB            	&DM     &BEB            	&DM     &BEB    \\
%\cline{2-12}						                                                                  
& Pass				&82     &49  			&64       &38               	&52       &30               	&35       &19       \\
& Fail				&42     &75  			&27       &53               	&29       &51               	&15       &31       \\
\cline{2-10}
& p-value Fisher test		&\multicolumn{2}{|c|}{$<0.001$} &\multicolumn{2}{|c|}{$<0.001$}  &\multicolumn{2}{|c|}{0.001}  	&\multicolumn{2}{|c|}{0.002} \\
& p-value Pearson $\chi^2$	&\multicolumn{2}{|c|}{$<0.001$} &\multicolumn{2}{|c|}{$<0.001$}  &\multicolumn{2}{|c|}{0.001}	&\multicolumn{2}{|c|}{0.001} \\
& p-value Yates $\chi^2$	&\multicolumn{2}{|c|}{$<0.001$} &\multicolumn{2}{|c|}{$<0.001$}  &\multicolumn{2}{|c|}{0.001}  	&\multicolumn{2}{|c|}{0.003} \\
\hline
\multirow{6}{*}{100-250 eV} 
& 				&DM	&BEB			&DM     &BEB            	&DM     &BEB            	&DM     &BEB    \\
%\cline{2-12}						                                                                  
& Pass				&89	&56 			&69       &47               	&54       &31               	&37       &23       \\
& Fail				&38	&71 			&24       &46               	&27       &50               	&14       &28       \\
\cline{2-10}
& p-value Fisher test		&\multicolumn{2}{|c|}{$<0.001$} &\multicolumn{2}{|c|}{0.001}  	&\multicolumn{2}{|c|}{$<0.001$} &\multicolumn{2}{|c|}{0.009} \\
& p-value Pearson $\chi^2$	&\multicolumn{2}{|c|}{$<0.001$} &\multicolumn{2}{|c|}{0.001}	&\multicolumn{2}{|c|}{$<0.001$}	&\multicolumn{2}{|c|}{0.005} \\
& p-value Yates $\chi^2$	&\multicolumn{2}{|c|}{$<0.001$} &\multicolumn{2}{|c|}{0.001}  	&\multicolumn{2}{|c|}{0.001}  	&\multicolumn{2}{|c|}{0.009} \\
\hline
\multirow{6}{*}{250 eV - 1 keV} 
& 				&DM	&BEB			&DM     &BEB            	&DM     &BEB            	&DM     &BEB    \\
%\cline{2-12}						                                                                  
& Pass				&62	&49 			&50      &44               	&31       &25               	&22      &21       \\
& Fail				&17	&30 			&8       &14               	&12       &18               	&4       &5       \\
\cline{2-10}
& p-value Fisher test		&\multicolumn{2}{|c|}{0.036} 	&\multicolumn{2}{|c|}{0.236}  	&\multicolumn{2}{|c|}{0.258}  	&\multicolumn{2}{|c|}{1} \\
& p-value Pearson $\chi^2$	&\multicolumn{2}{|c|}{0.024} 	&\multicolumn{2}{|c|}{0.155}  	&\multicolumn{2}{|c|}{0.175}  	&\multicolumn{2}{|c|}{not applicable} \\
& p-value Yates $\chi^2$	&\multicolumn{2}{|c|}{0.037} 	&\multicolumn{2}{|c|}{0.236}  	&\multicolumn{2}{|c|}{0.258}  	&\multicolumn{2}{|c|}{1} \\
\hline
\multirow{5}{*}{$>$ 1 keV} 
& 				&DM	&BEB			&DM     &BEB            	&DM     &BEB            	&DM     &BEB    \\		
%\cline{2-12}
& Pass				&22	&14 			&20	&14          		&11	&9               	&10	&9       \\
& Fail				&3	&11 			&2 	&8            		&1	&3               	&1  	&2       \\
\cline{2-10}
& p-value Fisher test		&\multicolumn{2}{|c|}{0.025} 	&\multicolumn{2}{|c|}{0.069}  	&\multicolumn{2}{|c|}{0.590}  	&\multicolumn{2}{|c|}{1} \\
%p-value Pearson $\chi^2$	&\multicolumn{2}{|c|}{0.} 	&\multicolumn{2}{|c|}{}  	&\multicolumn{2}{|c|}{}  	&\multicolumn{2}{|c|}{} \\
& p-value Yates $\chi^2$	&\multicolumn{2}{|c|}{0.027} 	&\multicolumn{2}{|c|}{0.072}  	&\multicolumn{2}{|c|}{0.584}  	&\multicolumn{2}{|c|}{1} \\
\hline
\end{tabular}
\end{center}
\end{table*}

\begin{table*}
\begin{center}
\caption{Contingency tables related to DM and EEDL cross section compatibility
with experimental data}
\label{tab_conteedl}
\begin{tabular}{|l|l|cc|cc|cc|cc|}
\hline
Energy 	&Goodness-of-fit test &\multicolumn{2}{|c|}{\bf All}	&\multicolumn{2}{|c|}{\bf Single}     &\multicolumn{2}{|c|}{\bf Absolute}  	&\multicolumn{2}{|c|}{\bf Single} \\
& 	&\multicolumn{2}{|c|}{}					&\multicolumn{2}{|c|}{\bf ionization} &\multicolumn{2}{|c|}{\bf measurement} 	&\multicolumn{2}{|c|}{\bf Absolute} \\
\hline
\multirow{6}{*}{$<$ 20 eV} 
& 				&DM	&EEDL			&DM     &EEDL            	&DM     &EEDL            	&DM     &EEDL    \\		
%\cline{2-10}						                                                                  
& Pass				&80	&38 			&56       &29               	&59       &28               	&37      &19       \\
& Fail				&27	&69 			&19       &46               	&14       &45               	&7       &25       \\
\cline{2-10}
& p-value Fisher test		&\multicolumn{2}{|c|}{$<0.001$} &\multicolumn{2}{|c|}{$<0.001$}  &\multicolumn{2}{|c|}{$<0.001$}  &\multicolumn{2}{|c|}{$<0.001$} \\
& p-value Pearson $\chi^2$	&\multicolumn{2}{|c|}{$<0.001$} &\multicolumn{2}{|c|}{$<0.001$}  &\multicolumn{2}{|c|}{$<0.001$}  &\multicolumn{2}{|c|}{$<0.001$} \\
& p-value Yates $\chi^2$	&\multicolumn{2}{|c|}{$<0.001$} &\multicolumn{2}{|c|}{$<0.001$}  &\multicolumn{2}{|c|}{$<0.001$}  &\multicolumn{2}{|c|}{$<0.001$} \\
\hline

\multirow{6}{*}{20-50 eV} 
& 				&DM	&EEDL			&DM     &EEDL            	&DM     &EEDL            	&DM     &EEDL    \\		
%\cline{2-10}						                                                                  
& Pass				&91	&22 			&68       &19               	&53       &11               	&34       &8       \\
& Fail				&38	&107 			&22       &71               	&30       &72               	&15       &41       \\
\cline{2-10}
& p-value Fisher test		&\multicolumn{2}{|c|}{$<0.001$} &\multicolumn{2}{|c|}{$<0.001$}  &\multicolumn{2}{|c|}{$<0.001$}  &\multicolumn{2}{|c|}{$<0.001$} \\
& p-value Pearson $\chi^2$	&\multicolumn{2}{|c|}{$<0.001$} &\multicolumn{2}{|c|}{$<0.001$}  &\multicolumn{2}{|c|}{$<0.001$}  &\multicolumn{2}{|c|}{$<0.001$} \\
& p-value Yates $\chi^2$	&\multicolumn{2}{|c|}{$<0.001$} &\multicolumn{2}{|c|}{$<0.001$}  &\multicolumn{2}{|c|}{$<0.001$}  &\multicolumn{2}{|c|}{$<0.001$} \\
\hline
\multirow{6}{*}{50-100 eV} 
& 				&DM	&EEDL			&DM     &EEDL            	&DM     &EEDL            	&DM     &EEDL    \\		
%\cline{2-10}						                                                                  
& Pass				&82	&22 			&64       &22               	&52       &7               	&35       &7       \\
& Fail				&42	&102 			&27       &69               	&29       &74               	&15       &43       \\
\cline{2-10}
& p-value Fisher test		&\multicolumn{2}{|c|}{$<0.001$} &\multicolumn{2}{|c|}{$<0.001$}  &\multicolumn{2}{|c|}{$<0.001$}  &\multicolumn{2}{|c|}{$<0.001$} \\
& p-value Pearson $\chi^2$	&\multicolumn{2}{|c|}{$<0.001$} &\multicolumn{2}{|c|}{$<0.001$}  &\multicolumn{2}{|c|}{$<0.001$}  &\multicolumn{2}{|c|}{$<0.001$} \\
& p-value Yates $\chi^2$	&\multicolumn{2}{|c|}{$<0.001$} &\multicolumn{2}{|c|}{$<0.001$}  &\multicolumn{2}{|c|}{$<0.001$}  &\multicolumn{2}{|c|}{$<0.001$} \\
\hline
\multirow{6}{*}{100-250 eV} 
& 				&DM	&EEDL			&DM     &EEDL            	&DM     &EEDL            	&DM     &EEDL    \\		
%\cline{2-10}						                                                                  
& Pass				&89	&49 			&69       &46               	&54       &23               	&37       &21       \\
& Fail				&38	&78 			&24       &47               	&27       &58               	&14       &30       \\
\cline{2-10}
& p-value Fisher test		&\multicolumn{2}{|c|}{$<0.001$} &\multicolumn{2}{|c|}{0.001}  	&\multicolumn{2}{|c|}{$<0.001$}  &\multicolumn{2}{|c|}{0.003} \\
& p-value Pearson $\chi^2$	&\multicolumn{2}{|c|}{$<0.001$} &\multicolumn{2}{|c|}{0.001}  	&\multicolumn{2}{|c|}{$<0.001$}	&\multicolumn{2}{|c|}{0.001} \\
& p-value Yates $\chi^2$	&\multicolumn{2}{|c|}{$<0.001$} &\multicolumn{2}{|c|}{0.001}  	&\multicolumn{2}{|c|}{$<0.001$}	&\multicolumn{2}{|c|}{0.003} \\
\hline
\multirow{6}{*}{250 eV - 1 keV} 
& 				&DM	&EEDL			&DM     &EEDL            	&DM     &EEDL            	&DM     &EEDL    \\							                                                                                                          
& Pass				&62     &53 			&50      &46               	&31       &28               	&22      &23       \\            
& Fail				&17     &26 			&8       &12               	&12       &15               	&4       &3       \\            
\cline{2-10}				                        
& p-value Fisher test		&\multicolumn{2}{|c|}{0.152} 	&\multicolumn{2}{|c|}{0.462}  	&\multicolumn{2}{|c|}{0.643}  	&\multicolumn{2}{|c|}{1} \\
& p-value Pearson $\chi^2$	&\multicolumn{2}{|c|}{0.108} 	&\multicolumn{2}{|c|}{0.326}  	&\multicolumn{2}{|c|}{0.486}  	&\multicolumn{2}{|c|}{not applicable} \\
& p-value Yates $\chi^2$	&\multicolumn{2}{|c|}{0.153} 	&\multicolumn{2}{|c|}{0.461}  	&\multicolumn{2}{|c|}{0.642}  	&\multicolumn{2}{|c|}{1} \\
\hline
\multirow{5}{*}{$>$ 1 keV} 
& 				&DM	&EEDL			&DM     &EEDL            	&DM     &EEDL            	&DM     &EEDL    \\		
%\hline						                                                                  
& Pass				&22	&18 		      	&20	&16	  		&11	&12           		&10	&11       \\
& Fail				&3	&7 		      	&2	&6        		&1	&0             		&1  	&0       \\
\cline{2-10}
& p-value Fisher test		&\multicolumn{2}{|c|}{0.289} 	&\multicolumn{2}{|c|}{0.240}  	&\multicolumn{2}{|c|}{1}  	&\multicolumn{2}{|c|}{1} \\
%& p-value Pearson $\chi^2$	&\multicolumn{2}{|c|}{0.}    	&\multicolumn{2}{|c|}{}  	&\multicolumn{2}{|c|}{}  	&\multicolumn{2}{|c|}{} \\
& p-value Yates $\chi^2$	&\multicolumn{2}{|c|}{0.289} 	&\multicolumn{2}{|c|}{0.241}  	&\multicolumn{2}{|c|}{1}  	&\multicolumn{2}{|c|}{1} \\
\hline
\end{tabular}
\end{center}
\end{table*}

\begin{table*}
\begin{center}
\caption{Contingency tables related to cross section compatibility with at least one experimental data set per element}
\label{tab_contz}
\begin{tabular}{|l|l|cc|cc|}
\hline
Energy 	&Goodness-of-fit test &\multicolumn{2}{|c|}{\bf Models}	&\multicolumn{2}{|c|}{\bf Models}     	 \\
\hline
\multirow{6}{*}{$<$ 20 eV} 
& 				&DM	&BEB                    	&DM     &EEDL    \\					                      
& Pass				&40     &43                     	&40      &21       \\
& Fail				&11     &8                     		&11      &30       \\
\cline{2-6}			                                
& p-value Fisher test		& \multicolumn{2}{|c|}{0.612}         	&\multicolumn{2}{|c|}{$<0.001$} \\
& p-value Pearson $\chi^2$	& \multicolumn{2}{|c|}{0.445}     	&\multicolumn{2}{|c|}{$<0.001$} \\
& p-value Yates $\chi^2$	& \multicolumn{2}{|c|}{0.611}         	&\multicolumn{2}{|c|}{$<0.001$} \\
\hline				                                				                                
\multirow{6}{*}{20-50 eV} 	                    
&				&DM     &BEB                            &DM     &EEDL    \\					                     
& Pass				&44    	&37                     	&44       &13       \\
& Fail				&9      &16                           	&9        &40       \\
\cline{2-6}			   
& p-value Fisher test		& \multicolumn{2}{|c|}{0.162}     		&\multicolumn{2}{|c|}{$<0.001$} \\
& p-value Pearson $\chi^2$	& \multicolumn{2}{|c|}{0.109}     	&\multicolumn{2}{|c|}{$<0.001$} \\
& p-value Yates $\chi^2$	& \multicolumn{2}{|c|}{0.170}           &\multicolumn{2}{|c|}{$<0.001$} \\
\hline				                                
\multirow{6}{*}{50-100 eV} 	                   
&				&DM     &BEB                           	&DM     &EEDL    \\						               
& Pass				&47     &32                     	&47       &17       \\
& Fail				&7      &22                             &7        &37       \\
\cline{2-6}			 
& p-value Fisher test		&\multicolumn{2}{|c|}{0.001}   	&\multicolumn{2}{|c|}{$<0.001$} \\
& p-value Pearson $\chi^2$	&\multicolumn{2}{|c|}{0.001}   	&\multicolumn{2}{|c|}{$<0.001$} \\
& p-value Yates $\chi^2$	&\multicolumn{2}{|c|}{0.002}         	&\multicolumn{2}{|c|}{$<0.001$} \\
\hline				                                
\multirow{6}{*}{100-250 eV} 	                    
&				&DM     &BEB                            &DM     &EEDL    \\					                  
& Pass				&50     &31                     	&50       &30       \\
& Fail				&4      &23                             & 4       &24       \\
\cline{2-6}			 
& p-value Fisher test		&\multicolumn{2}{|c|}{$<0.001$}  	&\multicolumn{2}{|c|}{$<0.001$} \\
%& p-value Pearson $\chi^2$     &\multicolumn{2}{|c|}{0.}               &\multicolumn{2}{|c|}{} \\
& p-value Yates $\chi^2$	&\multicolumn{2}{|c|}{$<0.001$}         &\multicolumn{2}{|c|}{$<0.001$} \\
\hline				                                
\multirow{6}{*}{250 eV - 1 keV}                     
&				&DM     &BEB                            &DM     &EEDL    \\			                                 
& Pass				&28     &24                     	&28      &24       \\            
& Fail				&7      &11                     	&7       &11      \\            
\cline{2-6}			                                
& p-value Fisher test		&\multicolumn{2}{|c|}{0.413}    		&\multicolumn{2}{|c|}{0.413} \\
& p-value Pearson $\chi^2$	&\multicolumn{2}{|c|}{0.274}    	&\multicolumn{2}{|c|}{0.274} \\
& p-value Yates $\chi^2$	&\multicolumn{2}{|c|}{0.412}    	&\multicolumn{2}{|c|}{0.412} \\
\hline				                                
\multirow{5}{*}{$>$ 1 keV} 	                                
& 				&DM     &BEB                    	&DM     &EEDL    \\						              
& Pass				&12     &8                      	&12	&10       \\
& Fail				&2      &6                      	&2  	&4       \\
\cline{2-6}			                                
& p-value Fisher test		&\multicolumn{2}{|c|}{0.209}    		&\multicolumn{2}{|c|}{0.648} \\
%& p-value Pearson $\chi^2$	&\multicolumn{2}{|c|}{0.}       	&\multicolumn{2}{|c|}{} \\
& p-value Yates $\chi^2$	&\multicolumn{2}{|c|}{0.209}    	&\multicolumn{2}{|c|}{0.645} \\
\hline
\end{tabular}
\end{center}
\end{table*}

%\section{Evaluation of possible sources of bias in the validation}
%\label{sec_systematic}

\subsection{Data used in the determination of DM parameters}

Some of the parameters in the formulation of the Deutsch-M\"ark model are
determined from a fit to experimental data.
The reuse of experimental data to which model parameters were fit should be
taken into account in the calculation of the number of degrees of freedom in the
goodness-of-fit tests concerning those experimental data sets.
Nevertheless, the  calculation of proper degrees of freedom is hindered by the difficulty 
of ascertaining which experimental data were used for the determination of the
weighting factors used in the DM model implementation, and what was actually fit.

In the earliest version of the model the fit was based on a few experimental
data sets for rare gases and uranium identified in \cite{dmMarg1994}, that were 
considered reliable by the original authors of the DM model.
A later revision
\cite{dmDeutsch1999}, which reports a subset of the weighting factors 
implemented in the software, mentions the inclusion of cross sections of small
molecules in the determination of the new model parameters: this suggests
a global fit for the determination of the weighting factors in
(\ref{eq_dms}) involving a set of molecules and atoms, which would have scarce
relation with the issue of the degrees of freedom in goodness-of-fit tests
concerning a single atom and a single experimental data sample.

According to goodness-of-fit tests, some of the cross sections calculated by the
software are incompatible with 0.01 significance with experimental data
exploited in the original fit of \cite{dmMarg1994}: this finding hints that they
retain weak memory of having been involved in a fit for the determination
of model parameters.
%The hypothesis of compatibility between DM cross sections calculated by the
%software and some of the experimental data sets exploited in the original fit of
%\cite{dmMarg1994} is rejected by goodness-of-fit tests with 0.01 significance;
%this hints that the correlation between the software calculations for elements
%involved in the original authors' fit and data used in the global fit may
%actually have a weak relation with the experimental data.
% the determination of weighting factors may have a weak relation
%with the compatibility of the cross sections calculated by the software with the
%experimental data involved in the original authors' fit.
A categorical analysis comparing the compliance with experimental data used in
the original authors' fits and with data samples excluding them produces
statistically equivalent results.
These observations suggest that the goodness-of-fit analysis for the validation
of the DM model applied in this paper is not significantly affected by the
inclusion of a small subset of experimental data, which may have been
somehow involved in the determination of the model parameters.

\subsection{Effect of BEB model parameters}

\begin{figure}
\centerline{\includegraphics[angle=0,width=8.5cm]{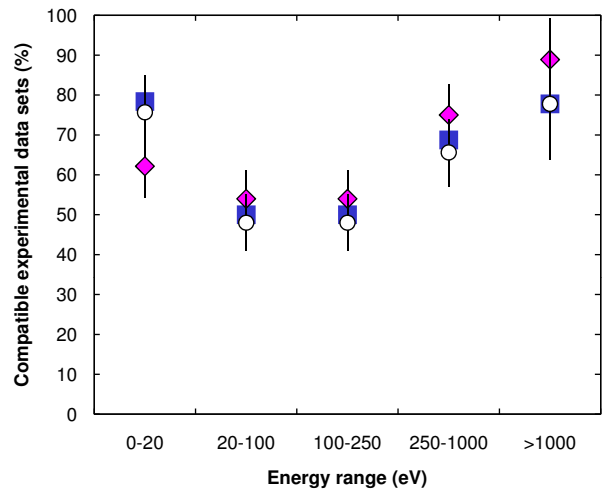}}
\caption{Fraction of test cases in which BEB cross sections calculated with
different atomic atomic parameters in (\ref{eq_beb}) are compatible
with experimental data at 0.05 significance level: with parameters as in the
software implementation described in section \ref{sec_implbeb} (blue squares),
as in the original model formulation (pink diamonds) and as in section
\ref{sec_implbeb} except for the average electron kinetic energies, which are as
in the original model formulation (white circles). The analysis of compatibility
reported in the plot is limited to the subset of elements for which the original
parameters are documented in the literature. }
\label{fig_beb_kim}
\end{figure}

As shown in section \ref{sec_bebveri}, the replacement of the atomic parameters
used in the original formulation affects the value of the  cross
sections calculated by the BEB model.
The effect of different values of the atomic parameters on the compatibility of the
model  with measurements is illustrated in Fig. \ref{fig_beb_kim},
which compares the fraction of test cases in which BEB cross sections calculated with
different atomic atomic parameters in (\ref{eq_beb}) are consistent
with experimental data at 0.05 significance level.

The influence of atomic parameters on the accuracy of the cross sections has
been estimated through a sensitivity analysis considering the compatibility with
experiment:.
Cross sections were calculated with parameters as in the original
model formulation, as in the software implementation, and with original average
electron kinetic energies along with the other parameters as described in
section \ref{sec_implbeb}.
The comparisons with experimental data concerning calculations based on original
parameters are necessarily limited to the elements for which original parameter
values could be retrieved in the literature.

\begin{figure}
\centerline{\includegraphics[angle=0,width=8.5cm]{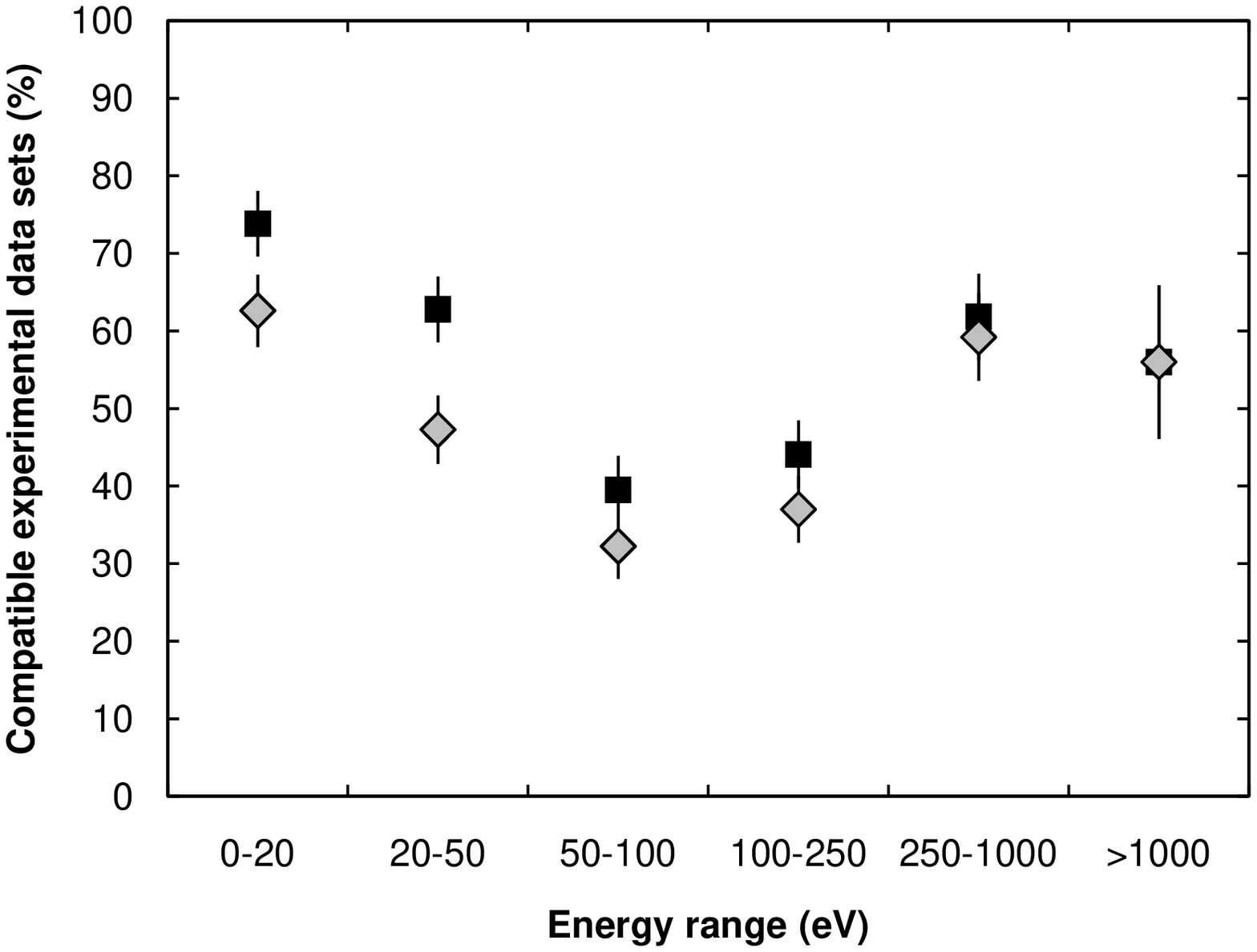}}
\caption{Fraction of test cases in which BEB cross sections calculated with NIST
ionization potentials (black squares) or with EADL ones (grey diamonds) in
(\ref{eq_beb}) are compatible with experimental data at 0.05
significance level.}
\label{fig_beb_eadl}
\end{figure}

The differences in compatibility with experiment are small, and a categorical
analysis based on contingency tables confirms the equivalence of the different
calculations.
Nevertheless, apart from the lowest energy range, the cross sections calculated
with the original parameters exhibit systematically greater compatibility with
experiment than those resulting from modified parameters. 
The difference in compatibility related to different values of the average
electron kinetic energies appears small; this observation suggests that the main
source of differences is related to atomic binding energies and occupation
numbers.
In particular, the value of the first ionization potential plays a significant
role in determining the accuracy of BEB cross sections, as one can observe in
Fig. \ref{fig_beb_eadl}, which compares cross sections calculated with NIST and
EADL ionization potentials.
An extensive evaluation of atomic binding energies and their effect in particle
transport can be found in \cite{tns_binding}.

The apparent overall better performance of the model with the original
parameters suggests that improved accuracy could be achieved by optimizing the
source of the atomic binding energies to be used in the calculation.
However, this is not a straightforward operation, since consistency should be
ensured with other atomic parameters, namely occupation numbers and electron
kinetic energies, involved in the formulation of the model.

\subsection{Effect of DM model parameters}

As discussed in section \ref{sec_dm_veri}, the cross sections calculated by the
software exhibit some differences with respect to those reported in
recent publications by the original authors of the model.

\begin{figure}
\centerline{\includegraphics[angle=0,width=8cm]{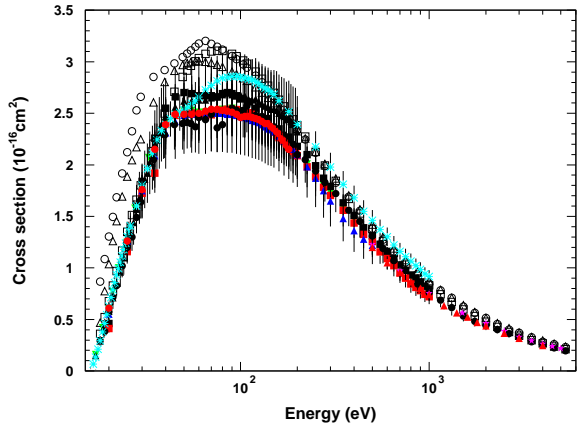}}
\caption{Cross section, Z=18: 
EEDL (empty circles), BEB model (empty squares), 
DM model (empty triangles)
and experimental data from 
\cite{expRejoub2002} (green upside-down triangles),
\cite{exp18S} (black squares),
\cite{wetzel} (black triangles),
\cite{krishnakumar} (red squares),
\cite{ma} (blue triangles),
\cite{expArMcCallion} (black circles),
\cite{expNagy1980} (pink stars),
\cite{rapp} (turquoise asterisks),
\cite{expArKrXeSchram} (red triangles) and
\cite{stephan} (red circles). }
\label{fig_beb18}
\end{figure}

The observed discrepancy is not a source of concern for the accuracy of the
software.
In fact, when subject to the validation process described in section
\ref{sec_validation}, the calculated DM cross sections are compatible with
experimental data in most of the test cases exhibiting relatively large discrepancies
with respect to recently published original values: apart from cerium,
gadolinium and dysprosium, all the elements exhibiting greater than 10\% average
difference with respect to original references are compatible with experimental
data with 0.05 significance over the energy range covered by measurements;
the calculations for gadolinium and dysprosium are compatible with measurements
above 20 eV.
It is worthwhile to note that also the cross sections for cerium, gadolinium and
dysprosium recently published by the original authors \cite{dmDeutsch2008L}
exhibit visible differences with respect to experimental data.
Regarding the discrepancy between calculated and original cross sections for
argon, the controversial experimental situation depicted in Fig. \ref{fig_beb18}
hinders the assessment of which calculation would produce more reliable cross
sections.

The verification and validation analysis suggests that, given the quality of the
available measurements, there is room for some flexibility in the determination
of the DM model parameters deriving from a fit to experimental data: different
parameters may modify the value of cross sections without affecting
substantially the overall accuracy of the model with respect to experimental
references.

It is worthwhile to note that, while a sensitivity analysis of the BEB model implementation to different values of atomic parameters appearing in its formulation, like electron binding energies, was feasible, a similar procedure would not be straightforward for the DM model, whose formulation is the result of a global fit performed by the original authors.

\subsection{Dependency on the type of cross section measurement}

The theoretical models considered in this paper concern the calculation of cross
sections for single ionization, while the experimental data to which they are
compared include both measurements of single and ``total counting'' or ``total
gross'' cross sections, that also account for multiple ionization.
In principle the former should be more reliable references for the validation
process, as the comparison would involve consistent physics quantities;
nevertheless, this assumption could be invalidated by the heterogeneous quality
of the experimental measurements discussed in section \ref{sec_exp}.

Some of the experimental data involved in the validation are relative cross
sections with respect to reference values taken from other theoretical or
experimental sources: for instance, several cross sections are reported relative
to the measurements of \cite{rapp}, while examples of normalization with respect
to theoretical calculations are the measurements of \cite{fujii} (relative to
Bray's calculations \cite{bray} at 15 eV) and of \cite{boivin} (relative to
McGuire's calculations \cite{mcguire} at 500 eV).
The normalization procedure is prone to introduce further uncertainties and
possible biases in the reference data.

\begin{figure}
\centerline{\includegraphics[angle=0,width=8.5cm]{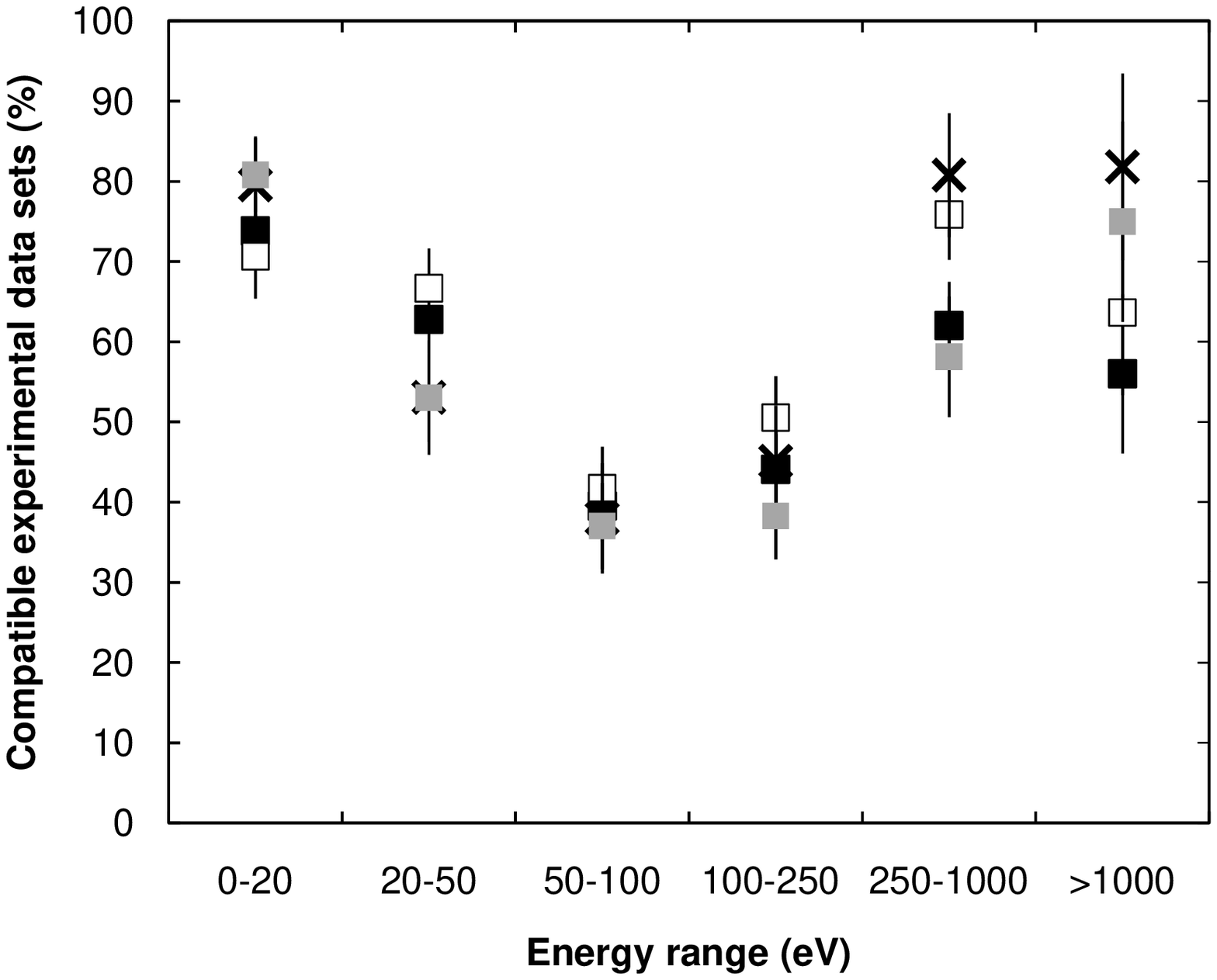}}
\caption{Fraction of test cases in which BEB cross sections are
compatible with experiment at 0.05 significance level, for different types of
measurements: all measurements (black squares), single ionization (white
squares), absolute cross section measurements (grey squares) and absolute 
measurements of single ionization (crosses).}
\label{fig_beb_syst}
\end{figure}

\begin{figure}
\centerline{\includegraphics[angle=0,width=8.5cm]{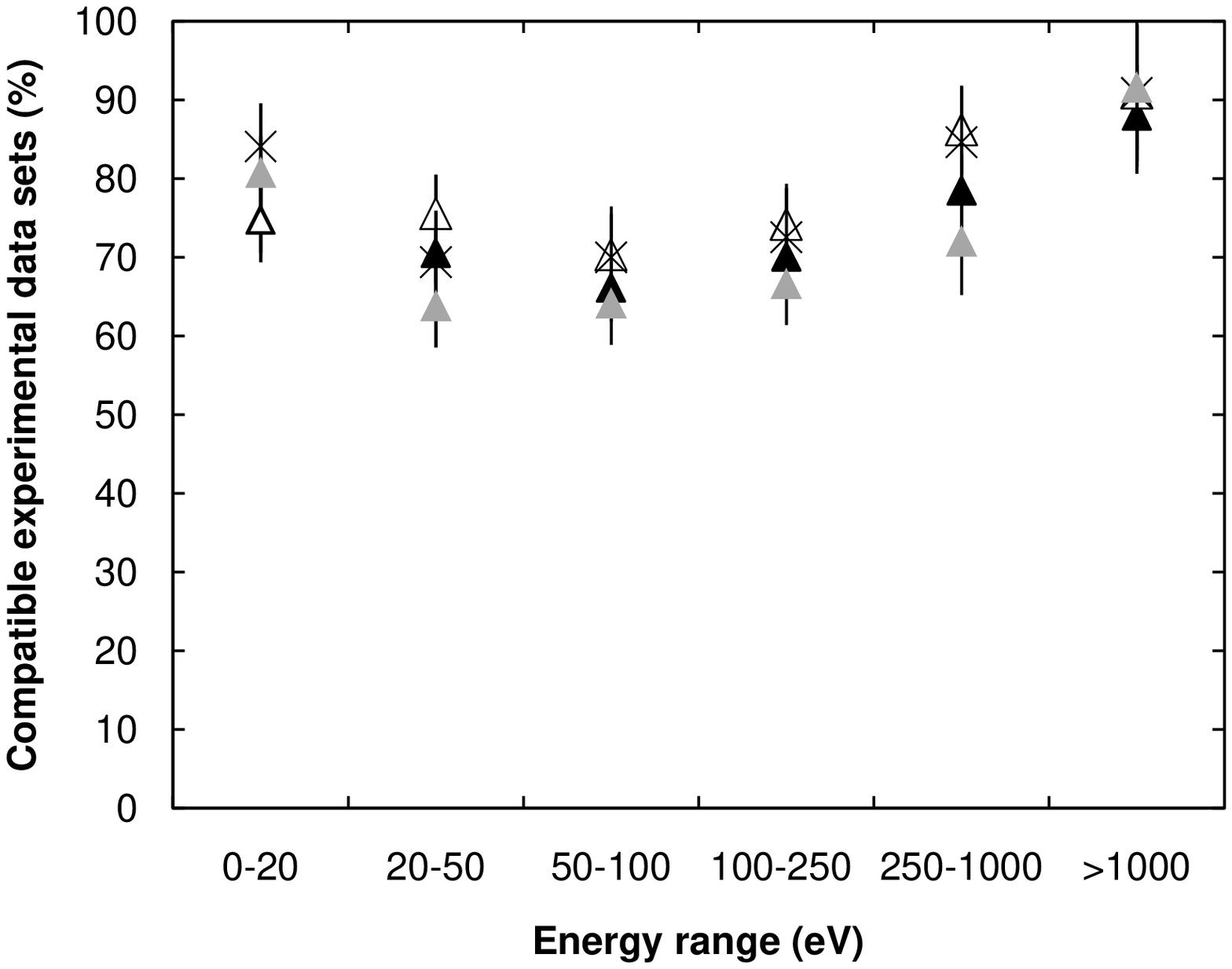}}
\caption{Fraction of test cases in which DM cross sections are
compatible with experiment at 0.05 significance level, for different types of
measurements: all measurements (black triangles), single ionization (white
triangles), absolute cross section measurements (grey triangles) and absolute 
measurements of single ionization (crosses).}
\label{fig_dm_syst}
\end{figure}

\begin{figure}
\centerline{\includegraphics[angle=0,width=8.5cm]{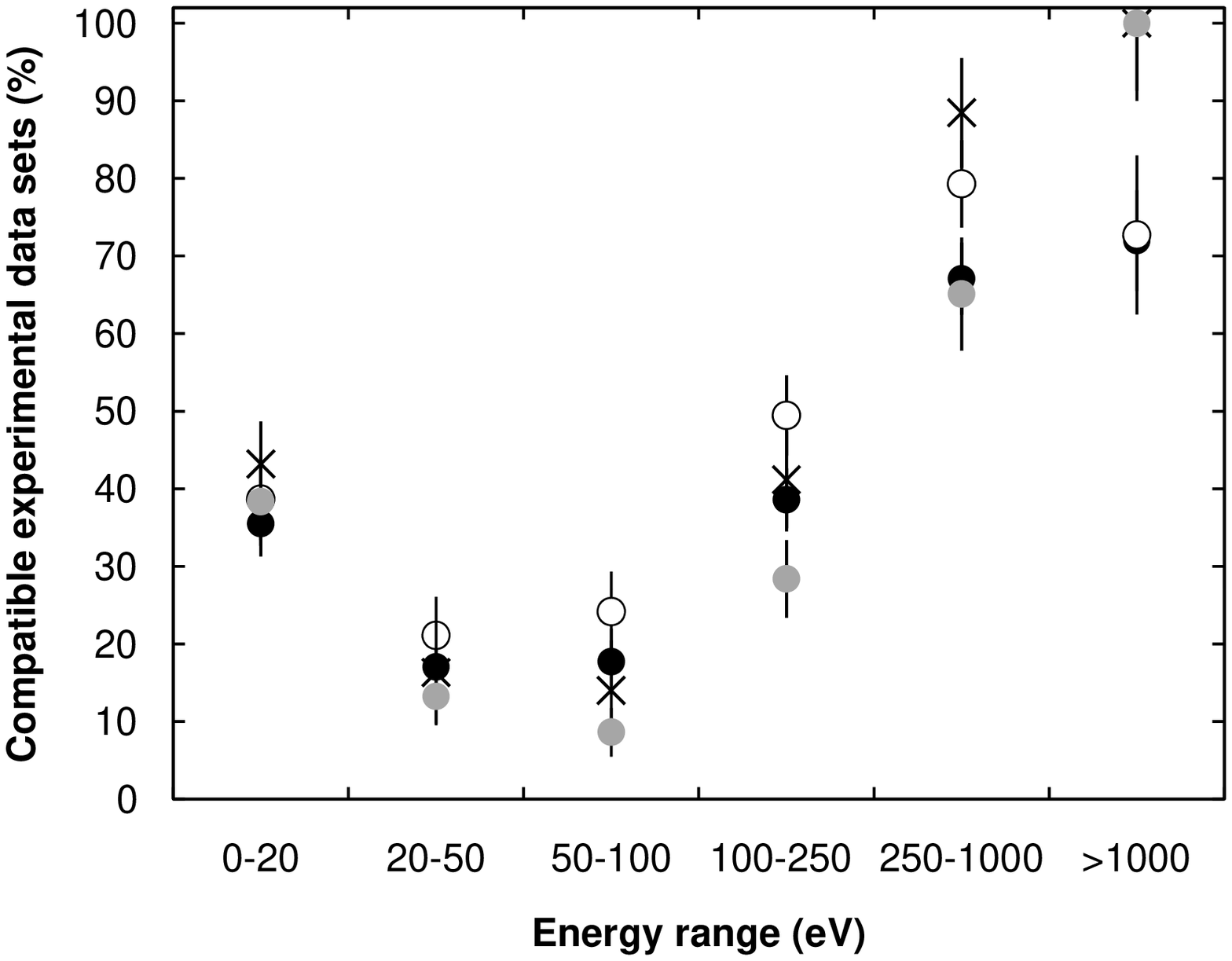}}
\caption{Fraction of test cases in which EEDL cross sections are
compatible with experiment at 0.05 significance level, for different types of
measurements: all measurements (black circles), single ionization (white
circles), absolute cross section measurements (grey circles) and absolute 
measurements of single ionization (crosses).}
\label{fig_eedl_syst}
\end{figure}

The fraction of test cases which are compatible with experiment is shown in
Figs. \ref{fig_beb_syst}-\ref{fig_eedl_syst} for different types of experimental
references: the whole sample, measurements of single ionization only, absolute
cross section measurements only, and absolute measurements of single ionization.
The consistency of the DM model with experiment appears independent from the
type of reference data, while for BEB and EEDL cross sections one can observe
some increased compatibility with measurements concerning single ionization.

\begin{figure}
\centerline{\includegraphics[angle=0,width=8.5cm]{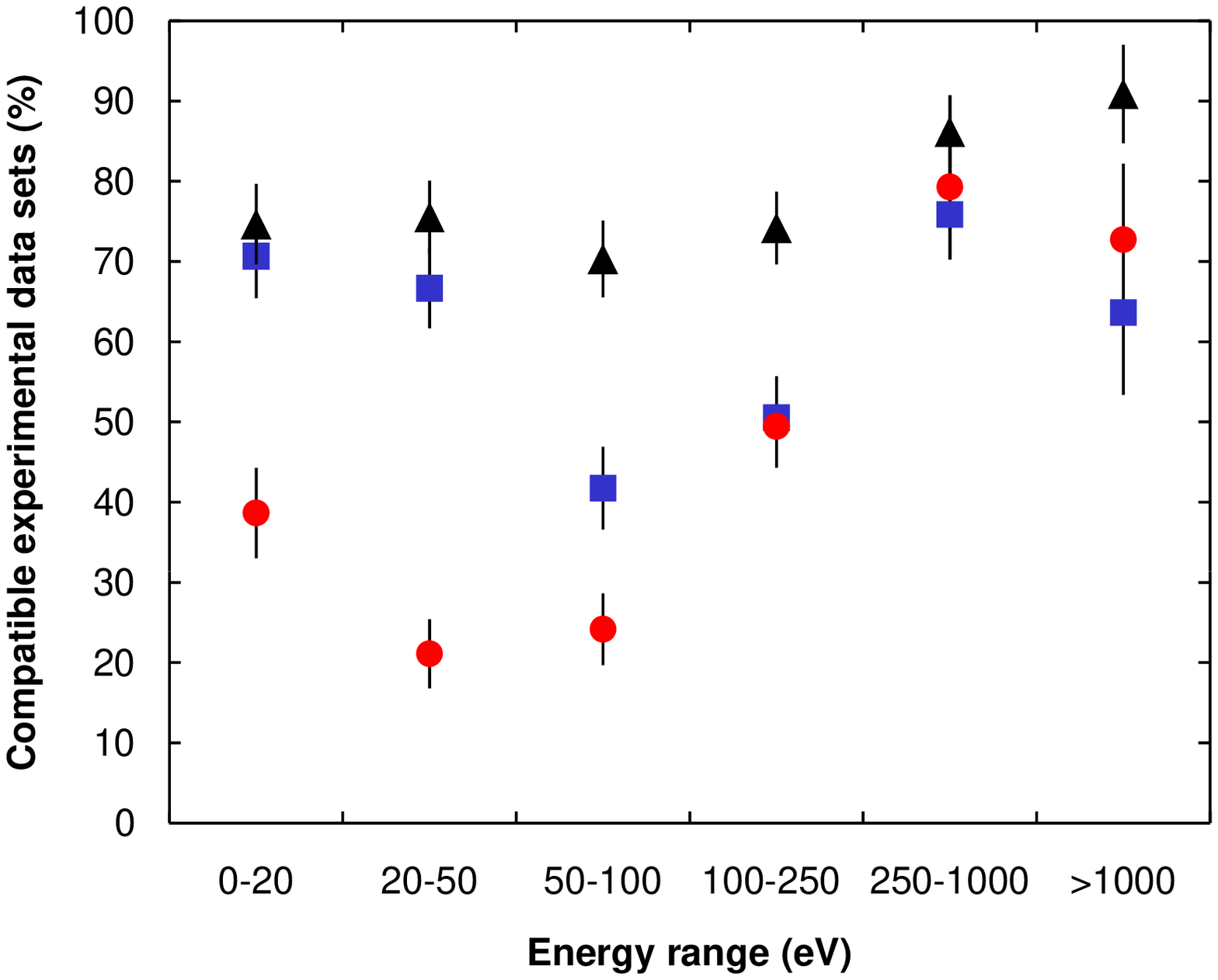}}
\caption{Fraction of test cases in which cross sections calculated by the three
models are compatible with single ionization measurements at 0.05 significance
level: BEB model (blue squares), DM model (black triangles) and EEDL (red
circles).}
\label{fig_models_single}
\end{figure}

\begin{figure}
\centerline{\includegraphics[angle=0,width=8.5cm]{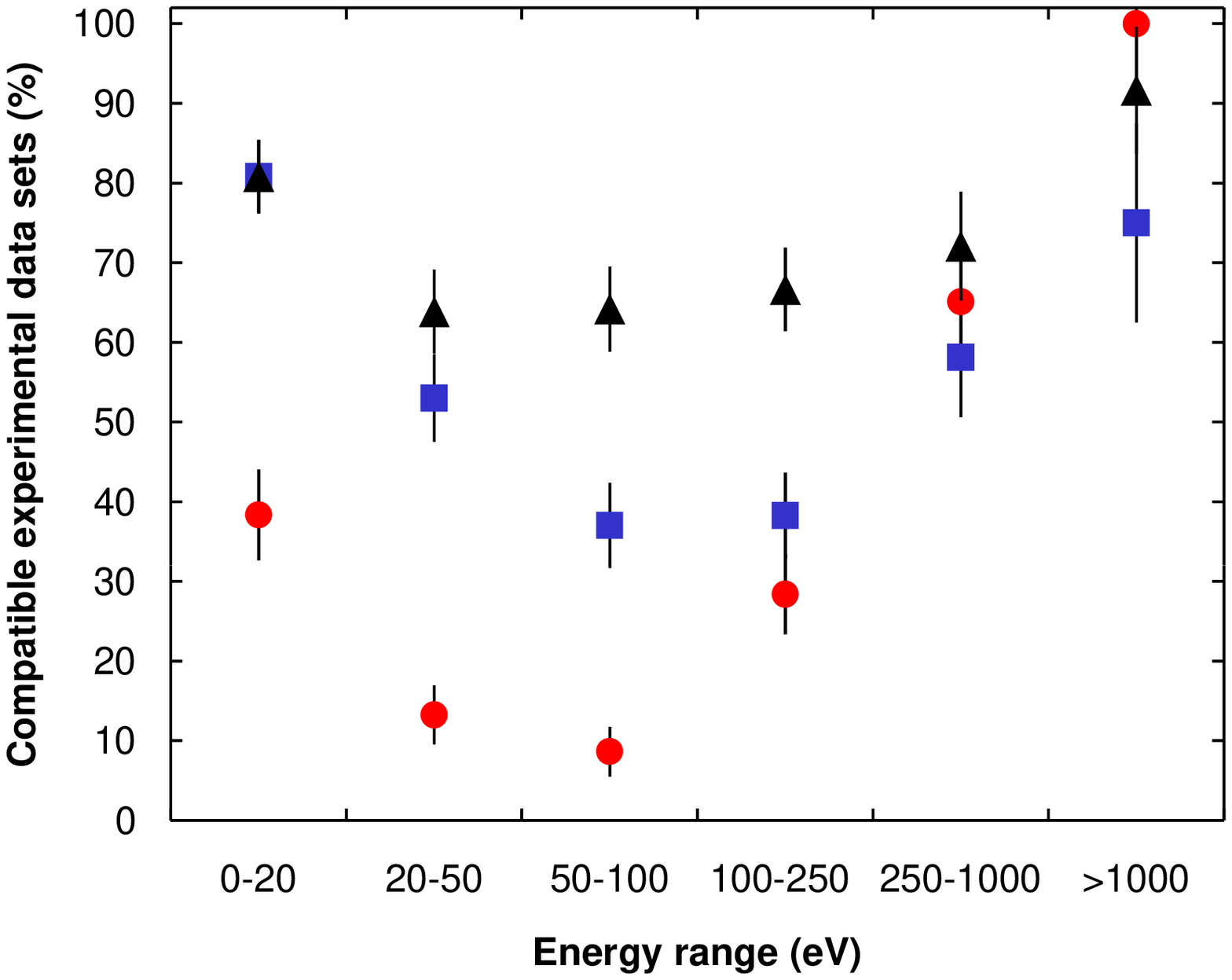}}
\caption{Fraction of test cases in which cross sections calculated by the three
models are compatible with absolute ionization cross section measurements at 0.05 significance
level: BEB model (blue squares), DM model (black triangles) and EEDL (red
circles).}
\label{fig_models_absolute}
\end{figure}

\begin{figure}
\centerline{\includegraphics[angle=0,width=8.5cm]{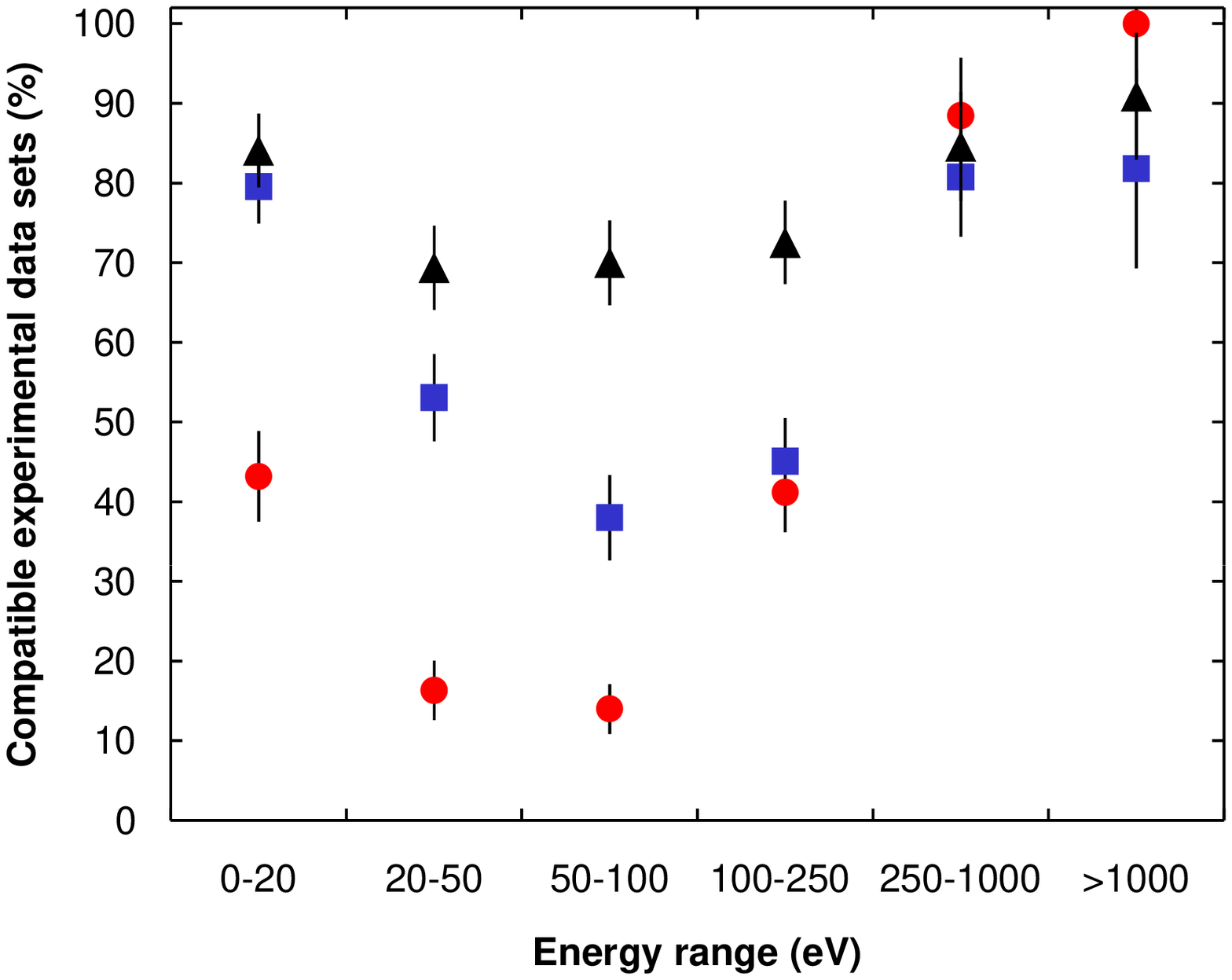}}
\caption{Fraction of test cases in which cross sections calculated by the three
models are compatible with absolute measurements of single ionization cross
sections at 0.05 significance level: BEB model (blue squares), DM model (black
triangles) and EEDL (red circles).}
\label{fig_models_singleabsolute}
\end{figure}

The relative trend of compatibility with experiment of the three cross section
models is scarcely affected by the nature of the reference experimental data, as
one can observe in Figs.
\ref{fig_models_single}-\ref{fig_models_singleabsolute}, to be compared with
Fig. \ref{fig_models_all} reporting the fraction of compatible test cases for all
types of measurement.

The influence of the type of measurements in assessing the accuracy of
a given model has been estimated through an analysis of compatibility with
alternative categories of experimental references: single ionization and total
counting or gross cross sections, absolute and relative measurements.
The analysis, summarized in Tables \ref{tab_contsingle}
and \ref{tab_contabsolute}, is based on contingency tables, which report the
``pass'' and ``fail'' outcome of goodness-of-fit tests associated with either
category of measurements in the various energy ranges.

%The possible contribution of systematic effects due to considering measurements
%of single ionization, or total counting or gross cross sections, is estimated by
%comparing the compatibility of each model with either type of experimental data.

The DM model exhibits equivalent compatibility with single ionization
measurements and total (counting or gross) experimental cross sections with 0.01
significance level over the whole energy range.
The results are similar for the BEB model, with the exception of energy range
between 250 eV and 1 keV, where the hypothesis of equivalent compatibility with
experiment over the categories of single and total (counting or gross) cross
section measurements is rejected with 0.01 significance.
Regarding EEDL, equivalent behavior with respect to the two categories of
experimental data is achieved at 0.01 significance level only in the low energy
range below 50 eV and above 1 keV.

The BEB and DM model exhibit statistically equivalent behavior at 0.01
significance level with respect to absolute and relative measurements, with the
exception of the energy range between 20 and 50 eV regarding the BEB model.
No clear trend can be identified in EEDL compatibility with either type
of measurements.

The compatibility of the individual models with respect to different types of
experimental measurements is reflected in their comparative analysis reported in
Table \ref{tab_contbeb}: the results of comparison with respect to the DM model
are consistent for all types of measurements up to 250 eV, but above 250 eV the
BEB model is statistically equivalent with 0.05 significance to the DM one with
respect to measurements of single ionization and absolute cross sections, while
their compatibility is limited to 0.01 significance over the whole collection of
data samples.
The sensitivity 
%of the comparative analysis 
to the type of measurements above
250 eV suggests that the equivalence of the BEB and DM models at reproducing
experimental data is somewhat marginal in this energy range, therefore small
variations in the experimental sample over which the two models are evaluated
are prone to perturb the outcome of their comparison.

From this analysis one can conclude that the evaluation of the validity of the
three models is scarcely affected by the type of experimental data taken as
references: the DM model exhibits the best overall compatibility with experiment
and its comparison with the other models produces consistent results at 0.01
significance level irrespective of the technique of measurement.

\begin{table*}
\begin{center}
\caption{Contingency tables comparing the compatibility of theoretical models with 
single and total (counting or gross) cross section measurements}
\label{tab_contsingle}
\begin{tabular}{|l|l|cc|cc|cc|}
\hline
Energy 	&Goodness-of-fit test &\multicolumn{2}{|c|}{\bf BEB}	&\multicolumn{2}{|c|}{\bf DM} 	&\multicolumn{2}{|c|}{\bf EEDL}         \\
\hline											                                      
\multirow{6}{*}{$<$ 20 eV} 								                                      
& 				&Single &Total			&Single     &Total  	      	&Single     &Total                \\
%\cline{2-12}						        		                                                
& Pass				&53	&26	 		&56	&24		      	&29         &9                      \\
& Fail				&22	&6 			&19	&8   		      	&46         &23                    \\
\cline{2-8}										                                      
& p-value Fisher test		& \multicolumn{2}{|c|}{0.339} 	&\multicolumn{2}{|c|}{1}       	&\multicolumn{2}{|c|}{0.379}           \\
& p-value Pearson $\chi^2$	& \multicolumn{2}{|c|}{0.254} 	&\multicolumn{2}{|c|}{0.971}  	&\multicolumn{2}{|c|}{0.297}    	\\
& p-value Yates $\chi^2$	& \multicolumn{2}{|c|}{0.368} 	&\multicolumn{2}{|c|}{1}  	&\multicolumn{2}{|c|}{0.411}    	\\
\hline											                                      
\multirow{6}{*}{20-50 eV} 								                                      
& 				&Single &Total                  &Single     &Total 	      &Single     &Total                \\
%\cline{2-12}						                                                                        
& Pass				&60	&21			&68         &23      	      &19       &3                        \\
& Fail				&30	&18			&22         &16		      &71       &36                        \\
\cline{2-8}										                                      
& p-value Fisher test		& \multicolumn{2}{|c|}{0.173} 	&\multicolumn{2}{|c|}{0.091}  &\multicolumn{2}{|c|}{0.076}           \\
& p-value Pearson $\chi^2$	& \multicolumn{2}{|c|}{0.167} 	&\multicolumn{2}{|c|}{0.058}  &\multicolumn{2}{|c|}{not applicable}    	\\
& p-value Yates $\chi^2$	& \multicolumn{2}{|c|}{0.236} 	&\multicolumn{2}{|c|}{0.092}  &\multicolumn{2}{|c|}{0.108}    	\\
\hline											                                      
\multirow{6}{*}{50-100 eV} 								                                      
& 				&Single &Total                  &Single     &Total 	      &Single     &Total                \\
%\cline{2-12}						                                                                        
& Pass				&38     &11  			&64         &18               &22      &0                         \\
& Fail				&53     &22  			&27         &15               &69      &33                         \\
\cline{2-8}										                                      
& p-value Fisher test		&\multicolumn{2}{|c|}{0.416} 	&\multicolumn{2}{|c|}{0.133}  &\multicolumn{2}{|c|}{0.001}           \\
& p-value Pearson $\chi^2$	&\multicolumn{2}{|c|}{0.396} 	&\multicolumn{2}{|c|}{0.101}  &\multicolumn{2}{|c|}{not applicable}    	\\
& p-value Yates $\chi^2$	&\multicolumn{2}{|c|}{0.522} 	&\multicolumn{2}{|c|}{0.154}  &\multicolumn{2}{|c|}{0.004}    	\\
\hline											                                      
\multirow{6}{*}{100-250 eV} 								                                      
& 				&Single &Total                  &Single     &Total 	      &Single     &Total                \\
%\cline{2-12}						                                                                        
& Pass				&47	&9 			&69         &20               &46         &3                      \\
& Fail				&46	&25 			&24         &14               &47         &31                      \\
\cline{2-8}										                                      
& p-value Fisher test		&\multicolumn{2}{|c|}{0.017} 	&\multicolumn{2}{|c|}{0.125}  &\multicolumn{2}{|c|}{$<0.001$}           \\
& p-value Pearson $\chi^2$	&\multicolumn{2}{|c|}{0.016} 	&\multicolumn{2}{|c|}{0.094}  &\multicolumn{2}{|c|}{not applicable}      \\
& p-value Yates $\chi^2$	&\multicolumn{2}{|c|}{0.027} 	&\multicolumn{2}{|c|}{0.145}  &\multicolumn{2}{|c|}{$<0.001$}    	\\
\hline											                                      
\multirow{6}{*}{250 eV - 1 keV} 							                                      
& 				&Single &Total                  &Single     &Total            &Single     &Total                \\
%\cline{2-12}						                                                                        
& Pass				&44	&5 			&50         &12               &46         &7                      \\
& Fail				&14	&16 			&8          &9                &12         &14                       \\
\cline{2-8}										                                      
& p-value Fisher test		&\multicolumn{2}{|c|}{$<0.001$} &\multicolumn{2}{|c|}{0.011}  &\multicolumn{2}{|c|}{$<0.001$}           \\
& p-value Pearson $\chi^2$	&\multicolumn{2}{|c|}{$<0.001$} &\multicolumn{2}{|c|}{0.005}  &\multicolumn{2}{|c|}{$<0.001$}    	\\
& p-value Yates $\chi^2$	&\multicolumn{2}{|c|}{$<0.001$} &\multicolumn{2}{|c|}{0.014}  &\multicolumn{2}{|c|}{$<0.001$}    	\\
\hline											                                      
\multirow{5}{*}{$>$ 1 keV} 								                                      
& 				&Single &Total                  &Single     &Total 	      &Single     &Total                 \\		
%\cline{2-12}										                                      
& Pass				&14	&0			&20	    &2      	      &16         &2                      \\
& Fail				&8 	&3			&2 	    &1       	      &6          &1                      \\
\cline{2-8}										                                      
& p-value Fisher test		&\multicolumn{2}{|c|}{0.072} 	&\multicolumn{2}{|c|}{0.330}       &\multicolumn{2}{|c|}{1}           \\
%p-value Pearson $\chi^2$	&\multicolumn{2}{|c|}{0.} 	&\multicolumn{2}{|c|}{}       &\multicolumn{2}{|c|}{}           \\
& p-value Yates $\chi^2$	&\multicolumn{2}{|c|}{0.143} 	&\multicolumn{2}{|c|}{0.791}  &\multicolumn{2}{|c|}{0.641}    	\\
\hline
\end{tabular}
\end{center}
\end{table*}

\begin{table*}
\begin{center}
\caption{Contingency tables comparing the compatibility of theoretical models with 
absolute and relative cross section measurements}
\label{tab_contabsolute}
\begin{tabular}{|l|l|cc|cc|cc|}
\hline
Energy 	&Goodness-of-fit test &\multicolumn{2}{|c|}{\bf BEB}	&\multicolumn{2}{|c|}{\bf DM} 	&\multicolumn{2}{|c|}{\bf EEDL}         \\
\hline											                                      
\multirow{6}{*}{$<$ 20 eV} 								                                      
& 					&Absolute &Relative			&Absolute     &Relative  	      	&Absolute     &Relative                \\
%\cline{2-12}						        		                                                
& Pass				&59	&20	 		&59	&21		      		&28         &10                      \\
& Fail					&14	&14 			&14	&13   		      	&45         &24                    \\
\cline{2-8}										                                      
& p-value Fisher test		& \multicolumn{2}{|c|}{0.020} 	&\multicolumn{2}{|c|}{0.054}       	&\multicolumn{2}{|c|}{0.395}           \\
& p-value Pearson $\chi^2$	& \multicolumn{2}{|c|}{0.016} 	&\multicolumn{2}{|c|}{0.035}  	&\multicolumn{2}{|c|}{0.368}    	\\
& p-value Yates $\chi^2$	& \multicolumn{2}{|c|}{0.030} 	&\multicolumn{2}{|c|}{0.061}  	&\multicolumn{2}{|c|}{0.494}    	\\
\hline											                                      
\multirow{6}{*}{20-50 eV} 								                                      
& 				&Absolute &Relative                  &Absolute     &Relative 	      &Absolute     &Relative                \\
%\cline{2-12}						                                                                        
& Pass				&44	&37			&53         &38      	      &11       &11                       \\
& Fail					&39	&9			&30         &8		      &72       &35                        \\
\cline{2-8}										                                      
& p-value Fisher test		& \multicolumn{2}{|c|}{0.002} 	&\multicolumn{2}{|c|}{0.028}  &\multicolumn{2}{|c|}{0.146}           \\
& p-value Pearson $\chi^2$	& \multicolumn{2}{|c|}{0.002} 	&\multicolumn{2}{|c|}{0.025}  &\multicolumn{2}{|c|}{0.123}    	\\
& p-value Yates $\chi^2$	& \multicolumn{2}{|c|}{0.004} 	&\multicolumn{2}{|c|}{0.042}  &\multicolumn{2}{|c|}{0.194}    	\\
\hline											                                      
\multirow{6}{*}{50-100 eV} 								                                      
& 				&Absolute &Relative                  &Absolute     &Relative 	      &Absolute     &Relative                \\
%\cline{2-12}						                                                                        
& Pass				&30     &19 			&52         &30               &7      &15                         \\
& Fail					&51     &24  			&29         &13               &74      &28                         \\
\cline{2-8}										                                      
& p-value Fisher test		&\multicolumn{2}{|c|}{0.448} 	&\multicolumn{2}{|c|}{0.557}  &\multicolumn{2}{|c|}{$<0.001$}           \\
& p-value Pearson $\chi^2$	&\multicolumn{2}{|c|}{0.438} 	&\multicolumn{2}{|c|}{0.533}  &\multicolumn{2}{|c|}{$<0.001$}    	\\
& p-value Yates $\chi^2$	&\multicolumn{2}{|c|}{0.561} 	&\multicolumn{2}{|c|}{0.671}  &\multicolumn{2}{|c|}{0.001}    	\\
\hline											                                      
\multirow{6}{*}{100-250 eV} 								                                      
& 				&Absolute &Relative                  &Absolute     &Relative 	      &Absolute     &Relative                \\
%\cline{2-12}						                                                                        
& Pass				&31	&25 			&54         &35               &23         &26                      \\
& Fail					&50	&21 			&27         &11               &58         &20                      \\
\cline{2-8}										                                      
& p-value Fisher test		&\multicolumn{2}{|c|}{0.095} 	&\multicolumn{2}{|c|}{0.316}  &\multicolumn{2}{|c|}{0.002}           \\
& p-value Pearson $\chi^2$	&\multicolumn{2}{|c|}{0.079} 	&\multicolumn{2}{|c|}{0.265}  &\multicolumn{2}{|c|}{0.002}      \\
& p-value Yates $\chi^2$	&\multicolumn{2}{|c|}{0.117} 	&\multicolumn{2}{|c|}{0.361}  &\multicolumn{2}{|c|}{0.003}    	\\
\hline											                                      
\multirow{6}{*}{250 eV - 1 keV} 							                                      
& 				&Absolute &Relative                  &Absolute     &Relative            &Absolute     &Relative                \\
%\cline{2-12}						                                                                        
& Pass				&20	&19 			&22         &22               &23         &20                     \\
& Fail					&12	&12 			&10          &9                &9           &11                       \\
\cline{2-8}										                                      
& p-value Fisher test		&\multicolumn{2}{|c|}{1} 			&\multicolumn{2}{|c|}{1}  	&\multicolumn{2}{|c|}{0.595}           \\
& p-value Pearson $\chi^2$	&\multicolumn{2}{|c|}{0.921} 		&\multicolumn{2}{|c|}{0.848}  	&\multicolumn{2}{|c|}{0.530}    	\\
& p-value Yates $\chi^2$	&\multicolumn{2}{|c|}{0.872} 		&\multicolumn{2}{|c|}{0.934}  	&\multicolumn{2}{|c|}{0.721}    	\\
\hline											                                      
\multirow{5}{*}{$>$ 1 keV} 								                                      
& 				&Absolute &Relative                  &Absolute     &Relative 	      &Absolute     &Relative                 \\		
%\cline{2-12}										                                      
& Pass				&9	&5			&11	    &11      	      &12         &6                      \\
& Fail					&3 	&8			&1 	    &2       	      &0          &7                      \\
\cline{2-8}										                                      
& p-value Fisher test		&\multicolumn{2}{|c|}{0.111} 	&\multicolumn{2}{|c|}{1}       &\multicolumn{2}{|c|}{0.005}           \\
%p-value Pearson $\chi^2$	&\multicolumn{2}{|c|}{0.} 	&\multicolumn{2}{|c|}{}       &\multicolumn{2}{|c|}{}           \\
& p-value Yates $\chi^2$	&\multicolumn{2}{|c|}{0.151} 	&\multicolumn{2}{|c|}{0.941}  &\multicolumn{2}{|c|}{0.011}    	\\
\hline
\end{tabular}
\end{center}
\end{table*}

\subsection{Excitation-autoionization}

Apart from direct ionization, which accounts for the ejection of a bound
electron directly into the continuum, additional indirect channels of ionization
may be important for open-shell atoms, such as the excitation of an inner-shell
electron to an upper bound state that leads to autoionization \cite{bebKim2002}.
Their contribution is generally included in the experimental measurements
of total cross section for single ionization, which do not distinguish direct and
indirect channels.

%Total ionization cross sections of atoms and molecules consist of two
%components, direct and indirect ionization. The direct ionization . The BEB/BED
%model is used to calculate direction ionization cross sections.
%
%BEB References 
%When the outermost orbital of an atom is not fully occupied, electrons in the
%core orbitals can be excited to the outermost orbital.
%Excitations of this kind may produce excited states which lie
%either below or above the first ionization limit. 
%The excited states above the first ionization limit must decay either by
%photoemission without producing an ion or by autoionization by ejecting an
%electron. When the core electron comes from an orbital with the same principal
%quantum number as the outermost orbital 􏳺e.g., 2 s 2 p m ) , cross sections of
%such excitation-autoionizations tend to be large, sometimes matching the
%magnitude of direct ionization cross section 􏳀13􏲣.
Contributions of indirect channels are not taken into account by the BEB
model, which describes only cross sections for direct ionization, while their
non explicit treatment could be partly mitigated by the semi-empirical nature of
the Deutsch-M\"ark model.
The results of the validation process in some way reflect this different
approach of the BEB and DM model towards the physics phenomena that contribute
to experimentally measured ionization cross sections.

The neglected contribution of excitation-autoionization could be also a reason for
the rather poor compatibility of EEDL with experiment below 250 eV; nevertheless,
no firm conclusion can be drawn in this respect due to the scarce documentation
of how EEDL tabulations have been calculated.

Methods to calculate cross sections for excitation-autoionization are documented
in the literature \cite{bebKim2002} and would be considered in future
development cycles to include this process among the interactions treated by
Geant4.
%It should be stressed, however, that accounting for excitation would imply
%introducing a new process of electron interaction with matter along with direct
%ionization.

% ------------------------------------------------------------------------
\section{Electron cross section data library}

The minimalist character of the software design and its minimal dependencies 
on other parts of Geant4 facilitate the exploitation of the
developments described in this paper.

The developed cross section classes can be used in association with the Geant4
toolkit for the simulation of electron ionization as a discrete process, through
the mechanism of a policy host class as described in
\cite{tns_dna,em_chep2009,em_nss2009}.
The BEB and DM cross section code can also be exploited for the creation of data
libraries to be used in the current Geant4 scheme, thus extending Geant4
simulation capabilities below the current 250 eV limit recommended for the use
of the EEDL library.

A data library consisting of tabulations of ionization cross section calculated
by the BEB and DM models has been produced exploiting the software developments
described in this paper; the cross sections are tabulated at the same energies
as in the EEDL data library.
This data library can be used at the place of the current EEDL data in
connection with existing implementations of Geant4 ionization process, thus
giving access to the extended energy coverage and the improved accuracy of the
new models in a transparent way.
Its possible use is not limited to Geant4; given its wider interest, it is intended to 
be publicly distributed independently from Geant4 through the Radiation Safety
Information Computational Center (RSICC) following the publication of this paper.

% ------------------------------------------------------------------------
%\section{Demonstration of application?}

% ------------------------------------------------------------------------
\section{Conclusion}

Two models for the calculation of the cross sections for the ionization of atoms
by electron impact, specialized in the low energy range, 
have been implemented: the Binary-Encounter-Bethe model and the
Deutsch-M\"ark model.
These models are intended to extend and improve the current capabilities of 
Geant4 for precision simulation of electron interactions.

The cross sections calculated by these models, as well as those included in the
Evaluated Electron Data Library, have been subject to extensive validation
in the energy range from a few eV to 10 keV.

Among the cross section models analyzed in this paper, the validation analysis
has identified the Deutsch-M\"ark model as the most accurate for modeling
electron ionization over the whole energy range considered in the test.
The EEDL cross sections exhibit statistically equivalent accuracy in the energy
range above 250 eV, in which they were originally recommended for use in Geant4;
they are not adequately accurate to extend their usage below this threshold.
In the lower energy end the Binary Encounter Bethe models exhibits statistically
equivalent accuracy with respect to the Deutsch-M\"ark one; nevertheless, its
performance appears to degrade at higher energies, presumably because it does
not account for other channels than direct ionization.

Possible sources of systematic effects, which could affect the accuracy of the 
implementation of the theoretical models or the outcome of the validation
process, have been analyzed.
The values of atomic parameters, namely atomic binding energies, play a
significant role in determining the accuracy of the calculated cross sections.

A cross section data library has been developed, containing tabulations of
ionization cross sections calculated by the software described in this paper; it
is meant for public release following the publication of this paper.
The availability of cross section tabulations in a publicly distributed data
library would extend the possibility of using them in other simulation systems
than Geant4.

The models investigated in this paper provide more extended capabilities, that
have not been exploited yet in the first development cycle described in this
paper: they can describe the ionization of molecules, which could be of interest
for the simulation of gaseous detectors and plasma interactions, and can
calculate cross section for the ionization of inner shells, thus enabling the
simulation of atomic relaxation determined by a vacancy in the shell occupation.
Such extensions and improvements, as well as the development of complementary
models for final state generation, are intended to be the object of further
developments.

The described developments for the first time endow a
general-purpose Monte Carlo simulation system of the capability of modeling
electron ionization down to the energy scale relevant to nanodosimetry, for
target elements spanning the whole periodic system.
Nevertheless, it is worthwhile to recall that other phenomena, apart from direct
ionization of atoms, should be taken into account for realistic simulation of particle
interactions at very low energies: further effort should be invested for Geant4
to achieve full functionality for particle transport at nano-scale.

%Further research activity is foreseen to investigate the use of these models
%across condensed and discrete simulation schemes, and in experimental use cases
%involving the transition between conventional dosimetry simulation and
%microdosimetry applications.

%Further improvements to more effectively exploit the simulation capabilities
%offered by the newly implemented cross section models in the lower energy range
%not yet covered by Geant4 are under investigation; ideally,
%the extension of cross section calculations down to lower energies should be complemented by
%improved final state models and transport methods.
Due to the already significant length of this paper and its focus on cross
section modeling and validation, applications of the models to real-life
experimental topics are meant to be covered in dedicated papers.

% ------------------------------------------------------------------------

\section*{Acknowledgment}

%This work was partly supported by the Nuclear Research and Development Program
%in Korea through the Radiation Technology Development Program and the Basic
%Atomic Energy Research Institute (BAERI). Support also was received from the
%Korean Ministry of Knowledge Economy (2008-P-EP-HM-E-06-0000)/the Sunkwang
%Atomic Energy Safety Co., Ltd.

The support of the CERN Library has been essential to this study;
the authors are especially grateful to Tullio Basaglia.

The authors thank Andreas Pfeiffer for his significant help with data analysis
tools throughout the project, Elisabetta Gargioni and Vladimir Grichine for
advice concerning the Binary-Encounter-Bethe model, H. Deutsch and K. Becker for
advice concerning the Deutsch-M\"ark model, Sergio Bertolucci and Simone
Giani for valuable discussions, and Matej Bati\v{c} for helpful comments.
The authors express their gratitude to CERN for the support received in the
course of this research project.

%The INFN Genova Computing Service
%(Alessandro Brunengo, Mirko Corosu, Paolo Lantero and Francesco Saffioti) 
%provided helpful technical assistance with the simulation production.

% ------------------------------------------------------------------------

%\clearpage

% ------------------------------------------------------------------------
\clearpage

% ------------------------------------------------------------------------
\clearpage

% ------------------------------------------------------------------------
\clearpage

% ------------------------------------------------------------------------
\clearpage

% ------------------------------------------------------------------------
%\clearpage

% ------------------------------------------------------------------------
%\clearpage

% ------------------------------------------------------------------------
\clearpage

% ------------------------------------------------------------------------
%\clearpage

\begin{figure}
\centerline{\includegraphics[angle=0,width=8cm]{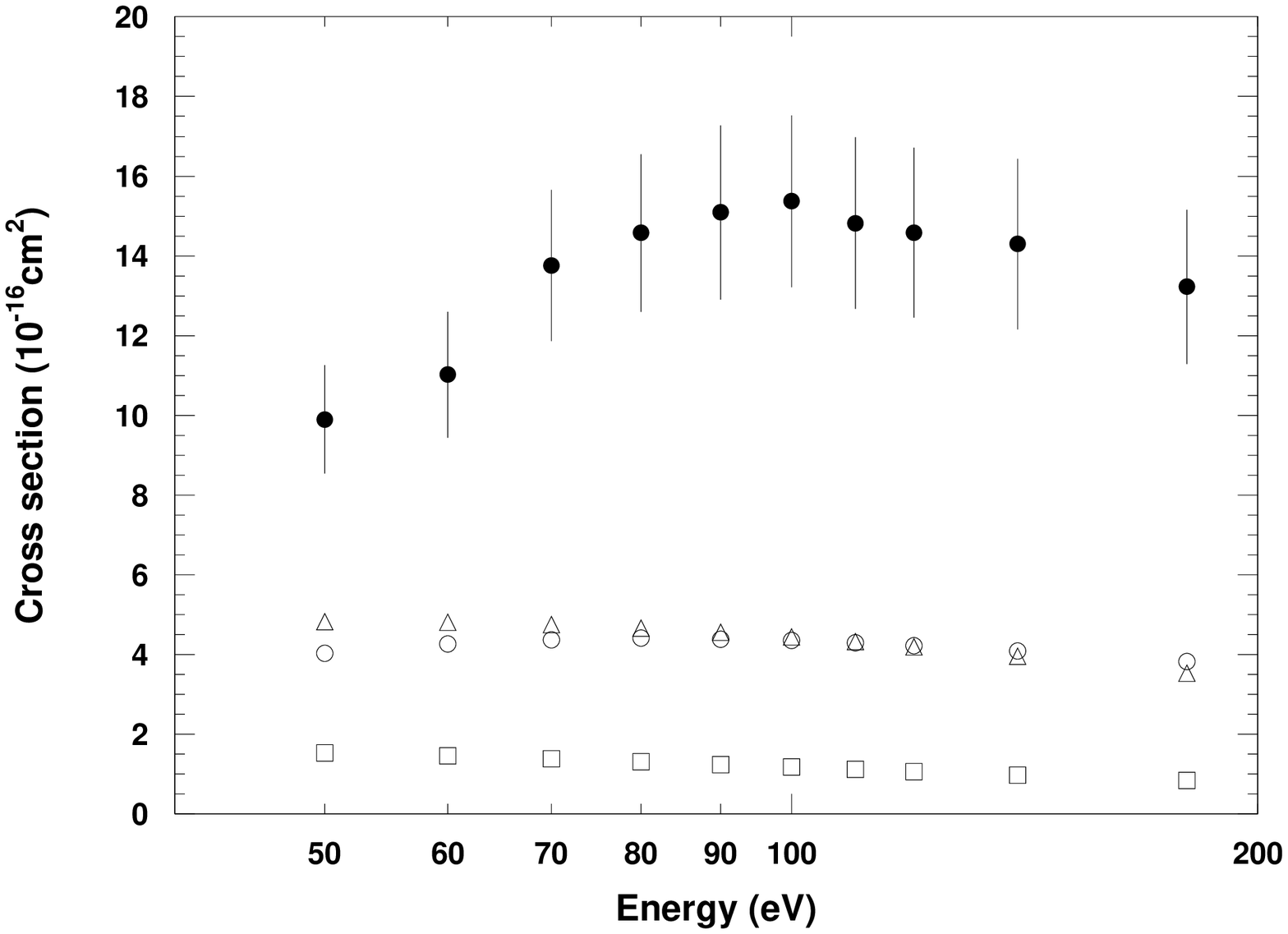}}
\caption{Cross section, Z=79: 
EEDL (empty circles), BEB model (empty squares), 
DM model (empty triangles)
and experimental data from \cite{schroeer} 
(black circles). }
\label{fig_beb79}
\end{figure}

\begin{figure}
\centerline{\includegraphics[angle=0,width=8cm]{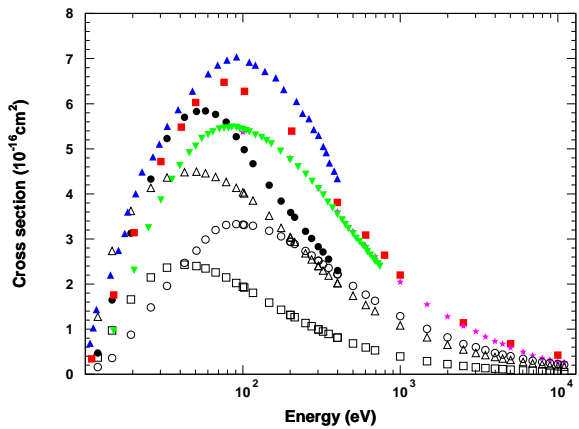}}
\caption{Cross section, Z=80: 
EEDL (empty circles), BEB model (empty squares), 
DM model (empty triangles)
and experimental data from 
\cite{exp80} (black circles),
Harrison (red squares),
Jones (blue triangles),
Liska (pink stars)
and Smith (green upside-down triangles) reported in \cite{expKieffer}. }
\label{fig_beb80}
\end{figure}

\begin{figure}
\centerline{\includegraphics[angle=0,width=8cm]{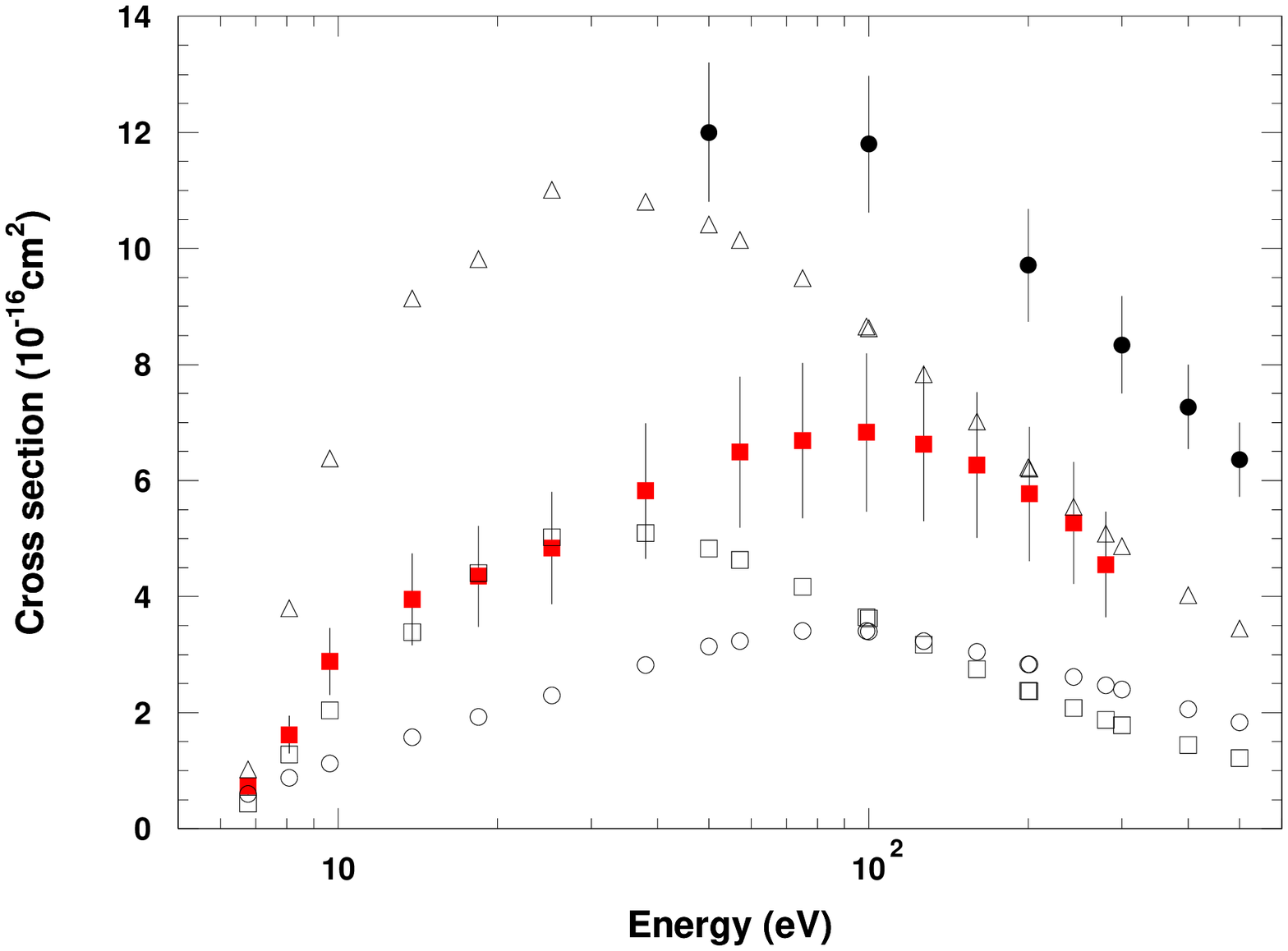}}
\caption{Cross section, Z=81: 
EEDL (empty circles), BEB model (empty squares), 
DM model (empty triangles)
and experimental data from \cite{expM1967} (black circles)
and \cite{shimonAlInTl} (red squares). }
\label{fig_beb81}
\end{figure}

\begin{figure}
\centerline{\includegraphics[angle=0,width=8cm]{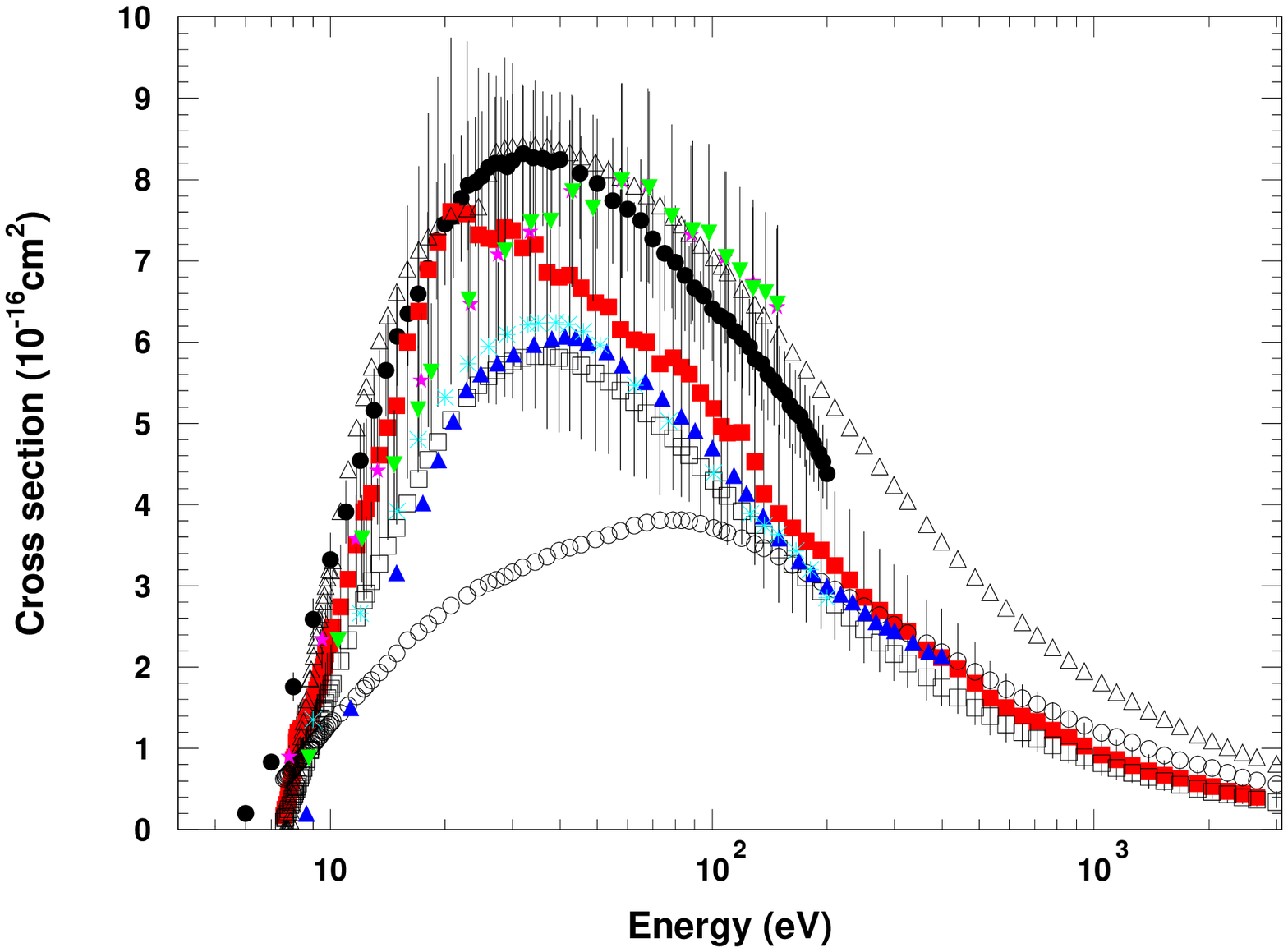}}
\caption{Cross section, Z=82: 
EEDL (empty circles), BEB model (empty squares), 
DM model (empty triangles)
and experimental data from 
\cite{expFreund} (black circles),
\cite{exp82M} (red squares),
\cite{pavlov}  (blue triangles),
\cite{beilina}  (pink stars),
\cite{golovachBaPb}  (turquoise asterisks) and
\cite{wareing} (green upside-down triangles). }
\label{fig_beb82}
\end{figure}

% ------------------------------------------------------------------------
%\clearpage

\begin{figure}
\centerline{\includegraphics[angle=0,width=8cm]{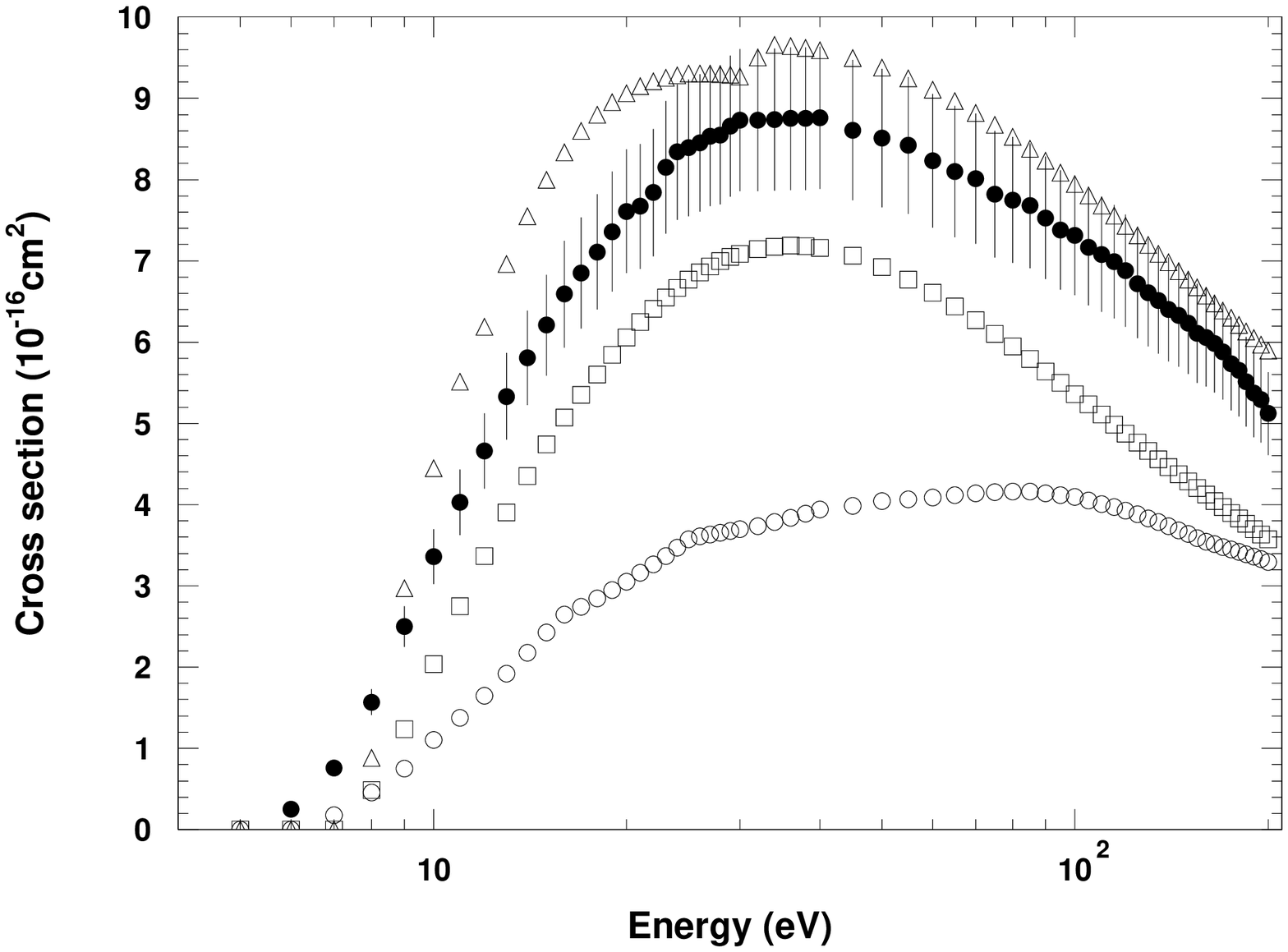}}
\caption{Cross section, Z=83: 
EEDL (empty circles), BEB model (empty squares), 
DM model (empty triangles) and 
experimental data from \cite{expFreund} 
(black circles). }
\label{fig_beb83}
\end{figure}

\begin{thebibliography}{199}

\bibitem{berger}
M. J. Berger, 
``Monte Carlo calculation of the penetration and diffusion of fast
charged particles'',
in \textit{Methods in Computational Physics}, vol. 1, Ed. B. Alder,
S. Fernbach and M. Rotenberg, Academic Press, New York, pp. 135-215,
1963.

%EGS
\bibitem{egs5}
H.~Hirayama, et al., ``The EGS5 code system'', Report SLAC-R-730,
Stanford Linear Accelerator Center, Stanford, CA, USA, 2006.

\bibitem{egsnrc}
I.~Kawrakow and D.W.O.~Rogers, 
``The EGSnrc Code System: Monte Carlo
Simulation of Electron and Photon Transport'', 
NRCC Report PIRS-701, Sep.\ 2006.

% FLUKA
\bibitem{fluka1}
G. Battistoni et al.,
``The FLUKA code: description and benchmarking",
in \textit{AIP Conf. Proc.}, vol. 896, pp. 31-49, 2007.
%A.~Fass{\`o} et al., ``The physics models of FLUKA: status and recent
%developments'', in {\sl Proc.\ Computing in High Energy and Nuclear
%Physics 2003 Conference (CHEP 2003)}, La Jolla, CA, USA, paper MOMT05.

\bibitem{fluka2}
A.~Ferrari et al., 
``Fluka: a multi-particle transport code'', 
Report CERN-2005-010, INFN/TC-05/11, SLAC-R-773, Geneva, Oct. 2005.

% Basic Geant4 references:
\bibitem{g4nim} 
S.~Agostinelli et al., 
``Geant4 - a simulation toolkit''
\textit{Nucl. Instrum. Meth. A}, vol. 506, no. 3, pp. 250-303, 2003.

\bibitem{g4tns}
J.~Allison et al., 
``Geant4 Developments and Applications'' 
\textit{IEEE Trans. Nucl. Sci.}, vol. 53, no. 1, pp. 270-278, 2006.

% MCNP
\bibitem{mcnpx}
J. S.~Hendricks et al., 
``MCNPX, Version 26c'', 
Los Alamos National Laboratory Report LA-UR-06-7991,  2006.

% Penelope
\bibitem{penelope}
J.~Baro, J.~Sempau, J. M.~Fern\'andez-Varea, and F.~Salvat,
``PENELOPE, an algorithm for Monte Carlo simulation of the
penetration and energy loss of electrons and positrons in
matter'', 
\emph{Nucl. Instrum. Meth. B}, vol. 100, no. 1, pp. 31-46, 1995.

%PHITS
\bibitem{phits}
K. Niita, T. Sato, H. Iwase, H. Nose, H. Nakashima and L. Sihver, 
``PHITS - a particle and heavy ion transport code system'',
\textit{ Radiat. Meas.}, vol. 41, pp. 1080-1090, 2006.

% Track structure codes

\bibitem{orec}
R. N. Hamm, H. A. Wright, R. H. Ritchie, J. E. Turner and T. P. Turner, 
``Monte Carlo calculation of transport of electrons through liquid water'', 
in \textit{Proc. 5th Workshop on Microdosim.}, pp. 1037-1053, 1975.

\bibitem{partrac}
S. Henss and H. G. Paretzke,
``Biophysical modeling of radiation induced damage in chromosome'',
in \textit{Proc. Biophysical modelling of radiation effects}, 
Ed.: Adam Hilger, New York, pp. 69-76, 1992.

\bibitem{ptbmc}
B. Grosswendt and S. Pszona,
``The track structure of a-particles from the point of view of
ionization-cluster formation in nanometric volumes of nitrogen'',
\textit{Radiat. Environ. Biophys.}, vol. 41, no. 2, pp. 91-102, 2002.

\bibitem{tramos}
T. Colladant, A. L'Hoir, J. E. Sauvestre and O. Flament,
``Monte-Carlo simulations of ion track in silicon and influence of its spatial
distribution on single event effects'',
\emph{Nucl. Instrum. Meth. B}, vol. 245, no. 2, pp. 464-474, 2006.

\bibitem{tns_dna}
S. Chauvie et al., 
"Geant4 physics processes for microdosimetry simulation: design foundation and
implementation of the first set of models",
\textit{IEEE Trans. Nucl. Sci.,} vol. 54, no. 6, pp. 2619-2628, 2007.

% Geant4 physics models
%\bibitem{em_mc2000}
%S. Chauvie et al.,
%``Geant4 electromagnetic physics'',
%in \emph{Proc. Monte Carlo Conference, Lisbon}, Nov. 2000.

\bibitem{eedl}
S. T. Perkins et al., 
``Tables and Graphs of Electron-Interaction Cross
Sections from 10 eV to 100 GeV Derived from the LLNL Evaluated
Electron Data Library (EEDL)'', 
UCRL-50400 Vol. 31, 1997.

\bibitem{lowe_chep}
S. Chauvie, G. Depaola, V. Ivanchenko, F. Longo, P. Nieminen and M. G. Pia,
``Geant4 Low Energy Electromagnetic Physics'',
in \textit{Proc. Computing in High Energy and Nuclear Physics}, 
Beijing, China, pp. 337-340, 2001.

\bibitem{lowe_nss}
S. Chauvie et al., ``Geant4 Low Energy Electromagnetic Physics'',
in \textit{2004 IEEE Nucl. Sci. Symp. Conf. Rec.}, pp. 1881-1885, 2004.

%\bibitem{lowe_e} 
%J. Apostolakis, S. Giani, M. Maire, P. Nieminen, M.G. Pia, L. Urban,
%``Geant4 low energy electromagnetic models for electrons and photons''
%\textit{INFN/AE-99/18}, Frascati, 1999. 

\bibitem{lowe_e} 
J. Apostolakis, S. Giani, M. Maire, P. Nieminen, M.G. Pia, and L. Urban,
``Geant4 low energy electromagnetic models for electrons and photons''
\textit{INFN/AE-99/18}, Frascati, 1999. 

\bibitem{standard}
H. Burkhardt et al., 
``Geant4 Standard Electromagnetic Package'',
in \textit{Proc. 2005 Conf. on Monte Carlo Method: Versatility Unbounded
in a Dynamic Computing World}, Am. Nucl. Soc., USA, 2005.

\bibitem{pai}
J. Apostolakis, S. Giani, L. Urban, M. Maire, A. V. Bagulya, and V. M. Grichine,
``An implementation of ionisation energy loss in very thin absorbers for the 
GEANT4 simulation package'',
\emph{Nucl. Instrum. Meth. A}, vol. 453, no. 3, pp. 597-605, 2000.

\bibitem{epdl97}	
D. Cullen et al., 
``EPDL97, the Evaluated Photon Data Library'', 
UCRL-50400, Vol. 6, Rev. 5, 1997.

%\bibitem{bote}
%D. Bote and F. Salvat,
%``Calculations of inner-shell ionization by electron impact with the
%distorted-wave and plane-wave Born approximations'',
%\textit{Phys. Rev. A}, vol. 77, no. 4, pp. 042701, 2008.

\bibitem{penelope2008}
F. Salvat, J. M. Fernandez-Varea, E. Costa, and J. Sempau, 
``PENELOPE - A Code System for Monte Carlo Simulation of Electron and Photon
Transport'',
NEA-06416, Barcelona, 2009. 

%\bibitem{penelope2001}
%F. Salvat, J. M. Fernandez-Varea, E. Costa, and J. Sempau, 
%``PENELOPE - A Code System for Monte Carlo Simulation of Electron and Photon 
%Transport'',
%NEA/NSC/DOC(2001)19, Issy-les-Moulineaux, 2001. 

%\bibitem{relax}
%S. Guatelli, A. Mantero, B. Mascialino, P. Nieminen, and M. G. Pia, 
%``Geant4 Atomic Relaxation'', 
%\emph{IEEE Trans. Nucl. Sci.}, vol. 54, no. 3, pp. 585-593, 2007.
%
%\bibitem{standard}
%H. Burkhardt et al., 
%``Geant4 Standard Electromagnetic Package'',
%in \textit{Proc. 2005 Conf. on Monte Carlo Method: Versatility Unbounded
%in a Dynamic Computing World}, Am. Nucl. Soc., USA, 2005.
%
%\bibitem{radphyschem}
%J. Apostolakis et al.,
%``Geometry and physics of the Geant4 toolkit for high and medium energy 
%applications'',
%\textit{Radiat. Phys. Chem.},  vol. 78, no. 10, pp. 859-873, 2009.

% Sandia test
\bibitem{tns_sandia}
A. Lechner, M.G. Pia, M. Sudhakar
``Validation of Geant4 low energy electromagnetic processes against 
precision measurements of electron energy deposit'',
\textit{IEEE Trans. Nucl. Sci.}, vol. 56, no. 2, pp. 398-416,  2009.

% SW process

\bibitem{up}
I. Jacobson, J. Booch, and J. Rumbaugh, 
``The Unified Software Development Process'', 
1st ed., Ed: Addison-Wesley, 1999.

% ---- Nano5
\bibitem{nano5_mc2009}
M. G. Pia et al.,
"R\&D for co-working condensed and discrete transport methods in Geant4 kernel",
in \textit{Proc. Int. Conf. on Mathematics, Computational Methods an Reactor Physics}, Saratoga Springs, NY, 2009.

\bibitem{nano5_nss2009}
M. Augelli et al.,  
"Geant4-related R\&D for new particle transport methods     
in \textit{2009 IEEE Nucl. Sci. Symp. Conf. Rec.}, pp. 173-176, 2009.

% SW process

\bibitem{waterfall}
W. W. Royce, 
``Managing the development of large software systems'', 
in \emph{Proc. IEEE WESCON}, Los Angeles, pp. 328-338, 1970.

% BEB, DM basic references

% BEB model:
\bibitem{bebKim1994} 
Y. K. Kim and M. E. Rudd, 
``Binary-encounter-dipole model for electron-impact ionization by electron impact,''
\textit{Phys. Rev. A}, vol. 50, pp. 3954-3967, 1994.

% DM model:
\bibitem{dmDeutsch1987} 
H. Deutsch and T. D. M\"ark, 
``Calculation of absolute electron impact ionization cross-section functions for 
single ionization of He, Ne, Ar, Kr, Xe, N and F,''
\textit{Int. J. Mass Spectrom. Ion Processes}, vol. 79, pp. R1-R8, 1987.

% ---- Software design

\bibitem{alexandrescu}
A. Alexandrescu, 
``Modern C++ Design'', Ed.: Addison-Wesley, 2001.

\bibitem{em_nss2009}
M. Augelli et al.,
"Research in Geant4 electromagnetic physics design, and its effects on
computational performance and quality assurance",
in \textit{2009 IEEE Nucl. Sci. Symp. Conf. Rec.}, pp. 177-180, 2009.

\bibitem{em_chep2009}
M. G. Pia et al.,
"Design and performance evaluations of generic programming techniques in a R\&D
prototype of Geant4 physics",
\textit{J. Phys.: Conf. Ser.}, vol. 219, pp. 042019, 2010.

% ---- Verification and Validation

\bibitem{ieee_vv}
IEEE Computer Society,
``IEEE Standard for Software Verification and Validation'', 
IEEE Std 1012-2004, Jun. 2005.

\bibitem{engauge}
M. Mitchell, 
Engauge Digitizer [Online]. 
Available: http://digitizer.sourceforge.net.

% ---- Theory BEB

\bibitem{mott}
N. F. Mott,
``The Collision between Two Electrons'',
\textit{Proc. Royal Soc. A}, vol. 126, no. 801, pp. 259-267, 1930.

%\bibitem{vriens}
%L. Vriens,
%``Binary-Encounter Electron-Atom Collision Theory'',
%\textit{Phys. Rev.}, vol. 141, pp. 88-92, 1966.

\bibitem{bethe}
H. Bethe,
``Zur Theorie des Durchgangs schneller Korpuskularstrahlen durch Materie'',
\textit{Ann. Phys.}, vol. 397, no. 3,  pp. 325–400, 1930.

\bibitem{jain}
D. K. Jain and S. P. Khare, 
``Ionizing collisions of electrons with CO$_2$, CO, H$_2$O, CH$_4$ and NH$_3$'',
\textit{J. Phys. B}, vol. 9, no. 8, pp. 1429-1438, 1976.

\bibitem{khare_meath}
S. P. Khare and W. J. Meath,
``Cross sections for the direct and dissociative ionisation of NH$_3$, H$_2$O and H$_2$S by electron impact '',
\textit{J. Phys. B}, vol. 20, pp. 2101-2116, 1987.

\bibitem{khare_kumar}
S. P. Khare and A. Kumar,
``Energy deposition by protons in molecular nitrogen'',
\textit{Physica}, vol. 100 C, no. 1, pp. 135-143, 1980.

\bibitem{kaushik}
Y. D. Kaushik, S. P. Khare and A. Kumar,
``Inelastic collisions of protons with water vapour'',
\textit{Physica}, vol. 1006 B-C, no.1, pp. 128-134, 1981.

\bibitem{bebKim1998} 
Y. K. Kim, J. Migdalek, W. Siegel, J. Bieron, 
``Electron-impact ionization cross section of rubidium,''
\textit{Phys. Rev. A}, vol. 57, pp. 246-254, 1998.

\bibitem{desclaux_metecc}
J. P. Desclaux,
in ``Methods and Techniques in Computational Chemistry: METECC'',
vol. A, pp. 253-274, ed. E. Clementi, Cagliari, 1993.

\bibitem{eadl}
S. T. Perkins et al., 
``Tables and Graphs of Atomic Subshell and
Relaxation Data Derived from the LLNL Evaluated Atomic Data Library (EADL)'', 
Z=1-100, UCRL-50400 Vol. 30, 1997.

% ---- Ionization potentials

\bibitem{nist_ionipot}
W. C. Martin, A. Musgrove, S. Kotochigova, and J. E. Sansonetti,
``Ground Levels and Ionization Energies for the Neutral Atoms'',
Online. Available: 
http://physics.nist.gov/PhysRefData/IonEnergy/ionEnergy.pdf.
%http://www.nist.gov/physlab/data/ion\_energy.cfm.

%\bibitem{martin_wiese}
%W. C. Martin and W. L. Wiese,
%``Atomic Spectroscopy'', 
%in Atomic, Molecular, and Optical Physics Handbook, 
%G. W. F. Drake ed., AIP, Woodbury, NY, 1996. 

%\bibitem{nistE2000} 
%S. G. Lias, 
%in Ionization Energy Evaluation, NIST Chemistry Webbook, 
%NIST Standard Reference Database No. 69, 
%edited by W. G. Mallard and P. J. Linstrom 
%(National Institute of Standards and Technology (NIST), Gaithersburg, MD, 2000),
%Online. Available: http://webbook.nist.gov.
%%[NIST tables: (http://webbook.nist.gov)].
%%Available: http://physics.nist.gov/PhysRefData/IonEnergy/ionEnergy.pdf

% ---- Theory DM

\bibitem{dmMarg1994} 
D. Margreiter, H. Deutsch, and T. D. M\"ark, 
``A semiclassical approach to the calculation of electron impact 
ionization cross-sections of atoms: from hydrogen to uranium,''
\textit{Int. J. Mass Spectrom. Ion Processes}, vol. 139, pp. 127-139, 1994.

\bibitem{dmDeutsch2000} 
H. Deutsch, K. Becker, S. Matt, T. D. M\"ark, 
``Theoretical determination of absolute electron-impact ionization cross 
sections of molecules,''
\textit{Int. J. Mass Spectrom.}, vol. 197, pp. 37-69, 2000.

\bibitem{dmDeutsch2004} 
H. Deutsch, P. Scheier, K. Becker, T. D. M\"ark, 
``Revised high energy behavior of the Deutsch-M\"ark (DM) formula for the 
calculation of electron impact ionization cross sections of atoms,''
\textit{Int. J. Mass Spectrom.}, vol. 233, pp. 13-17, 2004.

\bibitem{dmDeutsch2005} 
H. Deutsch, P. Scheier, S. Matt-Leubner, K. Becker, T. D. M\"ark, 
``A detailed comparison of calculated and measured electron-impact ionization 
cross sections of atoms using the Deutsch-M\"ark (DM) formalism,''
\textit{Int. J. Mass Spectrom.}, vol. 243, pp. 215-221, 2005.

\bibitem{dmDeutsch2005err} 
H. Deutsch, P. Scheier, S. Matt-Leubner, K. Becker, T. D. M\"ark, 
``Erratum to A detailed comparison of calculated and measured electron-impact ionization 
cross sections of atoms using the Deutsch-M\"ark (DM) formalism,''
\textit{Int. J. Mass Spectrom.}, vol. 246, p. 113, 2005.

\bibitem{thomson}
J. J. Thomson, 
``Ionization by moving electrified particles'',
\textit{Philos. Mag.}, vol. 23, no. 136, pp.  449-457, 1912.

\bibitem{gryzinski}
M. Gryzinski, 
``Classical Theory of Atomic Collisions. I. Theory of Inelastic Collisions'',
\textit{Phys. Rev.}, vol. 138, pp. 336-358, 1965.

\bibitem{desclaux} 
J. P. Desclaux, 
``Relativistic Dirac-Fock expectation values for atoms with Z=1 to Z=120,''
\textit{Atom. Data Nucl. Data Tables}, vol. 12, pp. 311-406, 1973.

\bibitem{lotz} 
W. Lotz, 
``Electron binding energies in free atoms,''
\textit{J. Opt. Sot. Am.}, vol. 60, pp. 206-210, 1970.

% --------------------
\bibitem{tns_binding}
M. G. Pia et al.,
``Evaluation of atomic electron binding energies for Monte Carlo particle transport'',
Submitted to IEEE Trans. Nucl. Sci., May 2011.

\bibitem{dm_privcomm}
H. Deutsch, e-mail: h.deutsch@rz.uni-greifswald.de,
Private communication, April 2010.

% --------------------
% DM model:

\bibitem{dmDeutsch2008} 
H. Deutsch, K. Becker, T. D. M\"ark, 
``Calculated absolute cross-sections for the electron-impact ionization of atoms 
with atomic numbers between 20 and 56 using the Deutsch-M\"ark (DM) formalism,''
\textit{Int. J. Mass Spectrom.}, vol. 271, pp. 58-62, 2008.

\bibitem{dmDeutsch2008L} 
H. Deutsch, K. Becker, H. Zhang, M. Probst, T. D. M\"ark, 
``Calculated absolute cross sections for the electron-impact ionization of the 
lanthanide atoms using the Deutsch-M\"ark (DM) formalism,''
\textit{Int. J. Mass Spectrom.}, vol. 271, pp. 63-67, 2008.

% ---- EEDL

\bibitem{seltzer1988}
S. M. Seltzer,
``Cross sections for Bremsstrahlung Production and Electron-Impact Ionization'',
in \textit{Monte Carlo Transport of Electrons and Photons}, T. M. Jenkins and W.
R. Nelson Ed., Plenum Press, New York, 1988.

\bibitem{moller}
C. M{\o}ller,
``Zur theorie des durchgangs schneller elektronen durch materie'',
\textit{Ann. Phys.}, vol. 406, no. 5,  pp. 531–585, 1932.

\bibitem{weizsacker}
C. F. von Weizsäcker,
``Ausstrahlung bei St\''ossen sehr schneller Elektronen'',
\textit{Z. Phys.}, vol. 88, no. 9-10, pp. 612-625, 1934.

\bibitem{williams}
E. J. Williams,
``Correlation of Certain Collision Problems with Radiation Theory'',
\textit{Mat. Fys. Medd.}, vol. 13, no. 4, pp. 1-50, 1935.

\bibitem{scofield1978}
J. H. Scofield,
``K and L shell ionization of atoms by relativistic electrons'',
\textit{Phys. Rev. A}, vol. 18, pp. 963-970, 1978.


%\bibitem{uml}
%G. Booch, J. Rumbaugh, and I. Jacobson,
%\relax \emph{``The Unified Modeling Language User Guide''},
%Ed. Boston: Addison-Wesley, 1999.

%\bibitem{aida}
%G. Barrand, P. Binko, M. Donszelmann, A. Johnson, and A. Pfeiffer, 
%``Abstract interfaces for data analysis: component architecture for data 
%analysis tools'', 
%in \emph{Proc. of CHEP Int. Conf. on Computing in High Energy and 
%Nucl. Phys.}, pp. 215-218, Science Press, Beijing, 2001.
%
%\bibitem{iAIDA}
%A. Pfeiffer, 
%``iAIDA - an implementation of AIDA in C++''.
%Online. Available: http://iaida.dynalias.net/

%\bibitem{pi}
%A. Pfeiffer, L. Moneta, V. Innocente, H. C. Lee, and W. L. Ueng, 
%``The LCG PI project: using interfaces for physics data analysis'', 
%\emph{IEEE Trans. Nucl. Sci.}, vol. 52, no. 6, pp. 2823 - 2826, 2005.

%\bibitem{fisz1960}
%M.~Fisz, 
%``On a result by M. Rosenblatt concerning the von Mises-Smirnov test'', 
%\relax \emph{Anls. Ma. St.}, vol. 31, pp. 427-429, 1960.

%\bibitem{kuiper1960}
%N.H.~Kuiper, 
%``Tests concerning random points on a circle'',
%\emph{Proc. Koninkl. Neder. Akad. van Wettensch. A}, vol.63, pp. 38-47, 1960.

%\bibitem{watson1961}
%G.~S.~Watson, 
%``Goodness-of-fit tests on a circle'', 
%\relax \emph{Biometrika}, vol. 48, pp. 109-114, 1961.

%\bibitem{sandia79}
%G. J.~Lockwood et al.,
%``Calorimetric Measurement of Electron energy deposition in Extented Media -
%Theory vs Experiment'',
%Sandia National Laboratories, Report S AND79-0414, Albuquerque, 1980.
%%Available: http://infoserve.sandia.gov/sand\_doc/1979/790414.pdf.

\bibitem{its}
J. A. Halbleib and T. A. Mehlhorn,
``ITS : The Integrated TIGER
Series of Coupled Electron/Photon Monte Carlo Transport Codes'',
Sandia National Laboratories Report No. SAND84-0573, Albuquerque,
November 1984.
%Available: http://infoserve.sandia.gov/sand\_doc/1991/911634.pdf.

%DM model

% BEB model:
\bibitem{bebAli1997} 
M. A. Ali, Y. K. Kim, W. Hwang, N. M. Weinberger, M. E. Rudd, 
``Electron-impact total ionization cross sections of silicon and germanium hydrides,''
\textit{J. Chem. Phys.}, vol. 106, no. 23, pp. 9602-9608, 1997.

\bibitem{bebKim2000} 
Y. K. Kim, W. R. Johnson, M. E. Rudd, 
``Cross sections for singly differential and total ionization of helium by electron impact,''
\textit{Phys. Rev. A}, vol. 61, p. 034702, 2000.

\bibitem{bebKim2000R} 
Y. K. Kim, J. P. Santos, F. Parente, 
``Extension of the binary-encounter-dipole model to relativistic incident electrons,''
\textit{Phys. Rev. A}, vol. 62, p. 052710, 2000.

\bibitem{bebKim2001} 
Y. K. Kim, P. M. Stone, 
``Ionization of boron, aluminum, gallium, and indium by electron impact,''
\textit{Phys. Rev. A}, vol. 64, p. 052707, 2001.

\bibitem{bebKim2002} 
Y. K. Kim, J. P. Desclaux, 
``Ionization of carbon, nitrogen, and oxygen by electron impact,''
\textit{Phys. Rev. A}, vol. 66, p. 012708, 2002.

\bibitem{bebKim2007} 
Y. K. Kim, P. M. Stone, 
``Ionization of silicon, germanium, tin and lead by electron impact,''
\textit{J. Phys. B: At. Mol. Opt. Phys.}, vol. 40, pp. 1597-1661, 2007.

\bibitem{bebAli2008} 
M. A. Ali, Y. K. Kim, 
``Ionization cross sections by electron impact on halogen atoms, diatomic halogen 
and hydrogen halide molecules,''
\textit{J. Phys. B: At. Mol. Opt. Phys.}, vol. 41, p. 145202, 2008.


% ?????

\bibitem{expKieffer} 
L. J. Kieffer, G. H. Dumm,
``Electron impact ionization cross-section data for atoms, atomic ions, and diatomic molecules: 
I. Experimental data,''
\textit{Rev. Mod. Phys.}, vol. 38, no. 1, pp. 1-35, 1966.

% Experimental data:
\bibitem{expHshah1987} 
M. B. Shah, D. S. Elliott, and H. B. Gilbody, 
``Pulsed crossed-beam study of the ionisation of atomic hydrogen 
by electron impact,''
\textit{J. Phys. B: At. Mol. Phys.}, vol. 20, pp. 3501-3514, 1987.

%\bibitem{boksenberg}
%A. Boksenberg as presented in Ref. \cite{expKieffer}.

\bibitem{fite}
W. L. Fite, R. T. Brackmann, 
``Collisions of electrons with hydrogen atoms. I. Ionization,''
\textit{Phys. Rev.}, vol. 112, no. 4, pp. 1141-1151, 1958.

\bibitem{rothe}
E. W. Rothe, L. L. Marino, R. H. Neynaber, S. M. Trujillo, 
``Electron impact ionization of atomic hydrogen and atomic oxygen,''
\textit{Phys. Rev.}, vol. 125, no. 2, pp. 582-583, 1962.

\bibitem{expRejoub2002}
R. Rejoub, B. G. Lindsay, and R. F. Stebbings, 
``Determination of the absolute partial and total cross sections for 
electron-impact ionization of the rare gases,''
\textit{Phys. Rev. A}, vol. 65, p. 042713, 2002.

\bibitem{exp2S}
M. B. Shah, D. S. Elliott, P. McCallion, and H. B. Gilbody, 
``Single and double ionisation of helium by electron impact,''
\textit{J. Phys. B: At. Mol. Opt. Phys.}, vol. 21, pp. 2751-2761, 1988.

\bibitem{rapp}
D. Rapp, P. Englander-Golden, 
``Total cross sections for ionization and attachment in gases by electron impact,''
\textit{Phys. Rev. A}, vol. 43, no. 5, pp. 1464-1479, 1965.

\bibitem{expHeNeSchram} 
B. L. Schram, A. J. H. Boerboom, J. Kistemaker, 
``Partial ionization cross sections of noble gases for electrons with energy 0.5-16 keV,''
\textit{Physica}, vol. 32, pp. 185-196, 1966.

\bibitem{stephan}
K. Stephan, H. Helm, and T. D. M\"ark, 
``Mass spectrometric determination of partial electron impact ionization 
cross sections of He, Ne, Ar and Kr from threshold up to 180 eV,''
\textit{J. Chem. Phys.}, vol. 73, no. 8, pp. 3763-3778, 1980.

\bibitem{krishnakumar}
E. Krishnakumar and S. K. Srivastava, 
``Ionisation cross sections of rare-gas atoms by electron impact,''
\textit{J. Phys. B: At. Mol. Opt. Phys.}, vol. 21, pp. 1055-1082, 1988.

\bibitem{expMontague1984} 
R. G. Montague, M. F. A. Harrison, and A. C. H. Smith,
``A measurement of the cross section for ionisation of helium by electron impact 
using a fast crossed beam technique,''
\textit{J. Phys. B: At. Mol. Phys.}, vol. 17, pp. 3295-3310, 1984.

\bibitem{expNagy1980} 
P. Nagy, A. Skutlartz, and V. Schmidt,
``Absolute ionisation cross sections for electron impact in rare gases,''
\textit{J. Phys. B: At. Mol. Phys.}, vol. 13, pp. 1249-1267, 1980.

\bibitem{wetzel}
R. C. Wetzel, F. A. Baiocchi, T. R. Hayes, and R. S. Freund, 
``Absolute cross sections for electron-impact ionization of the rare-gas atoms 
by the fast neutral beam method,''
\textit{Phys. Rev. A}, vol. 35, no. 2, pp. 559-577, 1987.

\bibitem{expM1965}
R. H. McFarland, J. D. Kinney, 
``Absolute cross sections of Li and other alkali metal atoms for ionization 
by electrons,''
\textit{Phys. Rev.}, vol. 137, no. 4A, pp. 1058-1061, 1965.

\bibitem{expZ1969}
I. P. Zapesochnyi, I. S. Aleksakhin, 
``Ionization of alkali metal atoms by slow electrons,''
\textit{Sov. Phys. JETP}, vol. 28, no. 1, pp. 41-45, 1969.

\bibitem{jalin}
R. Jalin, R. Hagemann, R. Botter, 
``Absolute electron impact ionization cross sections of Li in the energy range from 
100 to 2000 eV,''
\textit{J. Chem. Phys.}, vol. 59, no. 2, pp. 952-959, 1973.

\bibitem{expBrook1978} 
E. Brook, M. F. A. Harrison, and A. C. H. Smith, 
``Measurements of the electron impact ionisation cross sections 
of He, C, O and N atoms,''
\textit{J. Phys. B: Atom. Molec. Phys.}, vol. 11, no. 17, pp. 3115-3132, 1978.

\bibitem{smith}
A. C. H. Smith, E. Caplinger, R. H. Neynaber, E. W. Rothe, and S. M. Trujillo, 
``Electron impact ionization of atomic nitrogen,''
\textit{Phys. Rev.}, vol. 127, no. 5, pp. 1647-1649, 1962.

%Book???
\bibitem{expPeterson} 
J. R. Peterson, in Atomic Collision Processes, edited by M. R. C. McDowell, 
North-Holland Publishing Company, Amsterdam, 1964, pp. 465-473.
%%%%%%%%%%%%%%%%%%%%%%%%%%%%%%%%%%%%%%%%%%%%%%

\bibitem{exp8}
W. R. Thompson, M. B. Shah, and H. B. Gilbody, 
``Single and double ionization of atomic oxygen by electron impact,''
\textit{J. Phys. B: At. Mol. Opt. Phys.}, vol. 28, pp. 1321-1330, 1995.

\bibitem{expOfite1959} 
W. L. Fite, R. T. Brackmann, 
``Ionization of atomic oxygen on electron impact,''
\textit{Phys. Rev.}, vol. 113, pp. 815-816, 1959.

\bibitem{zipf}
E. C. Zipf, 
``The ionization of atomic oxygen by electron impact,''
\textit{Planet. Space Sci.}, vol. 33, no. 11, pp. 1303-1307, 1985.

\bibitem{adamczyk}
B. Adamczyk, A. J. H. Boerboom, B. L. Schram, and J. Kistemaker, 
``Partial ionization cross sections of He, Ne, H$_2$, and CH$_4$ for electrons from 20 to 500 eV,''
\textit{J. Chem. Phys.}, vol. 44, pp. 4640-4642, 1966.

\bibitem{almeida}
D. P. Almeida, A. C. Fontes, and C. F. L. Godinho, 
``Electron-impact ionization cross section of neon ($\sigma$$_{n+}$, n = 1-5),''
\textit{J. Phys. B: At. Mol. Opt. Phys.}, vol. 28, pp. 3335-3345, 1995.

\bibitem{fletcher}
J. Fletcher and I. R. Cowling, 
``Electron impact ionization of neon and argon,''
\textit{J. Phys. B: At. Mol. Phys.}, vol. 6, pp. L258-L261, 1973.

\bibitem{sorokin}
A. A. Sorokin, L. A. Shmaenok, S. V. Bobashev, 
``Measurements of electron-impact ionization cross sections of neon by 
comparison with photoionization,''
\textit{Phys. Rev. A}, vol. 58, no. 4, pp. 2900-2910, 1998.

\bibitem{brink}
G. O. Brink, 
``Absolute ionization cross sections of the alkali metals,''
\textit{Phys. Rev.}, vol. 134, no. 2A, pp. A345-A346, 1964.

\bibitem{fujii}
K. Fujii, S. K. Srivastava, 
``A measurement of the electron-impact ionization cross section of sodium,''
\textit{J. Phys. B}, vol. 28, pp. L559-L563, 1995.

\bibitem{johnston}
A. R. Johnston and P. D. Burrow, 
``Electron-impact ionization of Na,''
\textit{Phys. Rev. A}, vol. 51, no. 3, pp. R1735-R1737, 1995.

\bibitem{tan}
W. S. Tan, Z. Shi, C. H. Ying, and L. Vuskovic, 
``Electron-impact ionization of laser-excited sodium atom,''
\textit{Phys. Rev. A}, vol. 54, no. 5, pp. R3710-R3713, 1996.

\bibitem{expFreund} 
R. S. Freund, R. C. Wetzel, R. J. Shul, and T. R. Hayes, 
``Cross-section measurements for electron-impact ionization of atoms,''
\textit{Phys. Rev. A}, vol. 41, no. 7, pp. 3575-3595, 1990.

\bibitem{boivin}
R. F. Boivin and S. K. Srivastava, 
``Electron-impact ionization of Mg,''
\textit{J. Phys. B: At. Mol. Opt. Phys.}, vol. 31, pp. 2381-2394, 1998.

\bibitem{karstensen}
F. Karstensen and M. Schneider, 
``Absolute cross sections for single and double ionisation of Mg atoms by electron impact,''
\textit{J. Phys. B: At. Mol. Phys.}, vol. 11, no. 1, pp. 167-172, 1978.

\bibitem{expMgMcCallion} 
P. McCallion, M. B. Shah, and H. B. Gilbody, 
``Multiple ionization of magnesium by electron impact,''
\textit{J. Phys. B: At. Mol. Opt. Phys.}, vol. 25, no. 5, pp. 1051-1060, 1992.

\bibitem{expVainsh} 
L. A. Vainshtein, V. I. Ochkur, V. I. Rakhovskii, A. M. Stepanov, 
``Absolute values of electron impact ionization cross sections for magnesium, 
calcium, strontium and barium,''
\textit{Sov. Phys. JETP}, vol. 34, no. 2, pp. 271-275, 1972.

\bibitem{okunoMg}
Y. Okuno, K. Okuno, Y. Kaneko, I. Kanomata,
''Absolute measurement of total ionization cross section of Mg by electron impact,''
\textit{J. Phys. Soc. Japan}, vol. 29, pp. 164-172, 1970.

\bibitem{golovach}
D. G. Golovach, A. N. Drozdov, V. I. Rakhovskii, and V. M. Shustryakov, 
``Measurment of the ionization cross section of aluminum atoms by electronic impact,''
\textit{Meas. Tech. (USSR)}, vol. 30, pp. 587-589, 1987.

% page span???
\bibitem{shimonAlInTl}
L. L. Shimon, E. I. Nepiipov, I. P. Zapesochnyi,
``Effective total electron-impact ionization cross sections for aluminum, gallium, indium and thallium,''
\textit{Sov. Phys. Tech. Phys.}, vol. 20, no. 3, pp. 434-437, 1975.

\bibitem{ziegler}
D. L. Ziegler, J. H. Newman, L. N. Goeller, K. A. Smith, and R. F. Stebbings, 
``Single and multiple ionization of sulfur atoms by electron impact,''
\textit{Planet. Space Sci.}, vol. 30, no. 12, pp. 1269-1274, 1982.

\bibitem{expHayes} 
T. R. Hayes, R. C. Wetzel, and R. S. Freund, 
``Absolute electron-impact-ionization cross-section measurements of the 
halogen atoms,''
\textit{Phys. Rev. A}, vol. 35, no. 2 pp. 578-584, 1987.

\bibitem{exp18S} 
H. C. Straub, P. Renault, B. G. Lindsay, K. A. Smith, and R. F. Stebbings, 
``Absolute partial and total cross sections for electron-impact ionization of argon
from threshold to 1000 eV,''
\textit{Phys. Rev. A}, vol. 52, no. 2, pp. 1115-1124, 1995.

\bibitem{ma}
C. Ma, C. R. Sporleder, and R. A. Bonham, 
``A pulsed electron beam time of flight apparatus for measuring absolute electron impact 
ionization and dissociative ionization cross sections,''
\textit{Rev. Sci. Inst.}, vol. 62, pp. 909-923, 1991.

\bibitem{expArMcCallion} 
P. McCallion, M. B. Shah, and H. B. Gilbody, 
``A crossed beam study of the multiple ionization of argon by electron impact,''
\textit{J. Phys. B: At. Mol. Opt. Phys.}, vol. 25, no. 5, pp. 1061-1071, 1992.

\bibitem{expArKrXeSchram} 
B. L. Schram, 
``Partial ionization cross sections of noble gases for electrons with energy 0.5-18 keV,''
\textit{Physica}, vol. 32, pp. 197--208, 1966.

\bibitem{korchevoi}
Yu. P. Korchevoi, A. M. Przonski, 
``Effective electron impact excitation and ionization cross sections for cesium, 
rubidium, and potassium atoms in the pre-threshold region,''
\textit{Sov. Phys. JETP}, vol. 24, no. 6, pp. 1089-1092, 1967.

\bibitem{expKNygaard} 
K. J. Nygaard, 
``Electron impact autoionization in heavy alkali metals,''
\textit{Phys. Rev. A}, vol. 11, no. 4, pp. 1475-1478, 1975.

\bibitem{expM1967} 
R. H. McFarland, 
``Electron-impact ionization measurements of surface-ionizable atoms,''
\textit{Phys. Rev.}, vol. 159, no. 1, pp. 20--26, 1967.

\bibitem{rakhovskii}
V. J. Rakhovski, A. M. Stepanov, 
``Absolute values of the apparent cross section for calcium ionization by electron collision,''
\textit{High Temp.}, vol. 7, pp. 1001-1003, 1969.

\bibitem{okuno}
Y. Okuno,
``Ionization cross sections of Ca, Sr and Ba by electron impact,''
\textit{J. Phys. Soc. Japan}, vol. 31, no. 4, pp. 1189-1195, 1971.

\bibitem{schneider}
M. Schneider, 
``Measurement of absolute ionization cross sections for electron impact,''
\textit{J. Phys. D: Appl. Phys.}, vol. 7, pp. L83--L86, 1974.

%\bibitem{expTiVNi} 
%M. Koparnski as presented in Ref. \cite{dmMarg1994}.

\bibitem{expFeShah1993} 
M. B. Shah, P. McCallion, K. Okuno, and H. B. Gilbody, 
``Multiple ionization of iron by electron impact,''
\textit{J. Phys. B: At. Mol. Opt. Phys.}, vol. 26, pp. 2393-1401, 1993.

\bibitem{bolorizadeh}
M. A. Bolorizadeh, C. J. Patton, M. B. Shah, and H. B. Gilbody, 
``Multiple ionization of copper by electron impact,''
\textit{J. Phys. B: At. Mol. Opt. Phys.}, vol. 27, pp. 175-183, 1994.

% page span???
\bibitem{pavlovCuAg}
S. I. Pavlov, V. I. Rakhovskii, G. M. Fedorova, 
``Measurement of cross sections for ionization by electron impact at low vapor pressures,''
\textit{Sov. Phys. JETP}, vol. 25, p. 12-16, 1967.

\bibitem{schroeer}
J. M. Schroeer, D. H. Gunduz, S. Livingston,
``Electron impact ionization cross sections of Cu and Au between 40 and 250 eV, and 
the velocity of evaporated atoms,''
\textit{J. Chem. Phys.}, vol. 58, no. 11, pp. 5135-5140, 1973.

\bibitem{expZnCd} 
R. F. Pottie, 
``Cross sections for ionization by electrons. I. Absolute ionization cross sections 
of Zn, Cd, and Te$_2$. II. Comparison of theoretical with experimental values for 
atoms and molecules,''
\textit{J. Chem. Phys.}, vol. 44, no. 3, pp. 916-922, 1966.

\bibitem{expShul1989} 
R. J. Shul, R. C. Wetzel, and R. S. Freund, 
``Electron-impact-ionization cross section of the Ga and In atoms,''
\textit{Phys. Rev. A}, vol. 39, no. 11, pp. 5588-5596, 1989.

\bibitem{expGaInVainsh} 
L. A. Vainshtein, D. G. Golovach, V. I. Ochkur, V. I. Rakhovskii, N. M. Rumyantsev, 
V. M. Shustryakov, 
``Cross sections for ionization of gallium and indium by electrons,''
\textit{Sov. Phys. JETP}, vol. 66, no. 1, pp. 36-39, 1987.
 
\bibitem{patton}
C. J. Patton, K. O. Lozhkin, M. B. Shah, J. Geddes, and H. B. Gilbody, 
``Multiple ionization of gallium by electron impact,''
\textit{J. Phys. B: At. Mol. Opt. Phys.}, vol. 29, pp. 1409-1417, 1996.

\bibitem{expRbNygaard} 
K. J. Nygaard and Y. B. Hahn, 
``Total electron impact ionization cross section in rubidium from threshold
to 250 eV,''
\textit{J. Chem. Phys.}, vol. 58, no. 8, pp. 3493--3499, 1973.

\bibitem{schappe}
R. S. Schappe, T. Walker, L. W. Anderson, and C. C. Lin, 
``Absolute electron-impact ionization cross section measurements 
using a magneto-optical trap,''
\textit{Phys. Rev. Lett.}, vol. 76, no. 23, pp. 4328-4331, 1996.

\bibitem{crawford}
C. K. Crawford, K. I. Wang,
``Electron-impact ionization cross sections for silver,''
\textit{J. Chem. Phys.}, vol. 47, pp. 4667-4669, 1967.

\bibitem{franzreb}
K. Franzreb, A. Wucher, H. Oechsner,
``Absolute cross sections for electron impact ionization of Ag$_2$,''
\textit{Z. Phys. D}, vol. 19, pp. 77--79, 1991.

\bibitem{lin}
S. S. Lin, F. E. Stafford,
``Electron-impact ionization cross sections. IV. group IVb atoms,''
\textit{J. Chem. Phys.}, vol. 47, pp. 4664-4666, 1967.

% page span???
\bibitem{lyubimov}
A. P. Lyubimov, S. I. Pavlov, V. I. Rakhovskii, N. G. Zaitseva,
``Procedure for measuring the ionization cross sections and ionization coefficients of metal atoms,''
\textit{Bull. Acad. USSR. Phys. Ser.}, vol. 17, pp. 1033--1037, 1963.

\bibitem{mathur}
D. Mathur and C. Badrinathan, 
``Ionization of xenon by electrons: Partial cross sections for single, double, and triple ionization,''
\textit{Phys. Rev. A}, vol. 35, no. 3, pp. 1033-1042, 1987.

\bibitem{expStephan1984} 
K. Stephan, T. D. M\"ark, 
``Absolute partial electron impact ionization cross sections of Xe from threshold up to 180 eV,''
\textit{J. Chem. Phys.}, vol. 81, no. 7, pp. 3116-3117, 1984.

\bibitem{heil}
H. Heil, B. Scott, 
``Cesium ionization cross section from threshold to 50 eV,''
\textit{Phys. Rev.}, vol. 145, no. 1, pp. 279-284, 1966.

\bibitem{expCsNygaard} 
K. J. Nygaard, 
``Electron-impact ionization cross section in cesium,''
\textit{J. Chem. Phys.}, vol. 49, no. 5, pp. 1995-2002, 1968.

\bibitem{dettmann}
J. M. Dettmann and F. Karstensen, 
``Absolute ionisation functions for electron impact with barium,''
\textit{J. Phys. B: At. Mol. Phys.}, vol. 15, pp. 287-300, 1982.

\bibitem{expYagi2000} 
S. Yagi, T. Nagata,
``Absolute total and partial cross-sections for ionization of Ba and Eu atoms by electron impact,''
\textit{J. Phys. Soc. Japan}, vol. 69, no. 5, pp. 1374-1383, 2000.

\bibitem{golovachBaPb}
D. G. Golovach, V. I. Rakhovskii, V. M. Shustryakov,
``Apparatus for measurement of electronic-ionization cross sections of metal atoms,''
\textit{Instr. Exp. Tech.}, vol. 29, pp. 1396-1399, 1987.

\bibitem{expYagi2001} 
S. Yagi, T. Nagata,
``Absolute total and partial cross sections for ionization of free lanthanide atoms by 
electron impact,''
\textit{J. Phys. Soc. Japan}, vol. 70, no. 9, pp. 2559-2567, 2001.

\bibitem{shimon}
L. L. Shimon, P. N. Volovich, M. M. Chiriban,
``Multiple ionization of samarium, europium, thulium, and ytterbium atoms by electrons,''
\textit{Sov. Phys. Tech. Phys.}, vol. 34, no. 11, pp. 1264-1266, 1989.

\bibitem{exp80}
W. Bleakney,
``Probability and critical potentials for the formation of multiply charged ions in Hg 
vapor by electron impact,''
\textit{Phys. Rev.}, vol. 35, no. 2, pp. 139-148, 1930.

%\bibitem{harrison}
%H. Harrison as presented in Ref. \cite{expKieffer}.

%\bibitem{jones}
%T. J. Jones as presented in Ref. \cite{expKieffer}.

%\bibitem{liska}
%J. W. Liska as presented in Ref. \cite{expKieffer}.
%
%\bibitem{smith2}
%P. T. Smith as presented in Ref. \cite{expKieffer}.

\bibitem{exp82M}
P. C. E. McCartney, M. B. Shah, J. Geddes, and H. B. Gilbody,
``Multiple ionization of lead by electron impact,''
\textit{J. Phys. B: At. Mol. Opt. Phys.}, vol. 31, pp. 4821-4831, 1998.

\bibitem{pavlov}
S. I. Pavlov, G. I. Stotskii,
``Single and multiple ionization of lead atoms by electrons,''
\textit{Sov. Phys. JETP}, vol. 31, no. 1, pp. 61-64, 1970.

\bibitem{beilina}
G. M. Beilina, S. I. Pavlov, V. I. Rakhovskii, O. D. Sorokaletov,
``Measurement of electron impact ionization functions for metal atoms,''
\textit{J. Appl. Mechan. Tech. Phys.}, vol. 2, pp. 86--88, 1965.

\bibitem{wareing}
J. B. Wareing, K. T. Dolder,
``A measurement of the cross section for ionization of Li$^+$ to Li$^{2+}$ by electron impact,''
\textit{Proc. Phys. Soc.}, vol. 91, pp. 887-893, 1967.

\bibitem{exp92}
J. C. Halle, H. H. Lo, and W. L. Fite,
``Ionization of uranium atoms by electron impact,''
\textit{Phys. Rev. A}, vol. 23, no. 4, pp. 1708-1716, 1981.

\bibitem{wald}
A. Wald and J. Wolfowitz, 
``An exact test for randomness in the non-parametric case, based on serial
correlation,"
\textit{Ann. Math. Stat.}, vol. 14, pp. 378-388, 1943. 

\bibitem{gof1}
G. A. P. Cirrone et al., 
``A Goodness-of-Fit Statistical Toolkit'', 
\emph{IEEE Trans. Nucl. Sci.}, vol. 51, no. 5, pp. 2056-2063, 2004.

\bibitem{gof2}
B. Mascialino, A. Pfeiffer, M. G. Pia, A. Ribon, and P. Viarengo, 
``New developments of the Goodness-of-Fit Statistical Toolkit'', 
\emph{IEEE Trans. Nucl. Sci.}, vol. 53, no. 6, pp. 3834-3841,  2006.

\bibitem{bock}
R. K. Bock and W. Krischer,
``The Data Analysis BriefBook '',
Ed. Springer, Berlin, 1998. 

\bibitem{kolmogorov1933}
A.~N.~Kolmogorov, 
``Sulla determinazione empirica di una legge di distribuzione'', 
\relax \emph{Gior. Ist. Ital. Attuari}, vol. 4, pp. 83-91, 1933.

\bibitem{smirnov1939}
N.~V.~Smirnov, 
``On the estimation of the discrepancy between
empirical curves of distributions for two independent samples'',
\relax \emph{Bull. Math. Univ. Moscou}, 1939.

\bibitem{anderson1952}
T.~W.~Anderson and D.~A.~Darling, 
``Asymptotic theory of certain
goodness of fit criteria based on stochastic processes'',
 \relax \emph{Anls. Ma. St.}, vol. 23, pp. 193-212, 1952.

\bibitem{anderson1954}
T.~W.~Anderson and D.~A.~Darling, ``A test of goodness of fit'',
\relax \emph{JASA}, vol. 49, pp. 765-769, 1954.

\bibitem{cramer1928}
H.~Cram\'er, 
``On the composition of elementary errors. Second paper: statistical applications'', 
\relax \emph{Skand. Aktuarietidskr.}, vol. 11, pp. 13-74, pp. 141-180, 1928.

\bibitem{vonmises1931}
R.~von Mises, 
``Wahrscheinliehkeitsrechnung und ihre Anwendung 
in der Statistik und theoretischen Physik'', 
Leipzig: F. Duticke, 1931.


\bibitem{fisher}
R. A. Fisher,
``On the interpretation of  $\chi^2$ from contingency tables, 
and the calculation of P'',
\textit{J. Royal Stat. Soc.}, vol. 85, no. 1, pp. 87-94, 1922. 

\bibitem{yates}
F. Yates,
``Contingency table involving small numbers and the $\chi^2$ test'',
\emph{J. Royal Stat. Soc. Suppl.}, vol. 1, pp. 217-235, 1934.

\bibitem{pearson}
K. Pearson,
``On the $\chi^2$ test of Goodness of Fit'',
\emph{Biometrika}, vol. 14, no. 1-2, pp. 186-191, 1922.

\bibitem{dmDeutsch1999} 
H. Deutsch, K. Becker, T. D. M\"ark, 
``Application of the DM formalism to the calculation of electron-impact
ionization cross sections of alkali atoms'',
\textit{Int. J. Mass Spectrom.}, vol. 185-187, pp. 319-326, 1999.

\bibitem{bray} 
I. Bray,
``Calculation of electron impact total, ionization, and nonbreakup cross sections from the 
3S and 3P states of sodium,''
\textit{Phys. Rev. Lett.}, vol. 73, no. 8, pp. 1088-1090, 1994.

\bibitem{mcguire}
E. J. Mc Guire,
``Electron ionization cross sections in the Born approximation'',
\textit{Phys. Rev. A}, vol. 16, pp. 62–72, 1977.

\end{thebibliography}
\end{document}